\documentclass[12pt]{article}
\usepackage{epsfig}
\newskip\humongous \humongous=0pt plus 1000pt minus 1000pt

\newif\ifdtup

\def\VEV#1{\left\langle #1\right\rangle}

\def\pr#1{#1^\prime}

\def\beq{\begin{equation}}
\def\eeq{\end{equation}}

\def\beqn{\begin{eqnarray}}
\def\eeqn{\end{eqnarray}}
\relax

\def\dotx{\dotx{\dot\overline{x}}}

\relax

\catcode`\@=11

\@addtoreset{equation}{section}
\def\theequation{\arabic{section}.\arabic{equation}}

\def\@normalsize{\@setsize\normalsize{15pt}\xiipt\@xiipt
\abovedisplayskip 14pt plus3pt minus3pt%
\belowdisplayskip \abovedisplayskip
\abovedisplayshortskip \z@ plus3pt%
\belowdisplayshortskip 7pt plus3.5pt minus0pt}

\def\small{\@setsize\small{13.6pt}\xipt\@xipt
\abovedisplayskip 13pt plus3pt minus3pt%
\belowdisplayskip \abovedisplayskip
\abovedisplayshortskip \z@ plus3pt%
\belowdisplayshortskip 7pt plus3.5pt minus0pt
\def\@listi{\parsep 4.5pt plus 2pt minus 1pt
     \itemsep \parsep
     \topsep 9pt plus 3pt minus 3pt}}

\@twosidetrue

\long\def\@makecaption#1#2{
 \vskip 10pt 
 \leftskip .5 cm
 \rightskip .5 cm
 \setbox\@tempboxa\hbox{#1: #2}
 \ifdim \wd\@tempboxa >\hsize \unhbox\@tempboxa\par \else \hbox
to\hsize{\hfil\box\@tempboxa\hfil} 
 \fi}

\relax

\catcode`@=12

\catcode`\@=11

\def\section{\@startsection{section}{1}{\z@}{3.5ex plus 1ex minus
   .2ex}{2.3ex plus .2ex}{\bf \sc}}
\def\subsection{\@startsection{subsection}{2}{\z@}{3.25ex plus 1ex minus
   .2ex}{1.5ex plus .2ex}{\bf \boldmath} }
\def\subsubsection{\@startsection{subsubsection}{3}{\z@}{-3.25ex\@plus -1ex
     \@minus -.2ex}{1.5ex \@plus .2ex}{\it}}

\def\thesection{\arabic{section}}

\def\appendix{\setcounter{section}{0}
 \def\thesection{APPENDIX \Alph{section}:}
 \def\theequation{\Alph{section}.\arabic{equation}}}

\def\unsetheadings{
  \def\@oddfoot{}
  \def\@evenfoot{}
  \def\@evenhead{}
  \def\@oddhead{}
}

\def\setheadings#1#2{\ps@headings{#1}{#2}}

\def\ps@headings#1#2{
  \def\@oddfoot{}
  \def\@evenfoot{}
  \def\@evenhead{
       \makebox[\textwidth]{
           \thepage{}\ \  {#2} \hfill \hfill \hfill
       }
  }
  \def\@oddhead{
       \makebox[\textwidth]{
           \hfill \hfill \hfill {#1}\ \  \thepage{}
       }
 }
 \def\subsectionmark##1{\markboth{##1}{}}
}

\catcode`\@=12

\def\figcap{\section*{Figure Captions\markboth
 {FIGURECAPTIONS}{FIGURECAPTIONS}}\list
 {Fig. \arabic{enumi}:\hfill}{\settowidth\labelwidth{Fig. 999:}
 \leftmargin\labelwidth
 \advance\leftmargin\labelsep\usecounter{enumi}}}
 \relax
\def\tablecap{\section*{Table Captions\markboth
 {TABLECAPTIONS}{TABLECAPTIONS}}\list
 {Table \arabic{enumi}:\hfill}{\settowidth\labelwidth{Table 999:}
 \leftmargin\labelwidth
 \advance\leftmargin\labelsep\usecounter{enumi}}}
 \relax
\def\reflist{\section*{References\markboth
 {REFLIST}{REFLIST}}\list
 {[\arabic{enumi}]\hfill}{\settowidth\labelwidth{[999]}
 \leftmargin\labelwidth
 \advance\leftmargin\labelsep\usecounter{enumi}}}
 \relax

\relax

\relax
\def\pl#1#2#3{{\it Phys. Lett. }{\bf #1}(19#2)#3}
\def\zp#1#2#3{{\it Z. Phys. }{\bf #1}(19#2)#3}
\def\prl#1#2#3{{\it Phys. Rev. Lett. }{\bf #1}(19#2)#3}

\def\prep#1#2#3{{\it Phys. Rep. }{\bf #1}(19#2)#3}
\def\pr#1#2#3{{\it Phys. Rev. }{\bf #1}(19#2)#3}
\def\np#1#2#3{{\it Nucl. Phys. }{\bf #1}(19#2)#3}
\def\sjnp#1#2#3{{\it Sov. J. Nucl. Phys. }{\bf #1}(19#2)#3}
\def\app#1#2#3{{\it Acta Phys. Polon. }{\bf #1}(19#2)#3}

\relax

\catcode`\@=11

\def\@mbibitem#1{\item[{[#1]}]
\if@filesw {\let \protect \noexpand \immediate \write \@auxout
{\string \bibcite {#1}{#1}}}\fi \ignorespaces}

\def\@mlbibitem[#1]#2{\item{}{[{#2}]}
\if@filesw {\let \protect \noexpand \immediate \write \@auxout
{\string \bibcite {#2}{{\the \value {\@listctr }}-{#2}}}}\fi
\ignorespaces}

\def\bibitem{\@ifnextchar [ {\@mlbibitem }{\@mbibitem }}
\catcode`\@=12

\begin{document}
\renewcommand\topfraction{1}       
\renewcommand\bottomfraction{1}    
\renewcommand\textfraction{0}      
\setcounter{topnumber}{5}          
\setcounter{bottomnumber}{5}       
\setcounter{totalnumber}{5}        
\setcounter{dbltopnumber}{2}       
\hyphenation{parametri-za-tion}
\newcommand     \mub            {\mu{\rm b}}
\newcommand     \aop            {\frac{\as}{2 \pi}}
\newcommand     \asopi          {\frac{\as}{2\pi}}
\newcommand     \asopiof[1]     {\frac{\as(#1)}{2\pi}}
\newcommand     \auno{a^{(1)}}
\newcommand     \avr[1]         {\left\langle #1 \right\rangle}
\newcommand     \azero{a^{(0)}}
\newcommand     \B              {B }
\newcommand     \ba             {\begin{eqnarray}}
\newcommand     \be             {\begin{equation}}
\newcommand     \Ca             {{C_{\rm A}}}
\newcommand     \Cf             {{C_{\rm f}}}
\newcommand     \delr           {\mbox{$\Delta R$}}
\newcommand     \dhsigm         {\frac{d\hat\sigma}{dx}}
\newcommand     \dhsigma[1]     {\frac{d\hat\sigma_{#1}}{dx}}
\newcommand     \Dms{\hat{D}}
\newcommand     \dphi           {\mbox{$\Delta \phi$}}
\newcommand     \dsig[1]        {\frac{d\sigma_{#1}}{dy dk^2_t}}
\newcommand     \dsigma         {\frac{d\sigma}{dx}}
\newcommand     \dsigmal        {\frac{d\sigma_L}{dx}}
\newcommand     \dsigmat        {\frac{d\sigma_T}{dx}}
\newcommand     \dsign[1]       {\frac{d\sigma_{#1}}{dy d^2 l_t}}
\newcommand     \ea             {\end{eqnarray}}
\newcommand     \ee             {\end{equation}}
\newcommand     \ep             {\epsilon}
\newcommand     \epbar          {\bar{\epsilon}}
\newcommand     \epem           {\ifmmode{e^+e^-}\else{$e^+e^-$}\fi}
\newcommand     \hauno          {\hat{a}^{(1)}}
\newcommand     \hsigma         {\hat\sigma}
\newcommand     \lambdamsb     {\ifmmode
\Lambda_5^{\rm \scriptscriptstyle \overline{MS}} \else
$\Lambda_5^{\rm \scriptscriptstyle \overline{MS}}$ \fi}
\newcommand     \Lambdamsb      \lambdamsb
\newcommand     \LambdaQCD     {\ifmmode
\Lambda_{\rm \scriptscriptstyle QCD} \else
$\Lambda_{\rm \scriptscriptstyle QCD}$ \fi}
\newcommand     \lfr[2]         {{\log\frac{#1}{#2}\log\log\frac{#1}{#2}}}
\newcommand     \logfr[2]       {\log\frac{#1}{#2}}
\newcommand     \lomxplus       {\left(\frac{\log(1-x)}{1-x}\right)_+}
\newcommand     \lomxrho        {\left(\frac{\log(1-x)}{1-x}\right)_\rho}
\newcommand     \MSB            {\ifmmode {\overline{\rm MS}} \else
$\overline{\rm MS}$  \fi}
\newcommand     \muF            {\mu_{\rm F}}
\newcommand     \muR            {\mu_{\rm R}}
\newcommand     \nf             {n_{\rm f}}
\newcommand     \nlf            {n_{\rm lf}}
\newcommand     \nn             {\nonumber}
\newcommand     \omxplus        {\left(\frac{1}{1-x}\right)_+}
\newcommand     \omxrho         {\left(\frac{1}{1-x}\right)_\rho}
\newcommand     \ptg            {\mbox{$p^t_{Q\bar Q}$}}
\newcommand     \ptmin     {\ifmmode p_{\scriptscriptstyle T}^{\sss min} \else
$p_{\scriptscriptstyle T}^{\sss min}$ \fi}
\newcommand     \PUD            {p_1.p_2}
\newcommand     \PUT            {p_1.p_3}
\newcommand     \PUQ            {p_1.p_4}
\newcommand     \PUC            {p_1.p_5}
\newcommand     \PDT            {p_2.p_3}
\newcommand     \PDQ            {p_2.p_4}
\newcommand     \PDC            {p_2.p_5}
\newcommand     \PTQ            {p_3.p_4}
\newcommand     \PTC            {p_3.p_5}
\newcommand     \PQC            {p_4.p_5}
\newcommand     \refq[1]        {$^{[#1]}$}
\newcommand     \ret            {\\[0.3cm]}
\newcommand     \sigmah         {\hat\sigma}
\newcommand     \sigmaq         {\hat\sigma_Q}
\newcommand     \tauo           {\tau_1}
\newcommand     \taut           {\tau_2}
\newcommand     \taux           {\tau_x}
\newcommand     \Tf             {{T_{\rm f}}}
\newcommand     \txplus         {{\left[\frac{1}{\tau_x}\right]_+}}
\newcommand     \txlog          {{\left[\frac{\log\tau_x}{\tau_x}\right]_+}}
\newcommand     \ycm            {{y_{\scriptscriptstyle CM}}}
\newcommand\addsp[1]{\raisebox{0ex}[1.25\height][1.35\depth]{#1}}
\def\sethead#1{\setheadings{\it #1}{\it Heavy-Quark Production}}
\newcommand\smallcite[1]{\scriptsize \cite{#1}}
\newcommand{\ccaption}[2]{
\begin{center}
\parbox{0.85\textwidth}{
\caption[#1]{\small{\it{#2}}}
}
\end{center}
}
\newcommand\sss{\scriptscriptstyle}
\newcommand\mug{\mu_\gamma}
\newcommand\mue{\mu_e}
\newcommand\muf{\mu_{\sss F}}
\newcommand\mur{\mu_{\sss R}}
\newcommand\muo{\mu_0}
\newcommand\muww{\mu_{\sss WW}}
\newcommand\me{m_e}
\newcommand\as{\alpha_{\sss S}}
\newcommand\epb{\overline{\epsilon}}
\newcommand\aem{\alpha_{\rm em}}
\newcommand\qqb{{q\overline{q}}}
\newcommand\qb{\overline{q}}
\newcommand\qq{{\scriptscriptstyle Q\overline{Q}}}
\newcommand\DIG{{\rm DIS}_\gamma}
\newcommand\jetmsrex{dE_{\sss JT} d\eta_{\sss J}
d\Omega_{\sss J}^{(2-2\ep)}}
\newcommand\jetmsr{d\mu_{\sss J}}
\newcommand\jetmsrf{dE_{\sss JT} d\eta_{\sss J} d\varphi_{\sss J}}
\newcommand\EJT{E_{\sss JT}}
\newcommand\mq{\mbox{$m_{\sss \rm Q}$}}
\newcommand\Stwo{{\cal S}_2}
\newcommand\Sthree{{\cal S}_3}
\newcommand\Stilde{\widetilde{{\cal S}}_3}
\newcommand\Qb{\overline{Q}}
\newcommand\Js{{\sss J}}
\newcommand\Th{\theta}
\def \asopi{\mbox{$\frac{\alpha_s}{\pi}$}}
\def \oacube {\mbox{${\cal O}(\alpha_s^3)$}}
\def \oatwo {\mbox{${\cal O}(\alpha_s^2)$}}
\def \oas   {\mbox{${\cal O}(\alpha_s)$}}
\def \ppbar {\mbox{$p \bar p$}}
\def \bbbar {\mbox{$b \bar b$}}
\def \ccbar {\mbox{$c \bar c$}}
\def \pt   {\mbox{$p_{\scriptscriptstyle T}$}}
\newcommand\xf{x_{\scriptscriptstyle F}}
\def \et   {\mbox{$E_{\scriptscriptstyle T}$}}
\def \etsq {\mbox{$E_{\scriptscriptstyle T}^2$}}
\def \kt   {\mbox{$k_{\scriptscriptstyle T}$}}
\def \rap   {\mbox{$\eta$}}
\def \deltar {\mbox{$R$}}
\def \to   {\mbox{$\rightarrow$}}
\def    \mb             {\mbox{$m_b$}}
\def    \mc             {\mbox{$m_c$}}
\newcommand\gsim{\mathop{\mbox{\vbox{\hbox{$>$} \vskip -9pt \hbox{$\sim$}
\vskip -3pt  }}}}
%
\newcommand\ssmallfig{3cm}
\newcommand\smallfig{4cm}
\newcommand\mediumfig{5cm}
\newcommand\bigfig{6cm}
\newcommand\bbigfig{9cm}
\normalbaselineskip=14pt plus 2pt minus 2pt
\baselineskip=\normalbaselineskip
\nopagebreak
{\flushright{
        \begin{minipage}{4cm}
        CERN-TH/97-16  \hfill \\
        hep-ph/9702287\hfill \\
        \end{minipage}        }

}
\begin{center}
\vskip 1in
        { \Large  \bf \sc
        Heavy-Quark Production}\footnote{
        To appear in ``Heavy Flavours II'', eds. A.J. Buras and
        M. Lindner, Advanced Series on Directions in High Energy Physics,
        World Scientific Publishing Co.,
        Singapore}
\vskip 0.3in
{\bf Stefano FRIXIONE\footnote{Supported by the
  Swiss National Foundation}}

ETH, Zurich, Switzerland
\vskip .5cm
{\bf Michelangelo
L. MANGANO\footnote{On leave of absence from INFN, Pisa, Italy}
{\normalsize and}
Paolo NASON\footnote{On leave of absence from INFN, Milan, Italy}}

CERN, TH Division, Geneva, Switzerland

\vskip 0.3cm
{\bf Giovanni RIDOLFI}

INFN, Genoa, Italy
\vskip .3cm
\begin{abstract}
We review the present theoretical and experimental status of heavy quark
production in high-energy collisions. In particular, we cover hadro- and
photoproduction at fixed target experiments, at HERA and at the hadron
colliders, as well as aspects of heavy quark production in $e^+e^-$ collisions
at the $Z^0$ peak.
\end{abstract}
\vskip .6cm
\end{center}
\vskip 1cm
\vfill
CERN-TH/97-16  \hfill \\
February 1997 \hfill
\newpage

\tableofcontents
\section{Introduction}
\sethead{Introduction}

The theory of heavy-flavour production covers in reality a wide
variety of phenomena, which are at times only loosely related from a
theoretical point of view. In ref.~\cite{Nason92} an ample discussion
was given of the next-to-leading order (NLO) correction to the
hadroproduction of heavy flavours for the total and single-inclusive
cross section and for the double-differential distributions.  Both
hadroproduction and photoproduction have been considered in the
literature, and NLO calculations of heavy-flavour
production in deep-inelastic scattering have been available for a long
time.  Some very special topics, such as the multiplicity of heavy
flavours in jets, have also been discussed in the previous volume.

In this chapter we will discuss and review recent progress that took place
since the publication of
ref.~\cite{Nason92}. Many theoretical and experimental issues on heavy-flavour
production have been clarified, and new problems have arisen.
First of all, the top quark has been observed, 
and its cross section turns out to have roughly the predicted magnitude.
In the future, both at the Tevatron and at the LHC, it will be possible
to study more details of the $t\bar{t}$ production mechanism.
From the point of view of perturbative QCD this is a very interesting
possibility, since theoretical predictions for the top are very reliable.
This is in fact the only case in heavy-flavour production when
radiative corrections to the total cross section
are indeed small. They are of the order of
15\% at the Tevatron energy. The first studies on kinematical distributions
of the top have already appeared, showing a qualitative agreement with
QCD predictions. We will review in this work the present status
of the theoretical calculation of top cross section and distributions.
The importance of these studies is twofold. On the one hand one would like
to test the underlying strong interaction dynamics, and on the other hand,
deviations from QCD predictions may hint at the effects of
physics beyond the Standard Model.

The study of bottom production at the Tevatron has undergone a
considerable improvement, thanks to the introduction of vertexing
techniques.  The inclusive transverse-momentum distribution of $b$
mesons has been a long-standing problem, since the measured cross
section used to be several times higher than the theoretical
calculation.  Experimental refinements of the measurement have reduced
the cross section by a considerable amount. Nevertheless, the data
still tend to lie on the upper side of the theoretical prediction.
Although it is clear by now that the hard QCD production mechanism is
basically the correct one, it appears that some details remain to be
understood. Vertexing techniques have also allowed additional
measurements of correlations between the $b$ and the $\bar{b}$ quark,
adding evidence for the validity of the hard production mechanism.

Fixed-target studies of heavy-flavour production have provided a
wealth of data on charm. Total cross sections, single-inclusive
distributions, correlations between the quark and the antiquark have
been measured in both hadro- and photoproduction.  The theoretical
apparatus of perturbative QCD is in this case at its very limit of
applicability, because the charm mass is very close to
typical hadronic scales. Thus, effects of non-perturbative origin
will very likely play an important role. Conversely, it is hoped that
these effects may be better understood by studying charm production.
Although modern fixed-target experiments have considerably improved
the situation, many open problems remain in this field. All
experimental results are in qualitative agreement with perturbative
QCD calculations, thus supporting the ``hard'' nature of
charm-production phenomena. However, several quantitative deviations from
pure QCD are observed.  It is interesting to see whether simple models
of non-perturbative phenomena, such as fragmentation effects, intrinsic
transverse momenta, and dragging effects, may be sufficient to
explain these deviations. We will discuss these problems at length.
As of now, in our opinion, several different scenarios are possible,
and more experimental data may help to discriminate between them.

The electron--proton collider HERA has begun to produce data
on charm photoproduction and electroproduction. In this case,
thanks to the wide kinematical range potentially available,
the possibility exists of performing QCD studies on the
proton and photon structure functions.

Bottom and charm production studies at LEP have reached a remarkable
stage of precision. Accurate measurements are now available for the
$b$ production rate and for its fragmentation function. Furthermore,
measurements of the gluon splitting rate into $b$ and $c$ pairs have
been performed.

On the theoretical side, new developments have taken place, in
relation to the experimental progress that has been achieved. Thus,
top cross section calculations have been revisited, and the problem of
threshold effects has been re-examined and considerably clarified.  The
resummation of perturbative effects at large transverse momentum has
been investigated, because of its relevance to the interpretation of
the CDF and D0 data.  With a similar motivation, production of
heavy-flavoured jets has been considered.  Lastly, the computation of
heavy-flavour electroproduction has been extended to next-to-leading
order, in view of comparisons with HERA data.

We have limited the scope of the present work to open heavy-flavour
production. The problem of quarkonium production is of a rather
different nature, not directly related to the topics touched in our
review.  We have tried to be as complete as possible. However, we have
skipped topics that are, in our opinion, less important for the purpose
of testing the underlying production dynamics.

\section{Fixed-target production}\label{sec:ft}
\sethead{Fixed-target production}

Heavy-flavour production has been studied extensively in fixed-target
experiments, with both hadron and photon beams. The typical
centre-of-mass energy is in the range 10-40~GeV, where the bottom
cross section is rather small. Therefore most of the available data
are on charmed-hadron production.

In this section we will present a comparison between existing fixed-target
data for the hadroproduction and the photoproduction of heavy flavours
and NLO QCD predictions.
Because of the small value of the charm-quark mass, the perturbative
expansion may not be reliable, due to higher-order terms of the perturbative
expansion and to non-perturbative effects. We will discuss these problems
and the importance of such contributions.
More details are given in refs.~\cite{Mangano93} and \cite{Frixione94a}.
\subsection{Total cross sections}

A list of experimental results on total cross sections
for $D$-meson hadroproduction is presented
in table~\ref{totexp}. We used these results to estimate
the total charm-pair cross section $\sigma_{c\bar{c}}$.
\begin{table}
\begin{center}
\begin{tabular}{|l||c|c|c|c|} \hline
       & $E_b$ & $\sigma$ (total) or $\sigma_+$ ($x_{\sss F}>0$)
       & $\sigma_{D\bar{D}}$ & $D^+/D^0$
\\
      & (GeV) & ($\mu$b) & ($\mu$b)            &
\\ \hline\hline
$pN$  & 800
    & $\sigma(D^0/\bar{D}^0)=38\pm 3\pm 13 $ & $38\pm 10$ & $1.0\pm 0.6$
\\
\smallcite{Kodama91}   && $\sigma(D^+/D^-)=38\pm 9\pm 14$ & &
\\ \hline
$pN$  & 200 &
$ \sigma_+(D/\bar{D})=1.5\pm 0.7\pm 0.1$ & $1.5\pm 0.7$ & --- \\
\smallcite{Barlag88} &&&&
\\ \hline
 $pp$ & 800
    & $\sigma(D^0/\bar{D}^0)=22^{+9}_{-7}\pm 5$ & $24\pm 6$ & $1.2\pm 0.6$
\\
\smallcite{Ammar88}              &     & $\sigma(D^+/D^-)=26\pm 4\pm 6$ & &
\\ \hline
$pp$  & 400
    & $\sigma(D^0/\bar{D}^0)=18.3\pm 2.5$ & $15.1 \pm 1.5 $ & $0.7\pm 0.1$
\\
 \smallcite{Aguilar88}           &     & $\sigma(D^+/D^-)=11.9\pm 1.5$ &  &
\\ \hline
$pN$ & 250
    & $\sigma_+(D^0/\bar{D}^0)=5.6\pm 1.3\pm 1.5$ &
    $8.8 \pm 1.5 $ & $0.57\pm 0.22$
\\
\smallcite{Alves96} &     & $\sigma_+(D^+/D^-)=3.2\pm 0.4\pm 0.3 $ &  &
\\ \hline
$\pi^-p$ & 360
    & $\sigma_+(D^0/\bar{D}^0)=10.1\pm 2.2$ & $12.6\pm 2.2$
    & $0.6\pm 0.2$
\\
  \smallcite{Aguilar85a}   &     & $\sigma_+(D^+/D^-)=5.7\pm 1.6$ & &
\\ \hline
$\pi^-N$ & 230 &
$ \sigma_+(D^0/\bar{D}^0)=6.3\pm 0.3\pm 1.2$ & $7.6\pm 1.1$
& $0.5\pm 0.2$
\\
\smallcite{Barlag91}  &     & $\sigma_+(D^+/D^-)=3.2\pm 0.2\pm 0.7$ & &
\\ \hline
$\pi^-N$  & 200 &
$ \sigma_+(D^0/\bar{D}^0)=3.4^{+0.5}_{-0.4}\pm 0.3$ &
   $4.1^{+0.6}_{-0.5}$ & $ 0.5\pm 0.1$
\\
\smallcite{Barlag88}       &     &
      $\sigma_+(D^+/D^-)=1.7^{+0.4}_{-0.3}\pm 0.1$ & &
\\ \hline
$\pi^-N$ & 600 &
$ \sigma_+(D^0/\bar{D}^0)=22.05\pm 1.37\pm 4.82$ & $24.6\pm 4.3$
& $0.4\pm 0.1$
\\
\smallcite{Kodama92}       &     &
      $\sigma_+(D^+/D^-)=8.66\pm 0.46\pm 1.96$ & &
\\ \hline
$\pi^-N$ & 210 &
$ \sigma_+(D^0/\bar{D}^0)=6.3\pm 0.9\pm 0.3$ & $6.4\pm 0.8$
& $0.27\pm 0.06$
\\
\smallcite{Alves96}    &     &
      $\sigma_+(D^+/D^-)=1.7\pm 0.3\pm 0.1$ & &
\\ \hline
$\pi^-N$ & 250 &
$ \sigma_+(D^0/\bar{D}^0)=8.2\pm 0.7\pm 0.5$ & $9.4\pm 0.7$
& $0.44\pm 0.06$
\\
\smallcite{Alves96}    &     &
      $\sigma_+(D^+/D^-)=3.6\pm 0.2\pm 0.2$ & &
\\ \hline
$\pi^+N$ & 250 &
$ \sigma_+(D^0/\bar{D}^0)=5.7\pm 0.8\pm 0.4$ & $6.6\pm 0.8$
& $0.46\pm 0.09$
\\
\smallcite{Alves96}    &     &
      $\sigma_+(D^+/D^-)=2.6\pm 0.3\pm 0.2$ & &
\\ \hline
$\pi^-N$ & 350 &
$ \sigma_+(D^0/\bar{D}^0)=7.78\pm 0.14\pm 0.52$ & $8.8\pm 0.5$
& $0.42\pm 0.05$
\\
\smallcite{Adamovich96}  & & $\sigma_+(D^+/D^-)=3.28\pm 0.08\pm 0.29$ & &
\\ \hline
\end{tabular}
\caption{\label{totexp}\it
Experimental results on total cross sections for charm production.
The notation $\sigma_+$ represents the inclusive cross section for
positive rapidity. The pair cross section $\sigma_{D\bar{D}}$ includes
our corrections for possible $x_{\sss F}$ cuts, as discussed in the text.
The experimental results have not been
corrected for the updated $D\to K\pi$ branching ratios \cite{Barnett96}.
}
\end{center}
\end{table}
In most cases fixed-target experiments give their cross sections
with the Feynman-$x$ cut $x_{\sss F}>0$.
The ratio $\sigma_{c\bar{c}}/\sigma_{c\bar{c}}(x^{c}_{\sss F}>0)$
has been evaluated theoretically \cite{Mangano93}
and is approximately equal to $1.6$ in pion--nucleon collisions
and to $2$ in proton--nucleon collisions; it is
nearly independent of the heavy-quark mass and beam energy
(at least for $m_c$ between 1.2 and 1.8 GeV, and beam energy $E_b$
between 100 and 1000 GeV).
Therefore, the total $D\bar{D}$ cross section is obtained from
$\sigma\left(D/\bar{D},x_{\sss F}>0\right)$ by dividing
by 2 to get the pair cross section from the single-inclusive
one, and multiplying by 1.6 (for pion) or 2 (for proton) to account
for the partial coverage of the $x_{\sss F}$-range.

The contribution to $\sigma_{c\bar{c}}$
from $\Lambda_c$ and $D_s$ production must also be included.
We use (see ref.~\cite{Aoki92} and references therein)
\beqn
\label{ratiods}
\frac{\sigma\left(D_s\right)}{\sigma\left(D^0+D^+\right)}&\simeq&0.2\,,
 \\
\label{ratiolc}
\frac{\sigma\left(\Lambda_c\right)}{\sigma\left(D^0+D^+\right)}&\simeq&0.3
\eeqn
(here particle means also antiparticle).
Measurements of the cross sections for $D_s$ and $\Lambda_c$
have recently been presented
by the E769~\cite{Alves96} and WA92~\cite{Adamovich96} collaborations,
and values consistent with
eqs.~(\ref{ratiods}) and (\ref{ratiolc}) have been found.
Therefore, to obtain
$\sigma_{c\bar{c}}$ from the  total cross section for $D\bar{D}$
production, we have to multiply by a factor of $1.5$.

The WA75 collaboration \cite{Aoki92} does not present single-inclusive
cross sections for $D$ mesons, but quotes directly the $c\bar{c}$
cross section
\beq
\sigma_{c\bar{c}}\,=\,23.1\pm 1.3^{+4.0}_{-3.3}\,\mub\,,
\eeq
for a 350~GeV $\pi^-$ beam on emulsion,
assuming an atomic mass dependence $A^{0.87}$
(this result was not inserted in
table~\ref{totexp}). The E789 col\-la\-bo\-ra\-tion \cite{Leitch94},
using an 800~GeV proton beam colliding on a Be or Au target, studied
the nuclear dependence of the $D$-meson cross section, finding an
$A^\alpha$ behaviour, with $\alpha=1.02\pm 0.03\pm 0.02$.
As a by-product, they obtain
\beq
\sigma\left(D^0/\bar{D}^0\right)\,=\,17.7\pm 0.9\pm 3.4\,\mub\,.
\eeq
An analogous measurement for the charged-$D$ cross section is not
expected\footnote{M.~Schub, private communication.}.

From table~\ref{totexp} we see that most of the experimental collaborations
report separately the total cross sections for charged and
neutral $D$-meson production. In the last column of table~\ref{totexp}
we show the ratio $R$ between these two values.
For pion--nucleon collisions, the agreement between various collaborations
is fairly good (except for the result of ref.~\cite{Alves96} obtained with
a beam energy of 210~GeV), and the errors on $R$ are moderate; the $D^0$
cross section is found to be about twice as large as the $D^+$ cross section.
A simple model for estimating the charged-to-neutral $D$ cross section ratio
is the following.
One assumes isospin invariance in the $c\to D$ and $c\to D^*$ transition.
Furthermore, one assumes
that the $D$ cross section is one third
of the $D^*$ cross section, due to the counting
of polarization states. Using then the published values of the
$D^* \to D$ branching ratios \cite{Barnett96}, the result is roughly
\beq
\label{DDratio}
R\equiv\frac{\sigma(D^+)}{\sigma(D^0)}\simeq 0.32.
\eeq
Observe that this number has changed with respect to the value of 0.43
quoted in ref.~\cite{Frixione94}, because of changes in the measured
branching ratios.
This number is compatible with the values reported in
table~\ref{totexp} for $\pi N$ collisions, especially for the most recent data.
For $pN$ collisions, the same simple
argument should hold. Experimental measurements do not give
a unique indication; some results seem to give comparable total
cross sections for charged and neutral $D$. We do not find any
reasonable explanation of this fact.
For example, the HERWIG Monte Carlo program \cite{Marchesini88,Marchesini92}
gives roughly the same $R$ value for the proton and for the pion beams.
On the other hand, the experimental data for the proton beam are
much less clear than in the case of $\pi^- N$ collisions, the uncertainty
on $R$ being quite large. The two $R$ values with the smallest
uncertainty, namely those reported in ref.~\cite{Aguilar88} and in
ref.~\cite{Alves96}, indicate a behaviour similar to the one observed
in $\pi^- N$ collisions.

\begin{figure}[ht]
\begin{center}
\mbox{\psfig{file=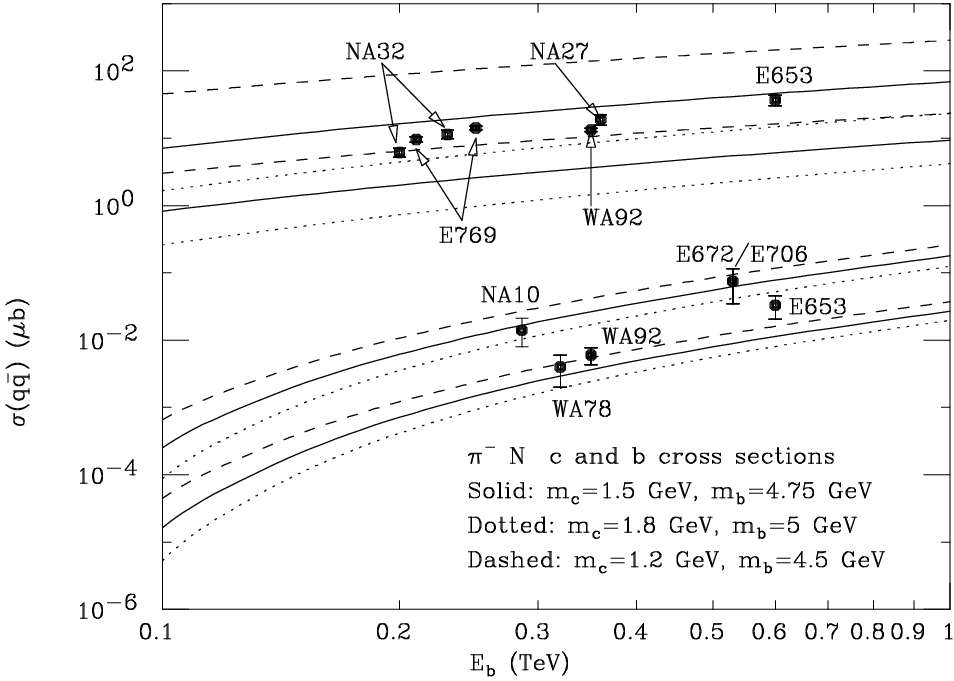,width=0.70\textwidth}}
\ccaption{}{\label{bcpion}
Pair cross sections for $b$ and $c$ production in $\pi^-N$
collisions versus experimental results.
}
\end{center}
\end{figure}
\begin{figure}[htb]
\begin{center}
\mbox{\psfig{file=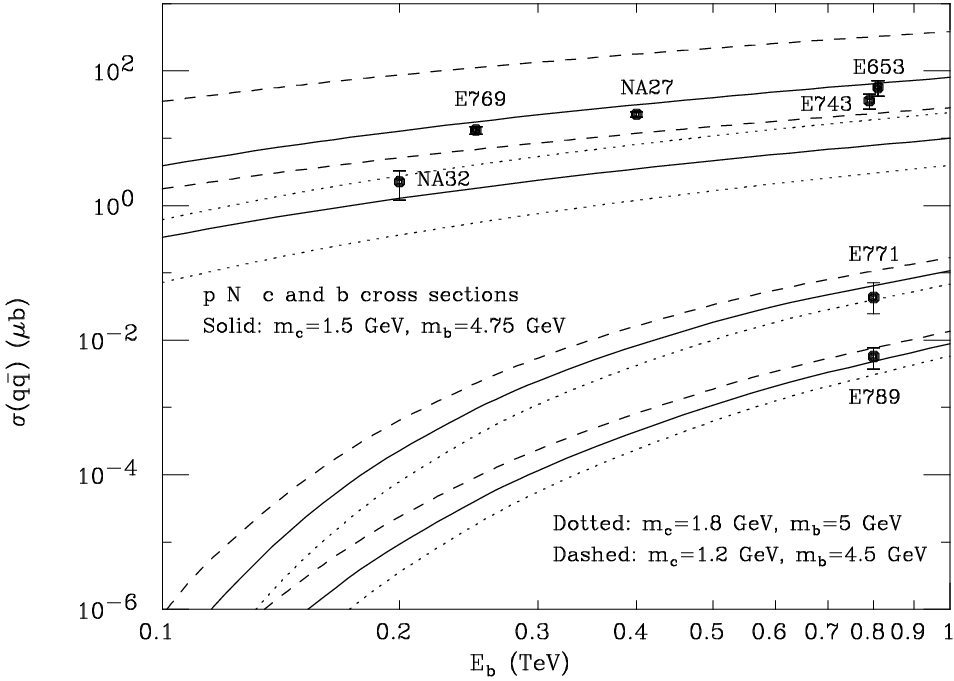,width=0.70\textwidth}}
\ccaption{}{\label{bcproton}
Pair cross sections for $b$ and $c$ production in $p N$ collisions versus
experimental results.
}
\end{center}
\end{figure}

In fig.~\ref{bcpion} we plot the $c\bar{c}$ and $b\bar{b}$ cross sections,
computed in QCD at NLO, as functions of the beam energy, for
$\pi^- N$ collisions.
The same quantities are shown in fig.~\ref{bcproton} for a proton beam.
The cross sections are calculated using the parton distribution sets
of ref.~\cite{Martin95a} for the nucleon, and the central set SMRS2
\cite{Sutton92} for the pion. The default values of the charm and bottom masses
are 1.5 and 4.75 GeV respectively, and the default choices for the
factorization scale $\muf$ and the renormalization scale $\mur$ are
\beq
\muf=2m_c,\quad \mur=m_c
\eeq
for charm and
\beq
\muf=\mur=m_b
\eeq
for bottom.

The bands in the figures are obtained as follows. We varied $\mur$ between half
the central value and twice this value. The factorization scale $\muf$ was also
varied between $m_b/2$ and $2m_b$ in the case of bottom,
while it was kept fixed
at $2m_c$ in the case of charm. This is because the adopted parametrizations of
parton densities are given for $Q^2$ larger than 5 GeV$^2$.
The bands shown in the figures are therefore
only an underestimate of the uncertainties involved in the computation of charm
production cross sections. We verified that considering independent variations
for the factorization and renormalization scales does not lead to a wider range
in the bottom cross section for the energies shown in the figures. We also show
the effect of varying $m_c$ between $1.2$ GeV and $1.8$ GeV, and $m_b$ between
4.5 and 5 GeV.

The proton parton densities of ref.~\cite{Martin95a} are available for a wide
range of $\LambdaQCD$ values, corresponding to
$\as(m_{\rm \sss Z})$
values between $0.105$ and $0.130$. The bands shown in figs.~\ref{bcpion} and
\ref{bcproton} for bottom production are obtained by letting $\LambdaQCD$
vary in this range. In the case of charm, values of $\LambdaQCD$
corresponding to $\as(m_{\sss Z})$ above 0.115 induce values of $\as(m_c)$ too
large to be used in a perturbative expansion. For this reason, the upper bounds
on charm production cross sections are obtained with $\as(m_{\sss Z})=0.115$.
We point out that, by varying $\LambdaQCD$, one is
forced to neglect the correlation between $\LambdaQCD$
and the pion parton densities, which were fitted in ref.~\cite{Sutton92} with
$\Lambdamsb=122$~MeV.

Experimental results on bottom production at fixed target have been reported in
refs.~\cite{Bari91,Basile81,Kodama93,Catanesi89,Bordalo88,Jesik95} for
pion--nucleon collisions.
Recently, the first measurement of the bottom cross section with a
proton beam has become available \cite{Jansen95}. Preliminary results
for $pN$ collisions have also been presented in ref.~\cite{Spiegel96}.
These results are shown in figs.~\ref{bcpion} and \ref{bcproton}.
We made no effort to
correct the data in order to get the $b\bar{b}$-quark cross section when
the $B$-hadron cross section was reported.

As can be seen, experimental results on total charm cross sections
are in reasonable agreement with theoretical
expectations, if the large theoretical uncertainties are taken into proper
account. We can see that the hadroproduction data are compatible with a value
of 1.5 GeV for the charm-quark mass.
In the case of bottom production, the spread of the experimental data
is almost as large as that of the theoretical predictions.
Both currently available data points lie in the theoretical band.

We remind the reader that many puzzling ISR
results in $pp$ collisions at 62~GeV cannot be currently explained
(see the review
in ref.~\cite{Tavernier87}), in particular the large $\Lambda_b$ production
rates reported in refs.~\cite{Bari91,Basile81}.

\begin{figure}[htb]
\begin{center}
\mbox{\psfig{file=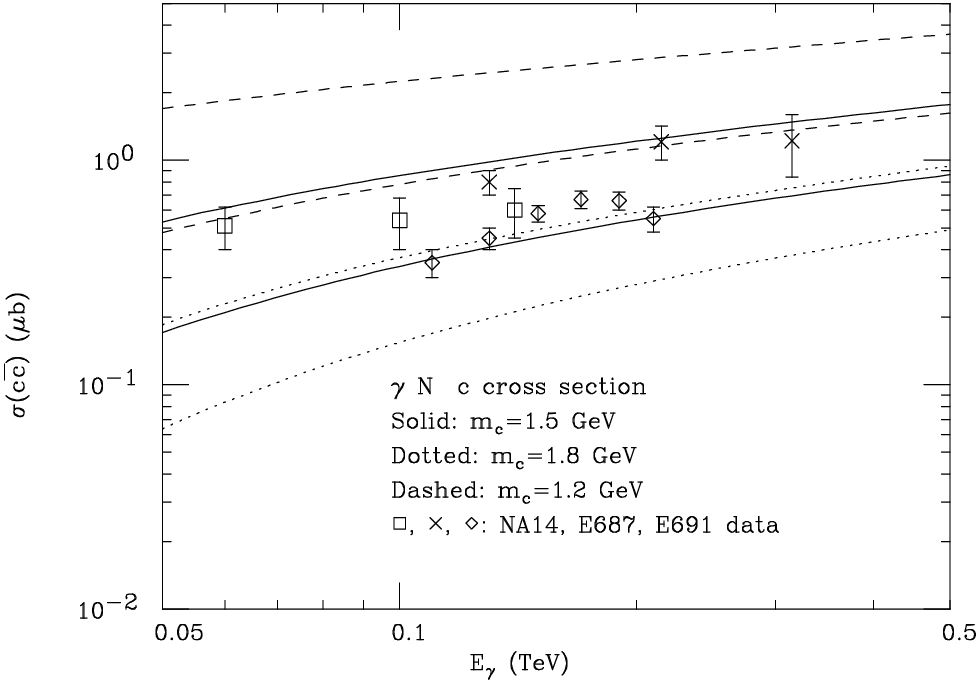,width=0.70\textwidth}}
\ccaption{}{\label{totgamma}
Pair cross sections for $c$ production in $\gamma N$ collisions versus
experimental results.
}
\end{center}
\end{figure}
Total cross sections for charm production have also been measured in
photoproduction experiments. In fig.~\ref{totgamma} the relevant experimental
results of refs.~\cite{Alvarez93,Anjos89,Anjos90,Bellini94} are shown in
comparison with NLO QCD predictions.
As can be seen, the theoretical uncertainties are smaller in this
case than in the hadroproduction case. Again, a charm mass of 1.5~GeV
is compatible with photoproduction data. It should be stressed
that some of the experimental results are totally incompatible
with one another. Until these discrepancies are resolved, it will not
be possible to use the data to constrain physical parameters.
For example, while the E687 data are inconsistent with a charm mass of
1.8~GeV, this mass value cannot be excluded because of the E691 data.
\subsection{Single-inclusive distributions}
\label{single}
Many hadroproduction and photoproduction experiments have measured
single-inclusive $x_{\sss F}$ and $\pt$ distributions for charmed hadrons
in $\pi N$ collisions \cite{Kodama92,Barlag88,Barlag91,Aguilar85a,
Alves92,Adamovich92,Aoki88,Aoki92a} and in $pN$ collisions
\cite{Kodama91,Barlag88,Ammar88,Aguilar88,Adamovich92}.
Distributions are expected to be more
affected by non-perturbative phenomena than total cross sections.
For example, an intrinsic transverse momentum of the incoming
partons, and the hadronization of the produced charm quarks, may
play an important r\^ole in this case. We will therefore try to
assess the impact of such phenomena by means of simple models.

Recent measurements of single-inclusive differential cross sections have
been performed at CERN by the WA92 collaboration~\cite{Adamovich96}, which
uses a $\pi^-$ beam of 350~GeV colliding with isosinglet nuclei, and at
FNAL by the E769 collaboration~\cite{Alves96a} with pion, proton and kaon
beams of 250~GeV on isosinglet targets.

\begin{figure}
\centerline{\epsfig{figure=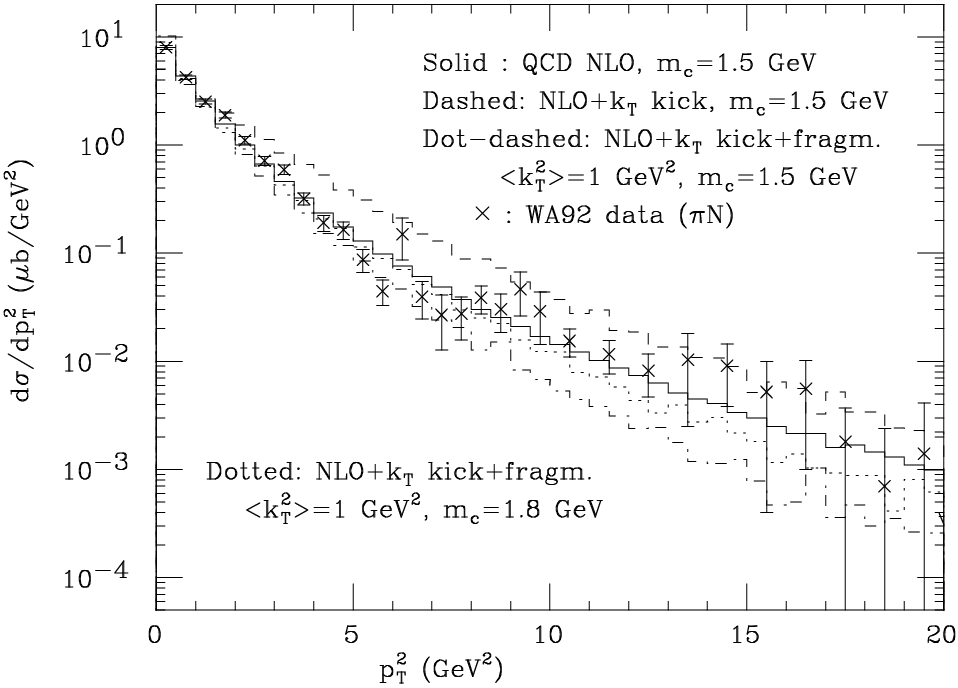,width=0.5\textwidth,clip=}
            \hspace{0.3cm}
            \epsfig{figure=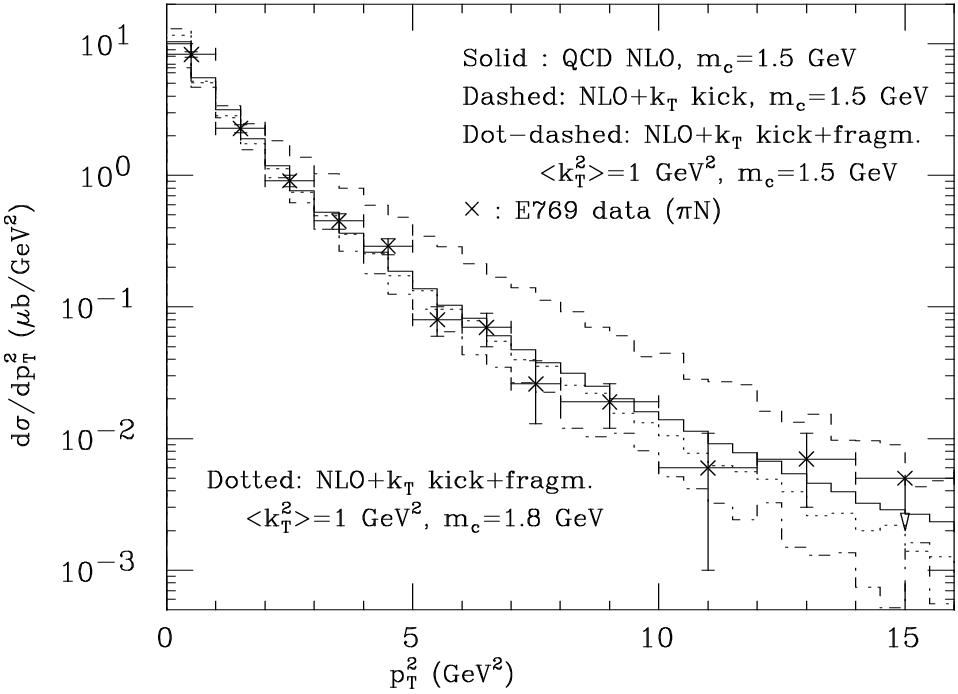,width=0.5\textwidth,clip=}}
\ccaption{}{ \label{hdrpt}
The single-inclusive $p_{\sss T}^2$ distribution measured by WA92 (left)
and E769 (right), compared to the NLO QCD predictions,
with and without the inclusion of non-perturbative effects.}
\end{figure}

In fig.~\ref{hdrpt} we show the comparison between the single-inclusive
$p_{\sss T}^2$ distributions measured by the WA92 and E769
collaborations in $\pi N$ collisions and the theoretical
predictions\footnote{In this subsection and in the following one, the relative
normalization of experimental distributions and theoretical curves
has been fixed in order to give the same total rate.}.
The solid curves represent the pure NLO QCD predictions
for charm quarks.
The dashed curves show the effect of adding to the perturbative results
an intrinsic transverse momentum of the incoming partons
($k_{\sss T}$ kick). This procedure is, to a large extent,
arbitrary. We implemented it in the following way.
We call $\vec{p}_{\sss T}(Q\overline{Q})$ the total transverse momentum
of the pair.
For each event, in the longitudinal centre-of-mass
frame of the heavy-quark pair, we boost the $Q\overline{Q}$
system to rest.
We then perform a second transverse boost, which gives
the pair a transverse momentum equal to
$\vec{p}_{\sss T}(Q\overline{Q})+ \vec{k}_{\sss T}(1)+\vec{k}_{\sss T}(2)$;
$\vec{k}_{\sss T}(1)$ and $\vec{k}_{\sss T}(2)$
are the transverse momenta of the incoming partons, which are
chosen randomly, with their moduli distributed according to
\beq
\frac{1}{N}\frac{dN}{dk_{\sss T}^2}=\frac{1}{\langle k_{\sss T}^2 \rangle}
      \exp(-k_{\sss T}^2/\langle k_{\sss T}^2 \rangle).
\eeq
Alternatively, one may
proceed as in the previous case, but giving the additional transverse
momentum $\vec{k}_{\sss T}(1)+\vec{k}_{\sss T}(2)$ to the whole final-state
system (at the NLO in QCD, this means the $Q\overline{Q}$ pair
plus a light parton), and not to the $Q\overline{Q}$ pair only.
We verified that the two methods give very similar results for
$\langle k_{\sss T}^2 \rangle$ smaller than about 2~GeV$^2$.

Another non-perturbative effect that must be accounted for is the hadronization
process. This effect can be described by
convoluting the partonic cross section with a fragmentation function, for
which we choose the parametrization proposed in ref.~\cite{Peterson83}:
\beq\label{Peterson-form}
D(x)\,=\,\frac{N}{x\left[1-1/x-\epsilon/(1-x)\right]^2},
\eeq
where $N$ is fixed by the condition $\int D(x) dx=1$.
We used $\epsilon=0.06$, which is the central value quoted in
ref.~\cite{Chrin87} for $D$ mesons. The fragmentation process degrades
the parent charm-quark momentum, and thus softens the
$p_{\sss T}^2$ distribution.

From fig.~\ref{hdrpt}, we see that the effect of the $k_{\sss T}$ kick
on the predictions for bare quarks results in a hardening of the
$p_{\sss T}^2$ spectrum, overshooting the data. On the other
hand, by combining the $k_{\sss T}$ kick with
$\langle k_{\sss T}^2 \rangle =1$ GeV$^2$ and the Peterson fragmentation,
the theoretical predictions undershoot the data (dot-dashed curves),
although no serious inconsistency can be inferred given the size
of the experimental uncertainty.
Taking the data at face value, one can notice that larger values
for $\langle k_{\sss T}^2\rangle$ and for $m_c$ would improve the
agreement.
We have checked that, in order to reproduce the WA92 and E769 data,
an average intrinsic transverse momentum
$\langle k_{\sss T}^2\rangle = 2\;$GeV$^2$ is needed.
This value is exceedingly large with respect to the typical
hadronic scale of a few hundred MeV.
On the other hand, the data are well described (dotted plots) by the
theoretical curves including fragmentation and $k_{\sss T}$ kick with
$\langle k_{\sss T}^2 \rangle =1$ GeV$^2$, if the larger value
of $m_c=1.8$~GeV is adopted.

We conclude that, when using the
central value of the charm mass, $m_c=1.5$~GeV, favoured by
the total cross section measurements, the theoretical
results for the $p_{\sss T}^2$ spectrum can
describe the data well, only if a large $k_{\sss T}$ kick
is applied to the fragmented curve. If a larger value of the charm
mass is adopted, a more moderate and physically acceptable
$k_{\sss T}$ kick is enough to get a good agreement with the
measurements.
Therefore, there seems to be a potential discrepancy between theory and
experiments in the $p_{\sss T}^2$ spectrum in charm hadroproduction.
From a different point of view, however, the discrepancy may be interpreted
as the signal that some of the other theoretical assumptions are not
totally sound. For example, the Peterson fragmentation function
may not be suitable to describe the hadronization process in hadronic
collisions; the data would suggest a function more peaked
towards the $x\simeq 1$ region. Moreover, higher-order
perturbative corrections may also play a r\^ole, especially in the
low-$p_{\sss T}$ region.

\begin{figure}
\centerline{\epsfig{figure=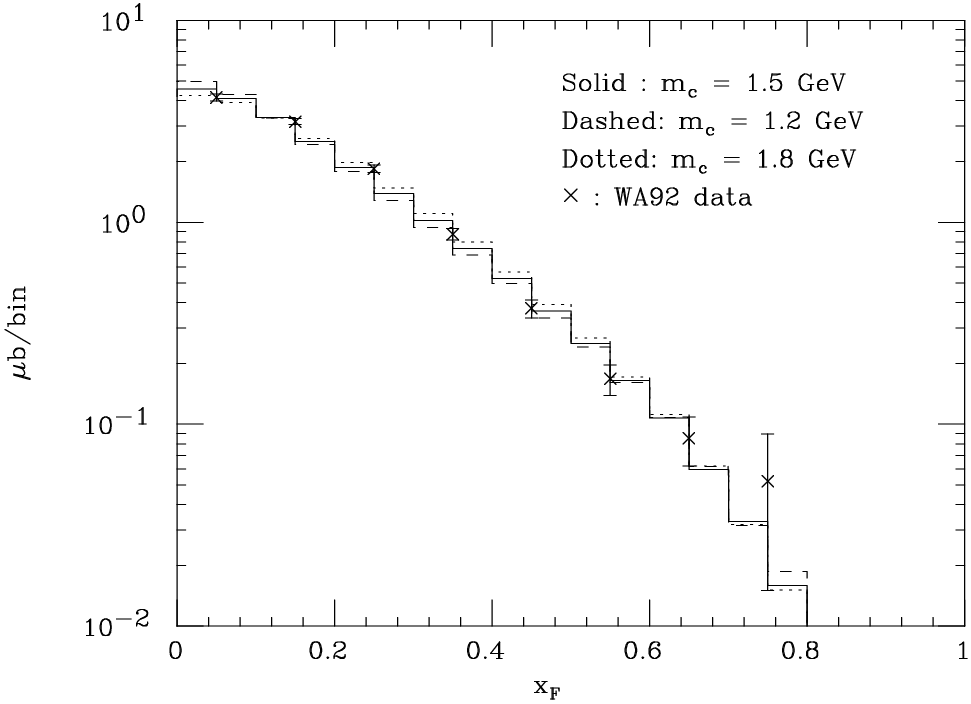,width=0.5\textwidth,clip=}
            \hspace{0.3cm}
            \epsfig{figure=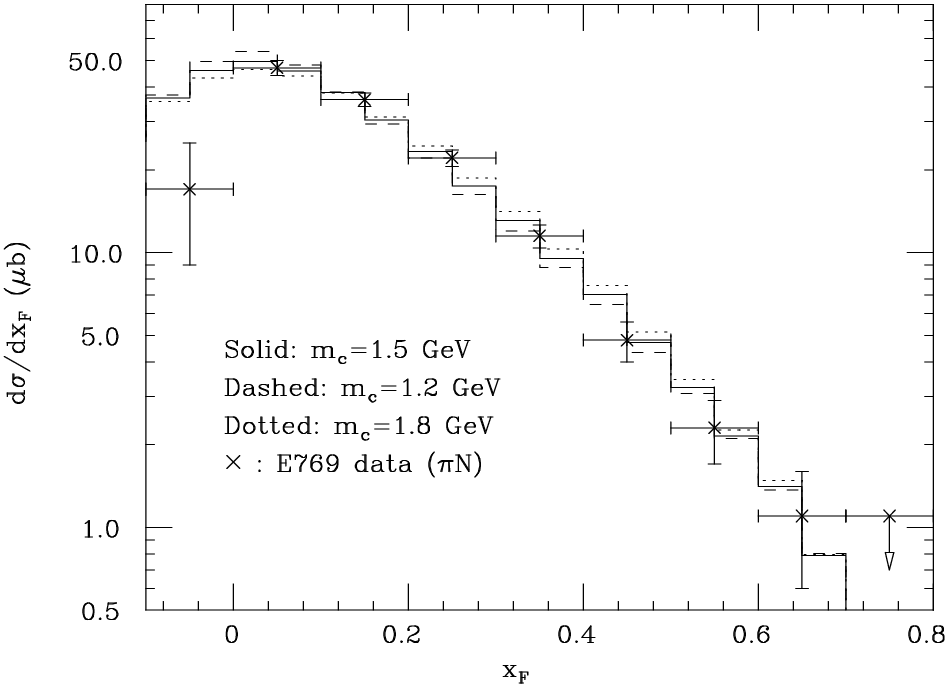,width=0.5\textwidth,clip=}}
\ccaption{}{\label{hdrxf}
Experimental $x_{\sss F}$ distributions for $D$ mesons,
compared to the NLO QCD prediction for charm quarks.}
\end{figure}
We now turn to the $x_{\sss F}$ distribution. The experimental
measurements of refs. \cite{Kodama92,Barlag88,Barlag91,Aguilar85a,
Alves92,Adamovich92,Aoki88,Aoki92a} (for $\pi N$ collisions) and of
refs.~\cite{Kodama91,Barlag88,Ammar88,Aguilar88,Adamovich92}
(for $pN$ collisions) find on average a behaviour that is harder
than the perturbative QCD result for bare quarks (see
ref.~\cite{Frixione94} for details). As in the case
of the $p_{\sss T}^2$ distribution, some description of the
hadronization phenomena should be added to the perturbative
calculation in order to compare it with the data. This problem
was considered in ref.~\cite{Mangano93},
where the hadronization phenomena were studied using the
parton shower Monte Carlo HERWIG. In ref.~\cite{Mangano93} the conclusion was
reached that the combined effects of perturbative higher orders and
non-perturbative (partonic intrinsic transverse momentum and hadronization)
contributions eventually result in a hardening of the $x_{\sss F}$
distribution for bare quarks. There it was also argued
that the usual approach of complementing the perturbative calculation with a
fragmentation function in order to describe the $x_{\sss F}$ distribution is
completely unjustified, since the factorization theorem only holds in the
large-$p_{\sss T}$ region.

The most recent experimental
results of WA92 \cite{Adamovich96} and E769 \cite{Alves96a} for
the $D$-meson $x_{\sss F}$ distribution
in $\pi N$ collisions are, instead, in agreement with
the perturbative QCD distributions for bare quarks,
as can be seen from
fig.~\ref{hdrxf}\footnote{The pion beam used by the
E769 collaboration~\cite{Alves96,Alves96a}
is a mixture of $\pi^-$ (70\%) and $\pi^+$ (30\%).
To produce the data appearing in fig.~\ref{hdrxf}, which
contain the contributions of both $D^+$ and $D^-$,
the distributions obtained with the different beams
have been combined together, since
no statistically significant discrepancy has been found
between them~\cite{Alves96a}.}.
This is also consistent with previous findings of E769 \cite{Alves92}.
The agreement is quite satisfactory in almost the whole range
considered for both experiments (which are performed at different
beam energies). The shape of the theoretical curve shows a mild
sensitivity with respect to the value of the charm mass (all curves have been
normalized to the data). Recent results for a proton beam are also
available~\cite{Alves96a},  and an agreement with theoretical
predictions similar to the one displayed in fig.~\ref{hdrxf} is found.

Almost all of the experimental
collaborations have observed, in pion--nucleon collisions,
the so-called leading-particle effect, that is, an enhanced production
at large $x_{\sss F}$ of those
$D$ mesons whose light valence quark is of the same flavour as one of the
valence quarks of the incoming pion.
In ref.~\cite{Frixione94} it was shown that the QCD predictions are in
better agreement with the data for non-leading particles than with the data
for the full $D$-meson sample which, as mentioned before, displays on
average a harder behaviour than the theory.
The difference between the leading and
the non-leading sample may be an indication that non-perturbative
phenomena (such as colour-drag effects) are present in the production
of leading particles. Recent experimental results on the asymmetry
between $D^-$ and $D^+$ production were presented in ref.~\cite{Aitala96},
for $\pi^- N$ collisions with $E_b=500$~GeV.
It was found that perturbative QCD, the instrinsic
charm hypotesis~\cite{Combridge79,Brodsky80}, and the standard Monte Carlo
simulations cannot describe the measured quantities. The
data can only be acceptably reproduced with a special tuning
of Monte Carlo models implementing beam-dragging effects.
Further studies of colour dragging and hadronization phenomena
have been presented in refs.~\cite{Aitala96a,Carter96}.

In the past, the $x_{\sss F}$ and $p_{\sss T}^2$ distributions have been fitted
with the functions
\beq
\label{formfit}
\frac{d\sigma}{dx_{\sss F}}=A\,(1-x_{\sss F})^n;\;\;\;\;\;
\frac{d\sigma}{dp_{\sss T}^2}=C\, e^{-b p_{\sss T}^2},
\eeq
and the data have been presented in the form of measured values for the
parameters $n$ and $b$.
A possible way of comparing the data with theoretical predictions
would be to fit
theoretical distributions in the same way, and then compare the
values of the parameters $n$ and $b$. This was done in ref.~\cite{Frixione94}.
However, as pointed out there, this procedure is
not satisfactory.
\begin{figure}
\centerline{\epsfig{figure=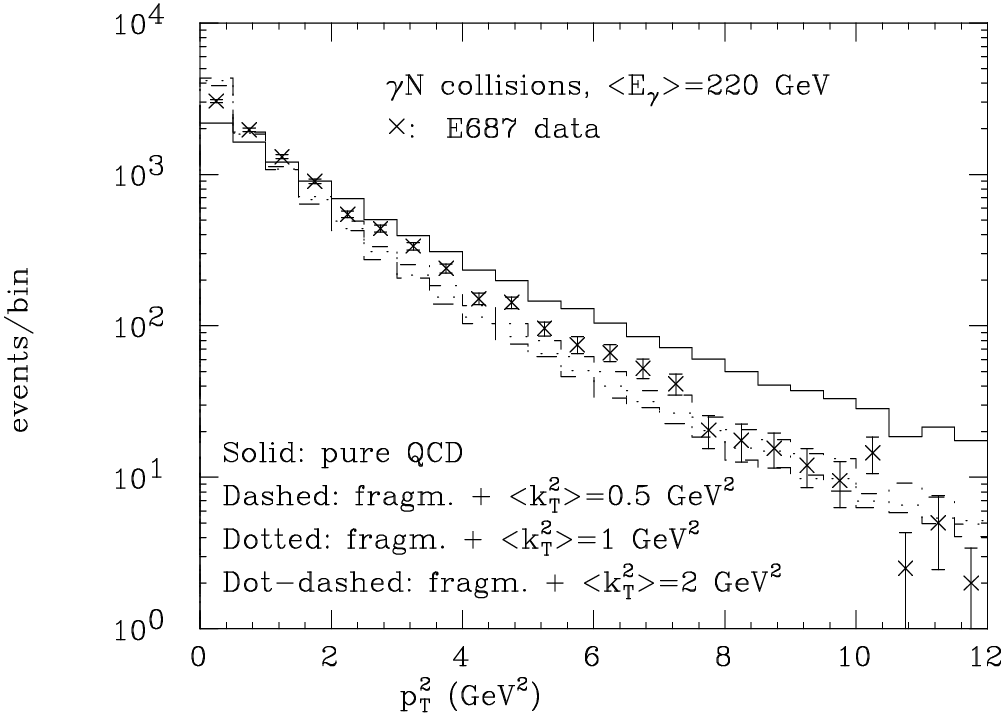,width=0.5\textwidth,clip=}
            \hspace{0.3cm}
            \epsfig{figure=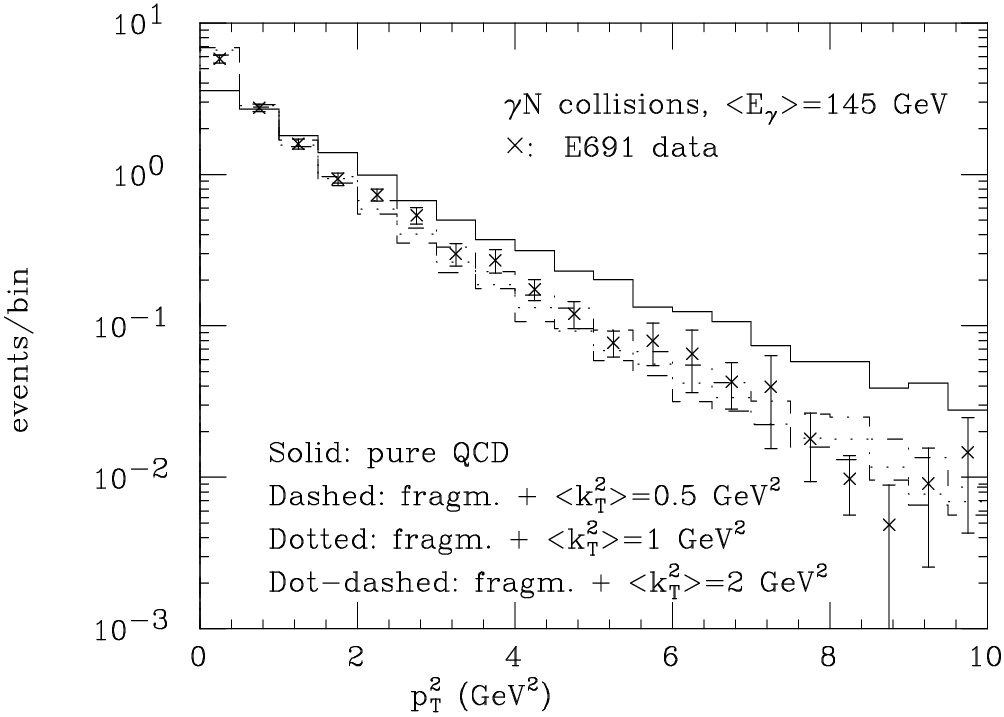,width=0.5\textwidth,clip=}}
\ccaption{}{ \label{phpt2}
Experimental $p_{\sss T}^2$ distribution compared to the NLO
QCD predictions ($m_c=1.5$~GeV), with and without the inclusion of
non-perturbative effects, in $\gamma N$ collisions.}
\end{figure}
In the case of the $p_{\sss T}^2$ distribution, for example,
it was found that the function
\beq
\label{newformpt}
\frac{d\sigma}{dp_{\sss T}^2}=
\left(\frac{C}{bm_c^2+p_{\sss T}^2}\right)^\beta
\eeq
provides an excellent fit to the theoretical curve.
Since, as discussed before, the data show the
same qualitative behaviour as the theoretical predictions over the
whole $p_{\sss T}^2$ range explored, the form of eq.~(\ref{newformpt})
is clearly preferable, as was explicitly shown more recently
in ref.~\cite{Alves96a}.

Single-inclusive distributions for charm production have also been
measured in photon--nucleon collision experiments. In the case
of photoproduction, we expect QCD predictions to be more reliable
than in the hadroproduction case, since only one
hadron is present in the initial state (see
refs.~\cite{Ellis89,Smith92,Frixione94a} for a detailed discussion).
In fig.~\ref{phpt2} we show the $p_{\sss T}^2$
distributions measured by the E687 \cite{Bellini94}
and by the E691~\cite{Anjos89,Anjos90} collaborations.
We also show the NLO QCD prediction for bare
quarks, and the QCD prediction supplemented with
Peterson fragmentation and an intrinsic transverse momentum for the incoming
partons with different values of $\langle k_{\sss T}^2 \rangle$.
It is interesting to notice that, in this case, the fragmentation effect,
combined with a moderate intrinsic transverse momentum of the
initial-state partons, is sufficient to reproduce the experimental
distributions. Contrary to what happens in the hadroproduction case,
the $p_{\sss T}^2$ distribution is now less sensitive to the choice of the
$\langle k_{\sss T}^2\rangle$. To show this fact clearly, we have
also presented the prediction for $\langle k_{\sss T}^2 \rangle =2$~GeV$^2$.

The E687 collaboration recently performed a measurement~\cite{Frabetti96}
of the asymmetries between charm and anticharm states in photon--nucleon
collisions, finding no compelling evidence for $D_s^+$ and $\Lambda_c^+$
asymmetries, and sizeable asymmetries for $D^+$, $D^0$ and $D^{*+}$ states.
The very tiny asymmetries predicted by NLO QCD cannot
account for these results, which are most likely to originated from
some non-perturbative effects, as discussed in the case of
hadroproduction. It was found that, by properly tuning Monte Carlo programs
based on the Lund string fragmentation model, the data could be reproduced.

%

In summary, a remarkable amount of experimental information
on the single-inclusive distributions for charm quarks is currently
available, both in hadroproduction and in photoproduction. Further data with
increased statistics are expected in the near future \cite{Kaplan94}.
Some modelling of non-perturbative effects is needed in order
to describe the transverse-momentum spectrum of charmed hadrons.
The inclusion of a fragmentation function tends to give too soft a $\pt$
spectrum. Assuming the presence
of a primordial transverse momentum of the colliding
partons, the $\pt$ spectrum becomes harder, and a better (albeit
not fully satisfactory) agreement with
the data is found.
Longitudinal ($\xf$) distributions are measured to be harder
than the theoretical predictions.
Similarly, asymmetries in the longitudinal distributions
between charm and anticharm states
are too large to be explained by perturbative QCD alone.
Simple models of hadronization, including
colour-dragging effects, can give a satisfactory description of the data.

We remind the reader that the E653 collaboration presented in
ref.~\cite{Kodama93} the first measurement of single-inclusive
distributions for bottom production at fixed target (the azimuthal
correlation of the $b\bar{b}$ pair was also studied). The data
appear to be in qualitative agreement with QCD predictions.
More investigations with improved statistics will be welcome
in this area.

\subsection{Double-differential distributions}

Many experimental results on correlations between charmed particles
in hadro- and photoproduction have been obtained by different
experiments (see for example
refs.~\cite{Aguilar85b,Adamovich87,Aguilar88,Barlag91a,
Kodama91a,Aoki92a,Alvarez92,Frabetti93});  these reported
distributions of the azimuthal distance
between the charmed hadrons, the rapidity difference, the invariant
mass and the transverse momentum of the pair.
A detailed comparison of these results with QCD predictions is performed in
ref.~\cite{Frixione94}. More recently, new measurements of the azimuthal
distance and pair transverse momentum
for charmed mesons have been presented by the WA92
collaboration~\cite{Adamovich96a}.
In what follows we will focus on the distribution of $\Delta\phi$,
defined as the angle between the projections of the momenta of the pair
onto the transverse plane, and of the transverse
momentum of the pair $p_{\sss T}(Q\overline{Q})$. We will discuss whether NLO
QCD predictions can describe the available experimental data.

In leading-order QCD the heavy-quark pair is produced in the
back-to-back configuration, corresponding to $\Delta\phi=\pi$ and
$p_{\sss T}(Q\overline{Q})=0$. NLO corrections, as well as
non-perturbative effects, can cause a broadening of these
distributions, as illustrated in refs.~\cite{Mangano93} and \cite{Frixione94a}.

\begin{figure}
\centerline{\epsfig{figure=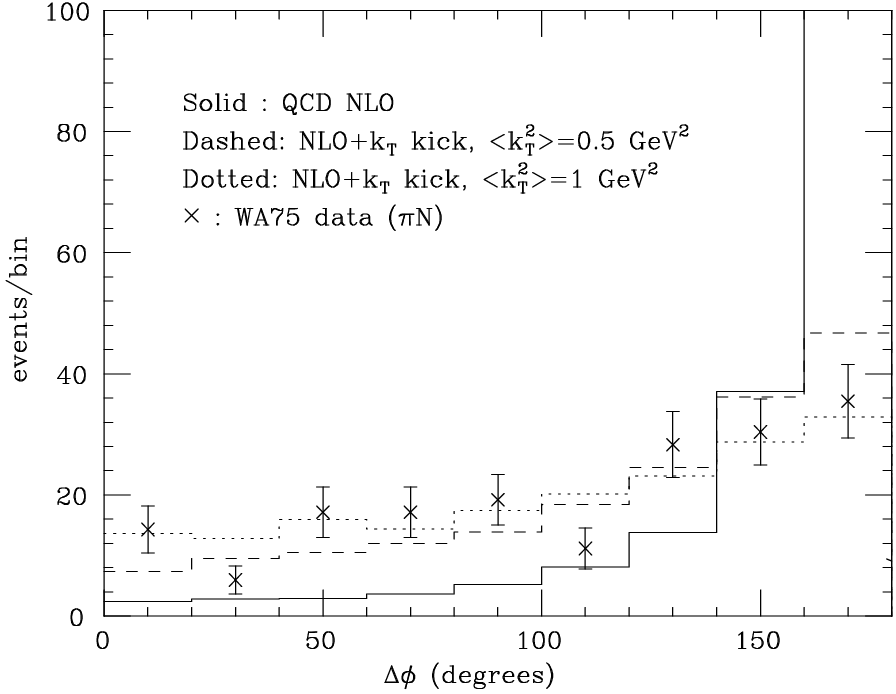,width=0.5\textwidth,clip=}
            \hspace{0.3cm}
            \epsfig{figure=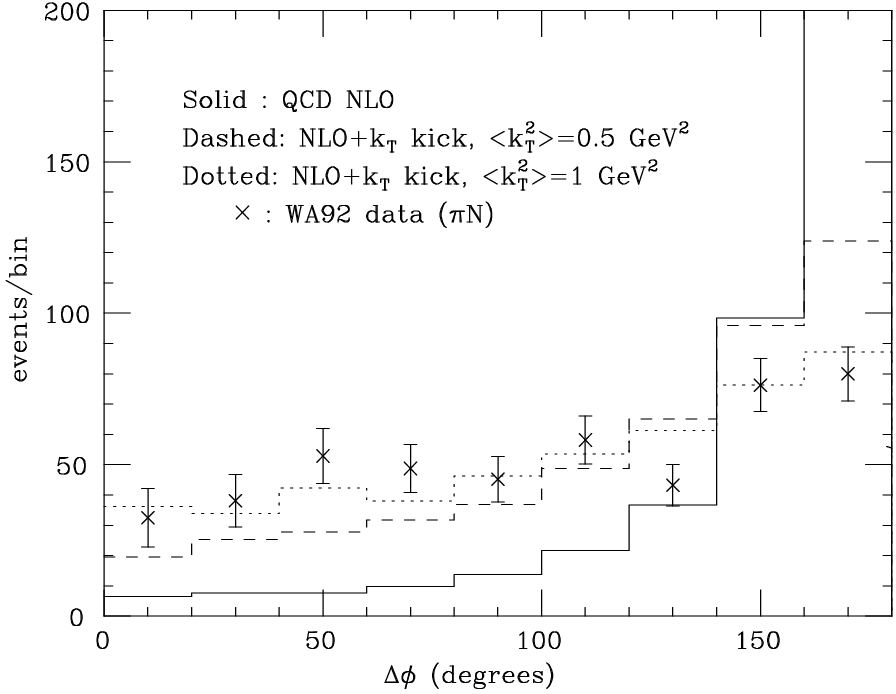,width=0.5\textwidth,clip=}}
\ccaption{}{\label{deltaphi}
Azimuthal correlation for charm production in $\pi N$ collisions:
NLO calculation versus the WA75 and WA92 data.
}
\end{figure}

We have chosen, as an illustration for hadroproduction, the cases of
the WA75 and the WA92 results, which have both been obtained in
$\pi^-N$ collisions at the same energy, $E_b=350$ GeV.
Let us first consider the $\Delta\phi$ distribution.
In fig.~\ref{deltaphi} we show (solid curves) the NLO
result superimposed on the data of two experiments.
The charm mass was set to its default value, $m_c=1.5$~GeV.
In both cases, we see that the experimental data favour a much broader
distribution than the pure NLO QCD result for charm quarks.

One should, however, take into account also non-perturbative effects,
as in the case of single-inclusive distributions. We have computed
the $\Delta\phi$ distribution in perturbative QCD with an intrinsic
transverse momentum of the incoming partons as described in
subsection~\ref{single} (the use of a fragmentation function has no effect
on the $\Delta\phi$ distribution, since it does not affect
momentum directions).
The dashed and dotted curves in fig.~\ref{deltaphi} correspond to
the NLO prediction, supplemented
with the effect of an intrinsic transverse momentum
with $\langle k_{\sss T}^2\rangle=0.5$ GeV$^2$
and $\langle k_{\sss T}^2\rangle=1$ GeV$^2$, respectively.
We see that with $\langle k_{\sss T}^2 \rangle=0.5\;$GeV$^2$
it is impossible to describe the WA75 and WA92 data.
This conclusion differs from the one of ref.~\cite{Frixione94}
for the WA92 result. This is because in
ref.~\cite{Adamovich96a} the WA92 collaboration has improved
the study of correlations with respect to ref.~\cite{Adamovich95}
by considering a wider set of correlation variables and by improving
the statistics by a factor of 5.
WA92 and WA75 data now appear to be consistent with each other.
As is apparent from fig.~\ref{deltaphi}, the acceptable
value of $\langle k_{\sss T}^2\rangle=1$ GeV$^2$ is required to
describe the data.

We point out that this result has to be interpreted in conjunction with
the adopted values of the input parameters entering the calculation.
We normally choose the following defaults:
\begin{itemize}
\item  $m_c=1.5\,$GeV,
\item $\muf=2\mu_0$, $\mur=\mu_0$, where
\beq
\mu_0=\sqrt{(p_{\sss T}^2+\bar{p}_{\sss T}^2)/2+m_c^2}\,,
\eeq
\item the MRSA$^\prime$ \cite{Martin95}
proton parton densities, with the corresponding
value of $\lambdamsb=152\,$MeV,
\item the SMRS2 \cite{Sutton92}
set for the pion parton densities.
\end{itemize}
Choosing smaller values for the charm mass, for example,
one would get a broader theoretical curve. Alternatively, one could keep
the default mass value, but choose $\mur=\mu_0/2$ for the renormalization
scale, instead of our default choice $\mur=\mu_0$. As shown in
ref.~\cite{Frixione94}, this would lead to a broadening of
the $\Delta\phi$ distribution similar to the one caused by an
average intrinsic transverse momentum with
$\langle k_{\sss T}^2\rangle=0.5$~GeV$^2$.

\begin{figure}[htb]
\centerline{\epsfig{figure=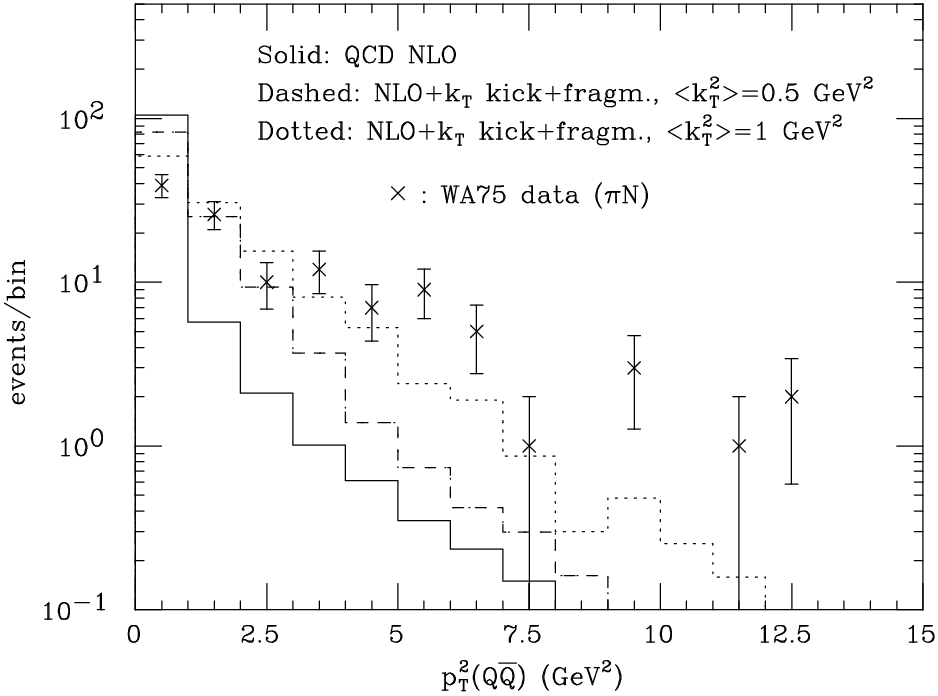,width=0.5\textwidth,clip=}
            \hspace{0.3cm}
            \epsfig{figure=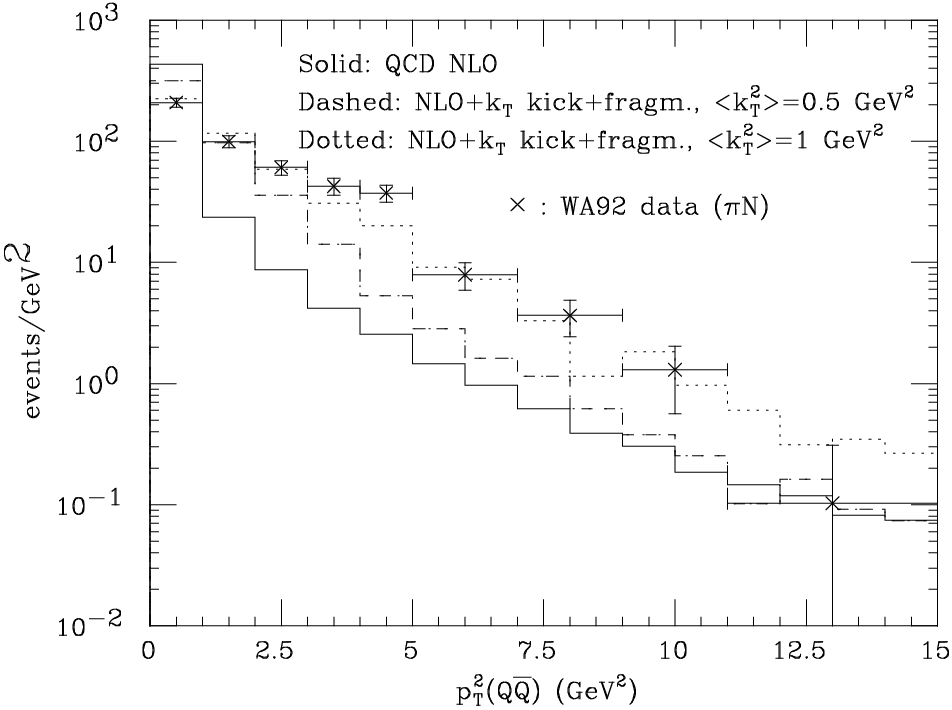,width=0.5\textwidth,clip=}}
\ccaption{}{\label{f:pt2qq}
NLO QCD result for the $p_{\sss T}^2(Q\overline{Q})$ supplemented
with an intrinsic transverse momentum for the incoming partons,
compared with the WA75 (left) and WA92 (right) data.
}
\end{figure}

The WA75 collaboration, and recently the WA92 collaboration, published
in refs.~\cite{Aoki92a,Adamovich96a} the distribution of the transverse
momentum of the heavy-quark pair.
The theoretical prediction supplemented with a
$k_{\sss T}$-kick with $\langle k_{\sss T}^2\rangle = 1\;$GeV$^2$
cannot reproduce the WA75 data, while it is in rough agreement
with the WA92 measurement, as displayed in fig.~\ref{f:pt2qq}.
Unlike the azimuthal correlation, the $p_{\sss T}^2(Q\overline{Q})$
distribution is
affected by fragmentation effects, since these can degrade the
momenta of the quark and antiquark by different amounts. Fragmentation
effects also moderate the pair transverse momentum arising from gluon
radiation or from an intrinsic parton transverse momentum.
We have verified that at the end, at $E_b=350$~GeV, the fragmentation
always tends to soften the $p_{\sss T}^2(Q\overline{Q})$ distribution.

\begin{figure}[htb]
\centerline{\epsfig{figure=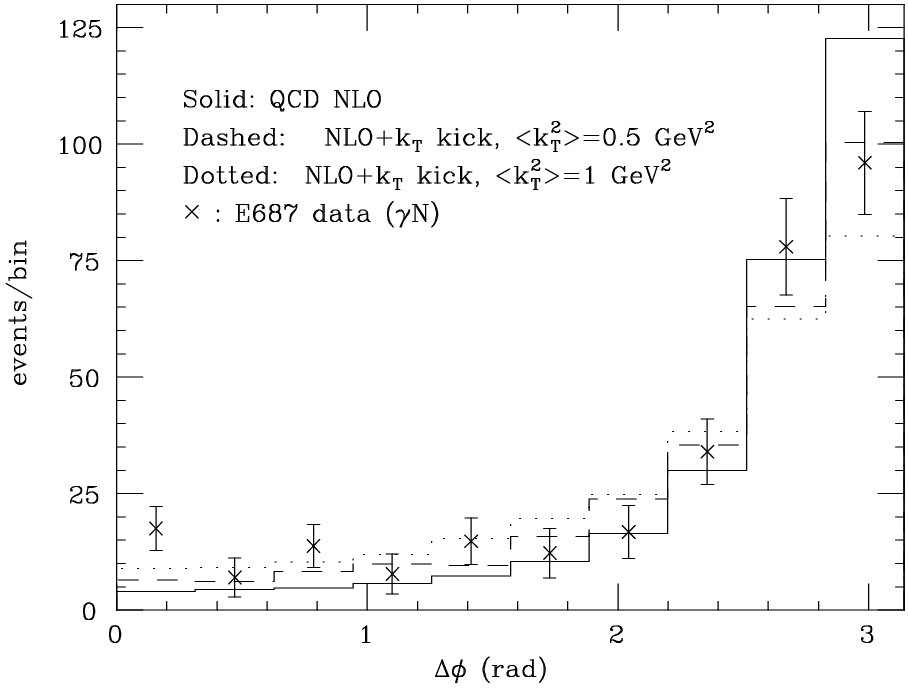,width=0.5\textwidth,clip=}
            \hspace{0.3cm}
            \epsfig{figure=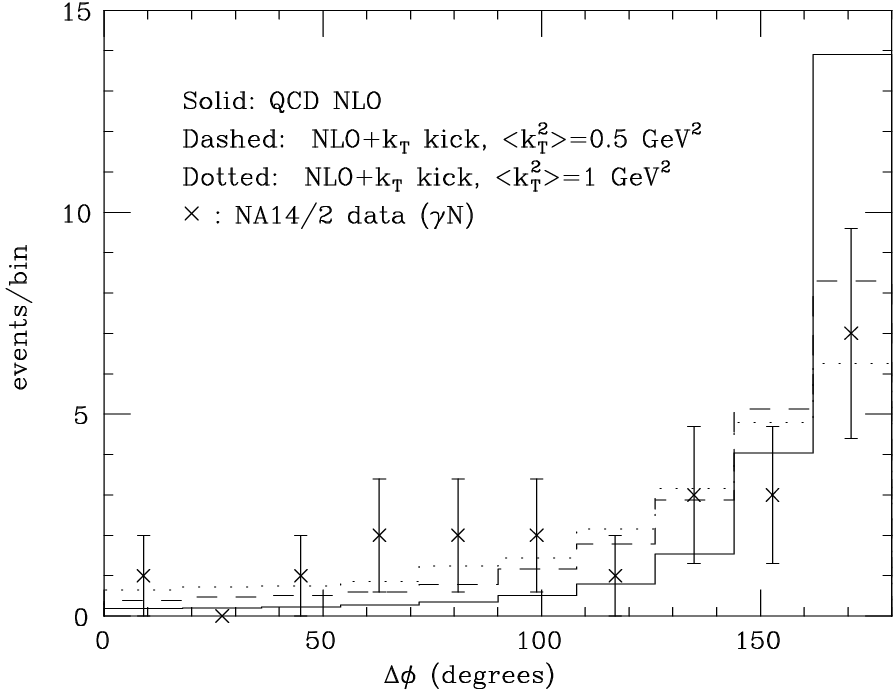,width=0.5\textwidth,clip=}}
\ccaption{}{\label{f:dphi_photo}
Azimuthal correlation of $D\bar{D}$ pair versus the perturbative
result in photoproduction for the E687 (left) and NA14/2 (right)
experiments.}
\end{figure}
Summarizing, the available experimental results for azimuthal
$c\bar{c}$ correlation in hadron--hadron collisions show
a tendency to peak in the back-to-back
region $\Delta\phi=\pi$, but the peak is less pronounced than
the one predicted by perturbative QCD. The addition of a $k_{\sss T}$ kick
of 1 GeV$^2$ on average gives a satisfactory description of the
data, if our default input parameters are chosen.
The data on the $p_{\sss T}^2(Q\overline{Q})$ distribution
do not give a unique indication. While the theoretical prediction
obtained with $\langle k_{\sss T}^2\rangle=1$ GeV$^2$ is in rough agreement
with the WA92 measurement, it is sizeably softer than the WA75 data.

\begin{figure}[htb]
\begin{center}
\mbox{\psfig{file=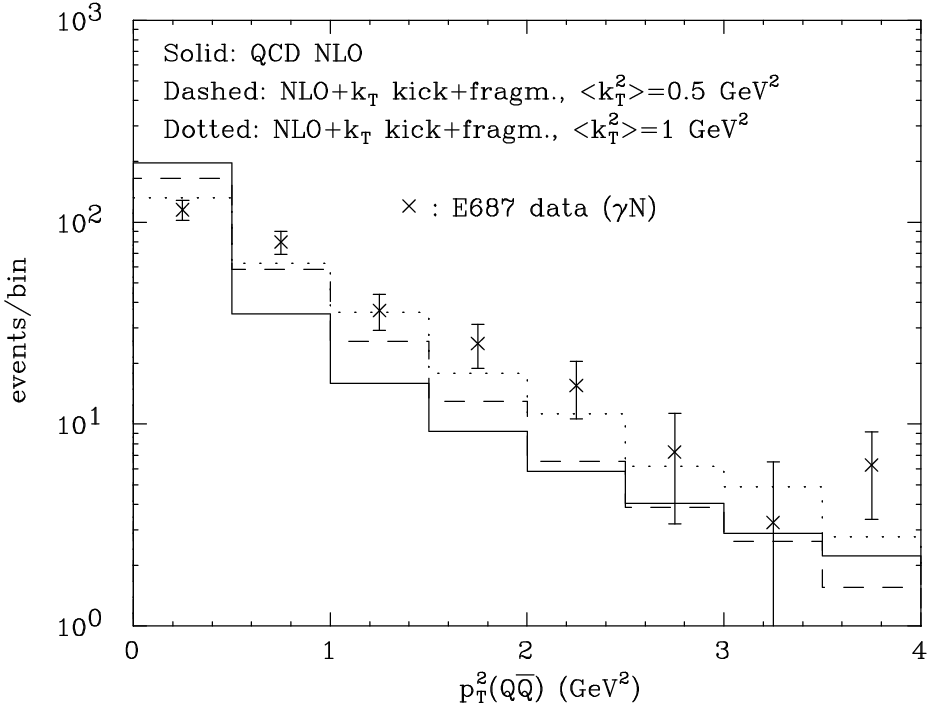,width=0.65\textwidth}}
\ccaption{}{\label{e687ptqq}
Transverse momentum distribution of the $D\bar{D}$ pair versus the
perturbative result for the E687 experiment.
}
\end{center}
\end{figure}

Photoproduction of heavy quarks \cite{Alvarez92,Frabetti93,Adamovich87}
is another example in which a $k_{\sss T}$ kick would induce
broader $\Delta\phi$ and $p_{\sss T}(Q\overline{Q})$ correlations.
On the left-hand side of fig.~\ref{f:dphi_photo}, the azimuthal correlation
measured by the E687 collaboration is given, together
with the NLO result. The NLO result
supplemented by an intrinsic $k_{\sss T}$ of the incoming partons is also
shown, for $\langle k_{\sss T}^2\rangle=0.5\;$GeV$^2$ and $\langle k_{\sss
T}^2\rangle=1\;$GeV$^2$.
As can be seen, the data do not require a large intrinsic transverse momentum.
All curves give a reasonable representation of the data, the one with $\langle
k_{\sss T}^2\rangle=0.5$ GeV$^2$ being slightly better. A similar conclusion
applies to the NA14/2 data (which are, however, affected by larger
uncertainties), as shown on the right-hand side of fig.~\ref{f:dphi_photo}.
The distribution in the transverse momentum of the heavy-quark
pair is displayed in fig.~\ref{e687ptqq}. In this case, we see that
the E687 data favour $\langle k_{\sss T}^2 \rangle=1$ GeV$^2$.

In conclusion, we have seen that the $\Delta\phi$ and
$p_{\sss T}(Q\overline{Q})$ distributions
are very sensitive to non-perturbative effects, especially
in the hadroproduction case.
We computed these distributions by assuming that the non-perturbative
effects are parametrized by a fragmentation function and
an intrinsic transverse momentum of the incoming partons.
In general, the azimuthal correlation of the pair is
well described by perturbative QCD supplemented by an
acceptable $k_{\sss T}$ kick.
The recent result of ref.~\cite{Adamovich96a} on the
$p_{\sss T}(Q\overline{Q})$ distribution can also be described
by the same choice of parameters. This is consistent
with what was found in the study of the single-inclusive $\pt$ distribution.
However, the $p_{\sss T}(Q\overline{Q})$ distribution reported
by WA75 \cite{Aoki92a} is much harder, and would require an unphysically
large $k_{\sss T}$ kick.
\section{Heavy-flavour production at HERA}\label{sec:HERA}
\sethead{Heavy-flavour production at HERA}
The $ep$ collider HERA offers new opportunities to study the
production mechanism of heavy quarks and to test the predictions
of perturbative QCD.
The dominant contribution to the cross section is due to those
events in which the virtuality of the photon exchanged between
the electron and the proton is very small. In this case, the electron
can be considered to be equivalent to a beam of on-shell photons, whose
distribution in energy (Weizs\"acker-Williams
function~\cite{Weizsaecker34,Williams34})
can be calculated in QED. The underlying production mechanism
is therefore a photoproduction one, which has been studied extensively
in fixed-target experiments.
At HERA, the available centre-of-mass energy is about one
order of magnitude larger than at fixed-target facilities
(200~GeV versus 20~GeV). This energy regime is totally
unexplored in photoproduction, and several new features have
to be taken into proper account. In particular, the large contribution
of the hadronic photon component introduces in the theoretical
predictions a source of uncertainty that is normally negligible
at fixed-target energies.

A complementary way of studying heavy-flavour production at HERA
is to retain only those events characterized by a large photon
virtuality (DIS).
Although the total rates are much
smaller than the photoproduction ones, the hadronic
component is completely eliminated, and more reliable theoretical
and experimental results can be obtained. Also, the dependence
of the data upon the photon virtuality can be used as a further
test of QCD predictions.
A significant improvement is expected with the planned luminosity upgrade of
the HERA collider~\cite{Ingelman96}, when the number of charm (bottom)
quarks produced will be of the order of $10^9$ ($10^6$).

In this section, we discuss some aspects of the perturbative QCD
calculations of the heavy-flavour
photoproduction cross sections of relevance for HERA. We review the
phenomenology of charm and bottom production, comparing the
available data with the theoretical predictions. We then discuss
future perspectives.

\subsection{Photoproduction cross sections}\label{subsec:phxsec}
It is well known that an on-shell photon has a finite probability
to fluctuate into a hadronic state before
undergoing a hard collision. In this case,
the photon is referred to as ``hadronic'' (or ``resolved''),
to contrast with those events in which it directly
interacts with the hadron (``point-like'' or ``direct'').
Therefore, a differential photon--hadron cross section can be written as
the sum of a point-like and a hadronic photon contribution:
\beq
d\sigma^{(\gamma {\sss H})}(P_\gamma,P_{\sss H})\,=\,
d\sigma^{(\gamma {\sss H})}_{\rm point}(P_\gamma,P_{\sss H})+
d\sigma^{(\gamma {\sss H})}_{\rm hadr}(P_\gamma,P_{\sss H})\,.
\label{fullxsec}
\eeq
Thanks to the factorization theorems in QCD~\cite{Ellis79,Collins89}, we have
\beqn
d\sigma_{\rm point}^{(\gamma {\sss H})}(P_\gamma,P_{\sss H})&=&\sum_j\int dx
f^{({\sss H})}_j(x,\muf)
d\hat{\sigma}_{\gamma j}(P_\gamma,xP_{\sss H},\as(\mur),\mur,\muf,\mug),
\label{pointcomp}
\\
d\sigma_{\rm hadr}^{(\gamma {\sss H})}(P_\gamma,P_{\sss H})
&=&\sum_{ij}\int dx dy
f^{(\gamma)}_i(x,\mug) f^{({\sss H})}_j(y,\muf^\prime)
\nonumber \\*&&\phantom{\sum_{ij}\int dx}\times
d\hat{\sigma}_{ij}(xP_\gamma,yP_{\sss H},
\as(\mur^\prime),\mur^\prime,\muf^\prime,\mug)\,.
\label{hadrcomp}
\eeqn
In eq.~(\ref{hadrcomp}), the $f^{(\gamma)}_i$ are the partonic densities
in the photon. Their physical meaning is analogous to the one
of the more familiar partonic densities in the hadron, $f^{({\sss H})}_j$.
They are universal, but not
calculable in perturbation theory. They satisfy
a renormalization group equation \cite{Witten77,Llewellyn78,Bardeen79,DeWitt79}
that can be obtained by slightly
modifying the usual Altarelli--Parisi \cite{Altarelli77} equation
\beq
\frac{\partial f^{(\gamma)}_i}{\partial\log\mu^2}=
\frac{\aem}{2\pi}P_{i\gamma}
+\frac{\as}{2\pi}\sum_j\,P_{ij}\otimes f^{(\gamma)}_j\,.
\label{gamma_AP}
\eeq
At the lowest order we have
\beq
P_{i\gamma}=N_c\,e_i^2\,\left(x^2+(1-x)^2\right)\,,
\eeq
where $N_c=3$ is the number of colours and $e_i$ is the electric charge
of the parton $i$ in units of the
charge of the positron (for gluons, $e_i=0$). The first term
on the RHS of eq.~(\ref{gamma_AP}), which is not present in the
evolution equation of the hadron densities, is due to the direct coupling
of the photon to the quarks.

To understand the strict interplay between the point-like and the
hadronic component in eq.~(\ref{fullxsec}), it is worth while to
sketch the derivation of the subtracted partonic cross sections
appearing in eqs.~(\ref{pointcomp}) and~(\ref{hadrcomp}).
As a first step, one performs a direct calculation of the partonic cross
sections for the processes \mbox{$\gamma+j\to Q+\overline{Q}+X$}
and \mbox{$i+j\to Q+\overline{Q}+X$}
(here $i$ and $j$ represent a generic parton).
After the ultraviolet
renormalization, one is left with the quantities $d\sigma_{\gamma j}$
and $d\sigma_{ij}$, which still display divergences due to the
collinear emission of a massless final-state parton from one of
the incoming partons. These divergences are subtracted by applying
the prescription of the factorization theorem.
Writing the expansion of the partonic cross sections at the
next-to-leading order in QCD as
\beqn
d\sigma_{\gamma j}&=&\aem\as d\sigma_{\gamma j}^{(0)}
+\aem\as^2 d\sigma_{\gamma j}^{(1)}\,,
\label{pointxsec}
\\
d\sigma_{ij}&=&\as^2 d\sigma_{ij}^{(0)}
+\as^3 d\sigma_{ij}^{(1)}\,,
\label{hadrxsec}
\eeqn
and an analogous expansion for the subtracted cross sections
$d\hat{\sigma}_{\gamma j}$ and $d\hat{\sigma}_{ij}$, we get,
using dimensional regularization ($d=4-2\ep$) and order by order
in perturbation theory,
\beqn
d\hat{\sigma}_{\gamma j}^{(0)}(p_1,p_2)&=&d\sigma_{\gamma j}^{(0)}(p_1,p_2),
\;\;\;
d\hat{\sigma}_{ij}^{(0)}(p_1,p_2)=d\sigma_{ij}^{(0)}(p_1,p_2),
\\
d\hat{\sigma}_{\gamma j}^{(1)}(p_1,p_2)&=&
d\sigma_{\gamma j}^{(1)}(p_1,p_2,\frac{1}{\epb})
+\frac{1}{2\pi}\sum_k\int dx \left(\frac{1}{\epb}P_{k\gamma}(x)
-H_{k\gamma}(x)\right)d\sigma_{kj}^{(0)}(xp_1,p_2)\,
\nonumber \\*
&&+\frac{1}{2\pi}\sum_k\int dx \left(\frac{1}{\epb}P_{kj}(x)-
K^{({\sss H})}_{kj}(x)\right) d\sigma_{\gamma k}^{(0)}(p_1,xp_2),
\label{subpoint}
\\
d\hat{\sigma}_{ij}^{(1)}(p_1,p_2)&=&
d\sigma_{ij}^{(1)}(p_1,p_2,\frac{1}{\epb})
+\frac{1}{2\pi}\sum_k\int dx \left(\frac{1}{\epb}P_{ki}(x)-
K_{ki}^{(\gamma)}(x)\right) d\sigma_{kj}^{(0)}(xp_1,p_2)
\nonumber \\*
&&+\frac{1}{2\pi}\sum_k\int dx \left(\frac{1}{\epb}P_{kj}(x)-
K_{kj}^{({\sss H})}(x)\right) d\sigma_{ik}^{(0)}(p_1,xp_2).
\label{subhadr}
\eeqn
On the RHS of eqs.~(\ref{subpoint}) and~(\ref{subhadr}) we explicitly
indicated the dependence of the partonic cross sections upon $1/\epb$
to recall that these quantities are still infrared-divergent.
The divergences are nevertheless properly cancelled by the terms
proportional to the Altarelli--Parisi kernels appearing in these equations.
The functions $K_{ij}^{({\sss H})}$, $K_{ij}^{(\gamma)}$ and $H_{k\gamma}$
are completely arbitrary, in that they define an extra finite part of the
subtraction; different choices correspond to different subtraction schemes.
In eqs.~(\ref{subpoint}) and~(\ref{subhadr}) the $\MSB$ scheme is equivalent
to $H=K\equiv 0$. For greater generality, we have allowed the possibility
to have different subtraction schemes on the photon and hadron legs.

The key feature of eqs.~(\ref{subpoint}) and~(\ref{subhadr}) is
the term proportional to $P_{k\gamma}(x)$ in eq.~(\ref{subpoint}).
The physical origin of this contribution is the direct coupling of
the photon with the quarks, which in turn implies the presence
of the inhomogeneous term in eq.~(\ref{gamma_AP}). In the language of the
Altarelli--Parisi equations, this means that the infrared divergences
due to the collinear emission of quarks from the incoming photon
in the point-like component, eq.~(\ref{pointcomp}), are re-absorbed
into the partonic densities in the photon, which appear in the
hadronic component, eq.~(\ref{hadrcomp}).
The same argument can be formulated in terms of renormalization
group (RG) equations. Equation~(\ref{pointcomp}) is RG-invariant with
respect to the variation of the scales $\mur$ and $\muf$; eq.~(\ref{hadrcomp})
is RG-invariant with respect to the variation of the scales $\mur^\prime$
and $\muf^\prime$. But neither eq.~(\ref{pointcomp}) nor
eq.~(\ref{hadrcomp}) are {\it separately} RG-invariant with respect
to the variation of the scale $\mug$; when varying $\mug$ in
eq.~(\ref{hadrcomp}), a residual dependence is left, due to the
inhomogeneous term in eq.~(\ref{gamma_AP}), which is cancelled only
by the explicit dependence of eq.~(\ref{pointcomp}) upon $\mug$.
A third way of understanding this issue is to consider a change
of the subtraction scheme. The partonic densities are modified
as follows:
\beqn
f_i^{(H)^\prime}&=&f_i^{(H)}
+\frac{\as}{2\pi}\sum_{j}K_{ij}^{(H)}\otimes f_j^{(H)},
\\
f_i^{(\gamma)^\prime}&=&f_i^{(\gamma)}+\frac{\aem}{2\pi}H_{i\gamma}
+\frac{\as}{2\pi}\sum_{j}K_{ij}^{(\gamma)}\otimes f_j^{(\gamma)}.
\eeqn
The term $H_{i\gamma}$, which is due to the change of scheme of the photon
densities entering eq.~(\ref{hadrcomp}), affects eq.~(\ref{pointcomp}),
since it defines the finite part of the infrared subtraction in the
photon leg via eq.~(\ref{subpoint}).

It should now be clear that the point-like and the hadronic components
of the photoproduction cross sections are very closely related, and
that only their sum is physically meaningful.
The separation of a cross section
into a point-like and a hadronic component is ambiguous
beyond leading order
(for a detailed discussion, see refs.~\cite{Schuler93,Schuler93a}),
as is shown by the discussion on the change of
subtraction scheme: finite terms can be subtracted to one
piece and added to the other, without affecting physical predictions.
One explicit example is presented in ref.~\cite{Gluck92}, where
a factorization scheme ($\DIG$) for the photon densities
is introduced, which uses $K^{(\gamma)}=0$ and $H\ne 0$.
That notwithstanding, we will keep on
talking about the point-like and the hadronic component. The reason
for this is twofold: we can use a leading-order approximation
to get a physical picture of the separation between the two
components; and the term ($H$) that can be exchanged between
the two components is numerically small.

The photon parton-densities are quite soft. Therefore, the
hadronic component is only important for large CM energies and
small masses for the produced system. We will see in the following
that it potentially affects
charm production at HERA. This process can therefore
be used to constrain the densities in the photon, which are
experimentally very poorly known at present.

As a final step, in order to obtain the $Q\overline{Q}$
cross sections in electron--proton
collisions, the photoproduction cross sections must be convoluted
with the Weizs\"acker--Williams distribution:
\beq
f_\gamma^{(e)}(y)=\frac{\aem}{2\pi}\frac{1+(1-y)^2}{y}
\log\frac{\mu_{\sss WW}^2(1-y)}{m_e^2 y^2}.
\label{wwhvq}
\eeq
The scale $\mu_{\sss WW}$ is, in general, a function of $y$ determined
by the kinematics of the process considered.
It has been pointed out in refs.~\cite{Budnev74,Olsen79,Bawa89,
Catani91,Frixione93}
that the choice of $\mu_{\sss WW}$ should also take into account the
dynamics of the production mechanism.
In the case of the production of heavy quarks it is reasonable to set
$\mu_{\sss WW}=\xi m_{\sss Q}$, where $\xi$ is of order 1.
It was shown in ref.~\cite{Frixione93} that for the point-like component
an appropriate choice for the parameter $\xi$
is $\xi\equiv 1$. The hadronic component case is more
involved, since the partonic densities in the {\it virtual} photon
introduce in the problem an additional mass scale. The presence
of this scale is due to the fact that the densities rapidly fall
to zero when the virtuality of the photon approaches the hard
scale of the process~\cite{Uematsu82,Borzumati93,Drees94},
which is of the order of the heavy-quark mass. It follows
that the scale $\mu_{\sss WW}$ must be chosen
smaller than $m_{\sss Q}$. In ref.~\cite{Frixione95a}
it has been argued that $\xi=0.6$--0.7 gives sensible
results. Although this is only a rough estimate,
the uncertainty due to the choice of the parameter $\xi$
is much smaller than the uncertainties coming from the
phenomenological parameters entering the calculations.

It is interesting to notice that a photoproduction event can be
experimentally defined by means of an anti-tag condition: all those
events in which the electron is scattered at an angle
larger than a given value $\theta_c$ are rejected.
The form of the Weizs\"acker--Williams function in the presence of
an anti-tag condition can be explicitly worked out~\cite{Frixione93}.
Indicating with $E$ the energy of the incoming electron
in the laboratory frame, we get
\beqn
f_\gamma^{(e)}(y)&=&\frac{\aem}{2\pi}\left\{
2(1-y)
\left[\frac{m_e^2y}{E^2(1-y)^2\theta_c^2 + m_e^2y^2}-\frac{1}{y}\right]\right.
\nonumber \\*&&\phantom{\frac{\aem}{2\pi}}
\left.+\frac{1+(1-y)^2}{y}
\log\frac{E^2(1-y)^2\theta_c^2 +m_e^2 y^2}{m_e^2 y^2}\right\}.
\label{ourww}
\eeqn
The error due to the approximation can be estimated to be
${\cal O}(\theta_c^2,m_e^2/E^2)$ in this case, and it is therefore
quite small for applications to HERA physics, where
$\theta_c\approx 5\times 10^{-3}$. Also notice that the
non-logarithmic term is singular in $y$ and therefore represents a
non-negligible correction.

\subsection{Charm photoproduction}
Taking into account all the relevant sources of uncertainty,
which have already been discussed in section~\ref{sec:ft}, we get
a theoretical prediction for the total rates with a large error. At the
NLO, the point-like cross section changes by
a factor of about 4 when varying the mass in the range
1.2~GeV~$<m_c<$~1.8~GeV, and by a factor of 2 when varying the
renormalization scale. The choice of the proton partonic densities
induces a 50\% uncertainty at $\sqrt{S_{\gamma p}}=30$~GeV, and a factor
of about 5 \cite{Frixione95b}
at $\sqrt{S_{\gamma p}}=300$~GeV. These effects are even more dramatic
in the case of the hadronic component, where the dominant source
of uncertainty is the choice of the partonic densities in the
photon. Results from dijet photoproduction at HERA give some indication
on the gluon density in the photon~\cite{Ahmed95}, but the very limited
statistics does not allow either
a distinction between different NLO
parametrizations~\cite{Gluck92a,Gordon92,Aurenche92,Aurenche94,Gordon96},
or a discrimination between the
LAC1~\cite{Abramowicz91} gluon and the flatter ones of the
NLO sets. Other LO sets~\cite{Hagiwara95}
are also consistent with the data. A reasonable way to estimate
the effect due to the uncertainty in the gluon density of the photon
is to take the sets GRV-HO~\cite{Gluck92a} and LAC1 as the two
extremes. At the highest photon--proton centre-of-mass energies available
at HERA, the prediction for the hadronic component with LAC1 is
one order of magnitude larger than the prediction with GRV-HO,
and much larger than the point-like component as well.

\begin{figure}
\centerline{\epsfig{figure=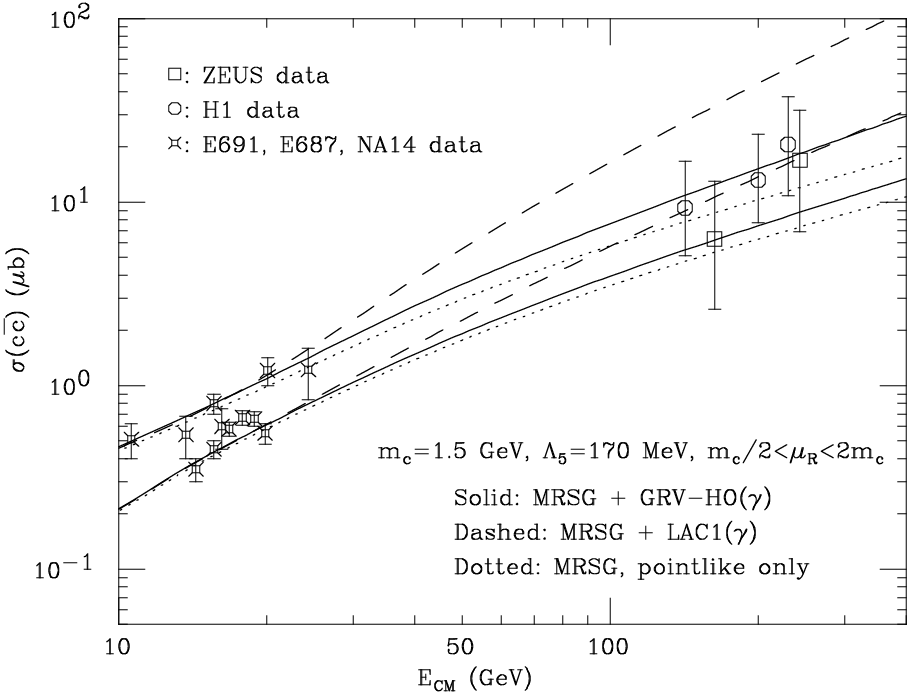,width=0.7\textwidth,clip=}}
\ccaption{}{ \label{f:sig_vs_ecm}
Total cross section for the photoproduction of $c\bar{c}$ pairs,
as a function of the $\gamma p$ centre-of-mass energy:
next-to-leading order QCD predictions versus experimental results.}
\end{figure}
The comparison between the theoretical predictions and the
experimental results \cite{Derrick95,Aid96}
is presented in fig.~\ref{f:sig_vs_ecm},
where only the uncertainty on the QCD result due to the choice of
the renormalization scale is displayed. Despite the fact that,
as discussed before, this only partially accounts for the overall
uncertainty on the theoretical predictions, the data show a
satisfactory agreement with the curves obtained by choosing
$m_c=1.5$~GeV and MRSG and GRV-HO for the partonic densities
in the proton and the photon respectively. It is important to notice that
a single choice of the input parameters allows us to describe the data
in the whole energy range considered; this is a non-trivial result,
the energies available at HERA being one order of magnitude larger
that those available at fixed-target experiments. The importance
of the hadronic component of the photon at fixed target and
at HERA is also clear from fig.~\ref{f:sig_vs_ecm}.

The experimental determination of the total charm cross section at HERA
deserves some further comment.
The experiments are sensitive to production in the central rapidity
region, typically $|y|<1.5$, where the cross section is far from
its maximum (see fig.~\ref{f:eta_HERA}).
Furthermore, a small-$p_{\sss T}$ cut is applied to the data, in
order to clearly separate the signal from the background.
In order to get the total cross section, one has therefore to extrapolate
to the full rapidity and transverse momentum range.
Large rapidities and small transverse momenta
typically involve small-$x$ values, and
the extrapolation is therefore subject to uncertainties due to our
ignorance of the small-$x$ behaviour of the parton densities.
Since these uncertainties are very large, we believe that
it would be useful to present also
the measurement of the cross section limited to the directly accessible
rapidity region.
The lack of sensitivity to large negative rapidities
is one of the reasons why it is at present
impossible to use the photoproduction
of charm quarks at HERA to distinguish among different parametrizations
of the gluon density in the proton.
An equally important reason (as pointed out in section~\ref{sec:ft})
lies in the presence of inconsistencies in the low-energy data.
From fig.~\ref{f:sig_vs_ecm} no definite conclusion can be drawn
on the photon densities either.

The first measurement of single-inclusive distributions
for charm photoproduction at HERA has been presented by the H1
Collaboration~\cite{Aid96}. The data are integrated over a
large range of photon--proton energies, and the
results should therefore be compared with the theoretical predictions
for electroproduction in the Weizs\"acker--Williams approximation.
\begin{figure}
\centerline{\epsfig{figure=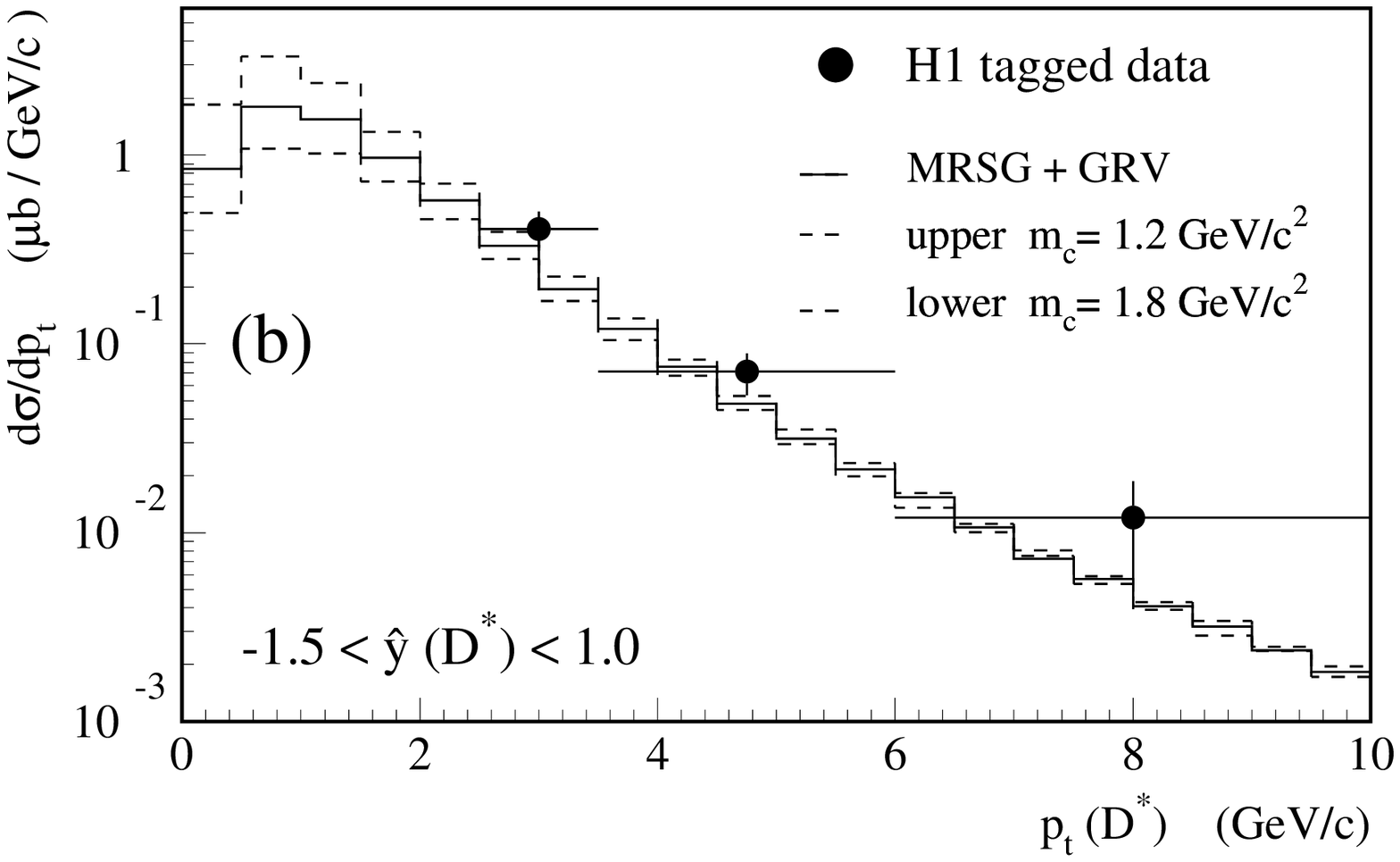,width=0.7\textwidth,clip=}}
\ccaption{}{ \label{f:pt_H1}
Transverse-momentum distribution of $D^*$ mesons in
$ep$ collisions at HERA, for $Q^2<0.01$~GeV$^2$.
Experimental results~\cite{Aid96}
are compared with next-to-leading order theoretical predictions.}
\end{figure}

Since the Weizs\"acker--Williams function grows in the small-$x$ region,
the bulk of the contribution to the cross section is due to photons
of relatively low energy. Therefore, the r\^ole of the hadronic component
of the photon
and of the small-$x$ region in the proton densities is marginal.
It turns out that, when cuts similar to the experimental
ones are applied and electroproduction in the Weizs\"acker--Williams
approximation is considered, the shape of the theoretically
predicted distributions is very stable~\cite{Frixione95a}.
 It follows
that single-inclusive charm electroproduction is of little help in
constraining the partonic densities of both the proton and
the photon, but can be used as a valuable test of the production mechanism.

Figure~\ref{f:pt_H1} presents the comparison between theory and H1
data~\cite{Aid96} for the $p_{\sss T}$ spectrum of the $D^*$ mesons.
The data are relative to the so-called ``tagged'' sample,
defined by the condition $Q^2<10^{-2}$~GeV$^2$.
The QCD predictions for charm quarks have been convoluted with the
Peterson fragmentation function~\cite{Peterson83}.
Given the limited statistics of the measurement, the comparison of
the data with the theory appears to be satisfactory, and qualitatively
analogous to similar results at fixed-target
experiments~\cite{Frixione94}. A comparison has also been carried out
by the H1 Collaboration \cite{Aid96} for the untagged sample
($\langle Q^2\rangle\simeq$~0.2~GeV$^2$);
the agreement is slightly worse, but still satisfactory.
As shown in fig.~\ref{f:pt_H1}, the uncertainty on the theoretical
prediction due to the value of the charm mass is completely negligible
when compared to the uncertainties in the experimental data.

\begin{figure}
\centerline{\epsfig{figure=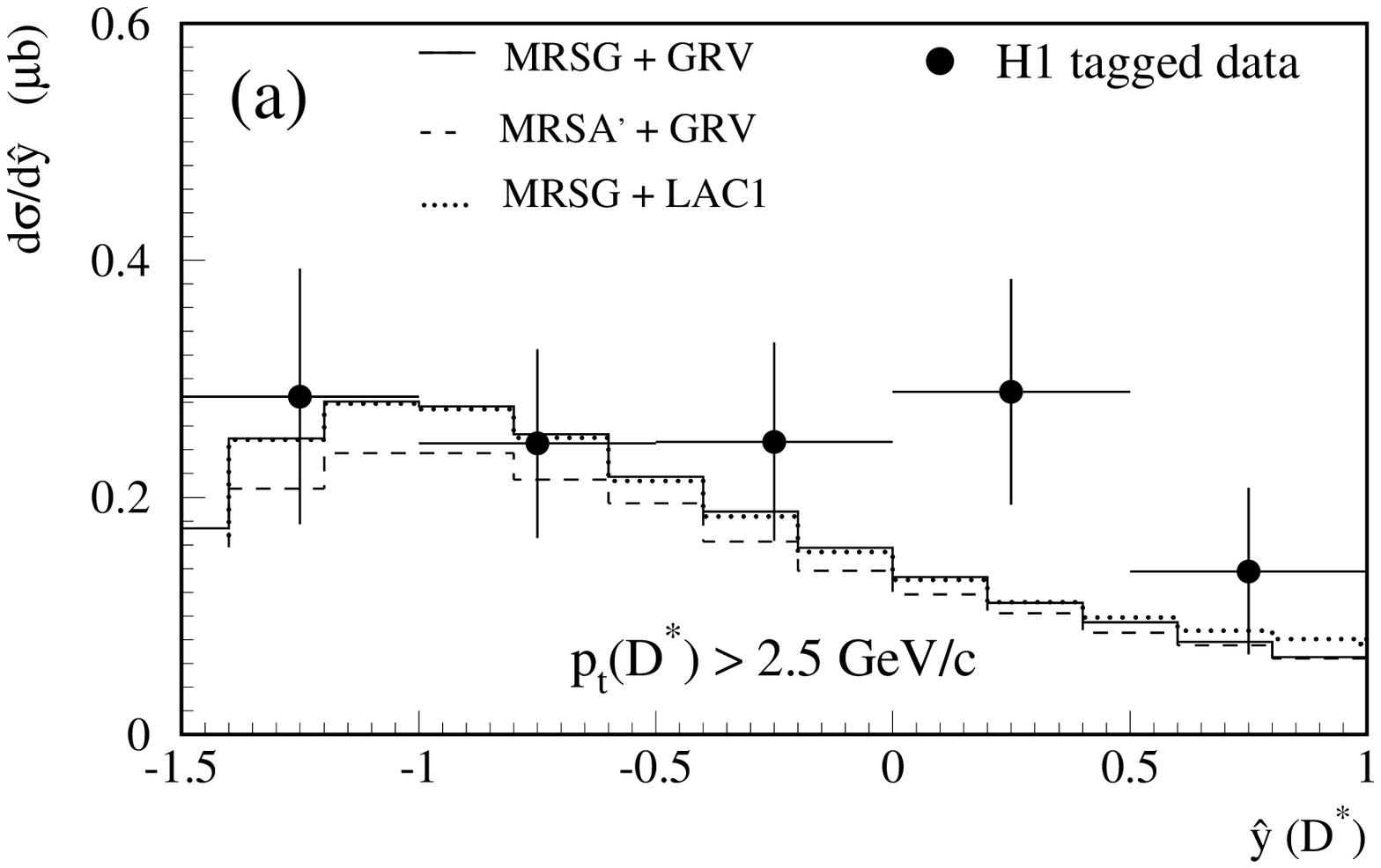,width=0.5\textwidth,
                    height=0.28\textheight,clip=}
            \hspace{0.3cm}
            \epsfig{figure=eta_286.eps,width=0.5\textwidth,
                    height=0.28\textheight,clip=}}
\ccaption{}{ \label{f:eta_HERA}
Left: rapidity distribution of $D^*$ mesons in $ep$ collisions at
HERA, for $Q^2<0.01$~GeV$^2$.
Experimental results~\cite{Aid96} are compared with
next-to-leading order QCD predictions. Right: pseudorapidity
distribution of fragmented charm quarks in monochromatic
photon--proton collisions.}
\end{figure}
The comparison between theory and H1 data for the rapidity distribution
is presented on the left-hand side of fig.~\ref{f:eta_HERA}.
Although the agreement is acceptable, only one point being more
than one standard deviation away from the theoretical predictions,
higher statistics are required in order to perform a more significant
comparison. Indeed, preliminary H1 results\footnote{C.~Grab, private
communication.} show
that, when improving the statistics, the data point at
$\hat{y}(D^\star)=0.25$ is also close to the theoretical curve.
We point out, however, that recent preliminary data from ZEUS \cite{Zeus96a}
are not in satisfactory agreement with QCD expectations.

If data with larger statistics were available, it would be possible
to consider the production processes initiated by very energetic
photons only. The right-hand side
of fig.~\ref{f:eta_HERA} shows the theoretical prediction for
the pseudorapidity distribution of fragmented charm quarks
in monochromatic photon--proton collisions, at a centre-of-mass
energy of $\sqrt{S_{\gamma p}}=286$~GeV, and for different choices of
the partonic densities in the photon. As is apparent from
the figure, the results obtained with the LAC1 and GRV-HO sets
are completely different in the forward region. Even at the moderate
pseudorapidity values covered by the present
configuration of the detectors, the large difference induced by
the two photon sets should have measurable effects.

Because of the large photon--proton centre-of-mass energy
available at HERA, the $\log(S/m_c^2)$ terms appearing in the
charm cross section may get large and spoil the convergence of the
perturbative series. The problem of resumming these terms
({\it small-x effects}) has been tackled by several
authors~\cite{Ellis90,Catani89,Catani90,Collins91}, mainly in
the context of $b$ production in hadronic collisions.
In ref.~\cite{Frixione95b} an estimate has been given of the
importance of the small-$x$ effect for charm physics in the
HERA energy range. It has been found that, resumming the $\log(S/m_c^2)$
terms, the total rate
can be increased by  by 20\% to 40\% with respect to the next-to-leading
order prediction for the point-like contribution, and by
20\% to 45\% for the hadronic contribution,
depending on the partonic densities used in the calculation
(as a rule of thumb, the flatter is the gluon density at
small $x$, the larger is the contribution expected from
the resummation of the small-$x$ effects). These effects are
therefore negligible with respect to the other sources of
uncertainty on the theoretical predictions.

We finally observe that the transverse-momentum distribution
is in principle affected by the presence of~\mbox{$\log(p_{\sss T}/m_c)$}
terms. These logarithms can be resummed by observing that,
at high $p_{\sss T}$, the heavy-quark mass is negligible, and
by using perturbative fragmentation functions~\cite{Cacciari96,Kniehl95}.
Remarkably enough, the fixed-order and the resummed results of
\cite{Cacciari96} agree
in a very wide range in $p_{\sss T}$. Recently, the massless approach
has been used in ref.~\cite{Kniehl96} to predict the $D^\star$
spectrum at HERA.

\subsection{Bottom photoproduction}
\begin{figure}
\centerline{\epsfig{figure=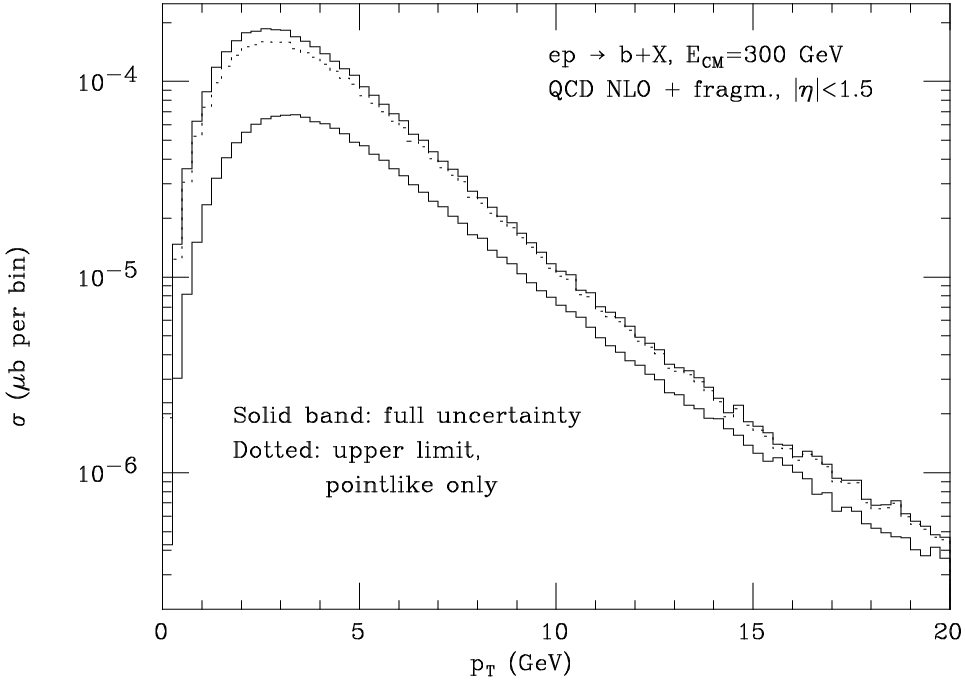,width=0.7\textwidth,clip=}}
\ccaption{}{ \label{f:b_pt_at_HERA}
Full uncertainty on the transverse-momentum distribution for bottom
electroproduction (Weizs\"acker-Williams approximation)
with Peterson fragmentation and a pseudorapidity cut.}
\end{figure}
Thanks to the higher value of the quark mass, perturbative QCD predictions
for bottom production are more reliable than those for charm. In fact,
all the uncertainties we have discussed for charm are in this case
strongly reduced. The resummation of the small-$x$ effects
has been estimated in ref.~\cite{Frixione95b} to increase the
next-to-leading order result by $\approx 5$\%.
In monochromatic photon--proton collisions, the point-like component
has an uncertainty of a factor of 2 if all the parameters are varied
{\it together} in the direction that makes the cross section larger
or smaller. At $\sqrt{S_{\gamma p}}=100$~GeV, the lower and upper limits of
the point-like component are 16~nb and 35~nb respectively, while
at $\sqrt{S_{\gamma p}}=280$~GeV we get 41~nb and
101~nb~\cite{Frixione95b}. The hadronic component
has larger uncertainties, but much smaller than in charm
production. As discussed above, the main source of
uncertainty is the parton density set
for the photon. Nevertheless, in bottom production the small-$x$
region is probed to a lesser extent than in charm production,
and the sensitivity of the result to the photon densities
is therefore milder; we get an uncertainty of a factor of 3
(to be compared with a factor of 10 in the case of charm).
The hadronic component can still be the dominant contribution
to the photoproduction cross section, if the gluon in the photon
is as soft as the LAC1 parametrization suggests.

The bottom rates are about a factor of 200 smaller than the
charm ones. To perform a statistically significant study of bottom
production, the luminosity upgrade at HERA is necessary.
In any case, it is very likely that a comparison with the theory
could only be done by considering electroproduction
in the Weizs\"acker--Williams limit.
In this case, consistently with what was discussed for charm,
the sensitivity of the theoretical predictions to the input
parameters is sizeably reduced, and a reliable
comparison between theory and data can be performed.
For example, in electroproduction the hadronic component contribution
to the total cross section is at most 75\% of the point-like contribution,
even if the LAC1 set is used. The most interesting results
are, however, obtained when considering more exclusive quantities,
such as single-inclusive distributions. In particular, as
shown in fig.~\ref{f:b_pt_at_HERA}, the transverse momentum
of the bottom quark at HERA can be predicted by
perturbative QCD quite accurately. It is clear that
even with the LAC1 set the hadronic component affects the prediction
only marginally; this fact is basically a consequence of the
applied pseudorapidity cut. We can therefore regard
fig.~\ref{f:b_pt_at_HERA} as a reliable prediction of QCD for
the $p_{\sss T}$ spectrum of $b$ hadrons at HERA.
The comparison of this prediction with the data would be extremely
useful in the light of the status of the comparison between theory and data
for $b$ production at the Tevatron (see section~\ref{sec:colliders}).

\subsection{Deep-inelastic production}
The first data on charm production in the deep-inelastic regime
at HERA have recently become available~\cite{Zeus96,Adloff96}.
On the theoretical side, NLO QCD calculations for total and single-inclusive
deep-inelastic production have been performed in
refs.~\cite{Laenen92a,Laenen93,Laenen93a}. Recently, fully exclusive
cross sections have been computed~\cite{Harris95,Harris95a}.
In this context, it is possible to perform comparisons
between theory and experiment in different aspects.

To start with, one can consider the contribution of charm quarks
to the proton structure functions $F_2$ and $F_L$. It turns out
that NLO corrections to these quantities are non-negligible.
The full NLO results display a mild dependence upon the renormalization
and factorization scales (well below 10\% for $Q^2$ larger than 10~GeV$^2$,
when varying scales between half and twice the default value
$\sqrt{Q^2+m_c^2}$). The dominant uncertainty in the calculation
is the small-$x$ behaviour of the gluon density in the proton.
For this reason, it is possible to conclude that structure-function
measurements may provide useful information about the gluon distribution
in the proton (see ref.~\cite{Laenen96a} for a discussion of this issue).
For $Q^2\gg 4m_c^2$ the calculation of $F_2$ and $F_L$ requires
the inclusion of large logarithmic effects arising from evolution.
Approaches to this problem can be found in refs.~\cite{Aivazis94,
Aivazis94a,Lai97} and \cite{Martin96}.

\begin{figure}
\centerline{\epsfig{figure=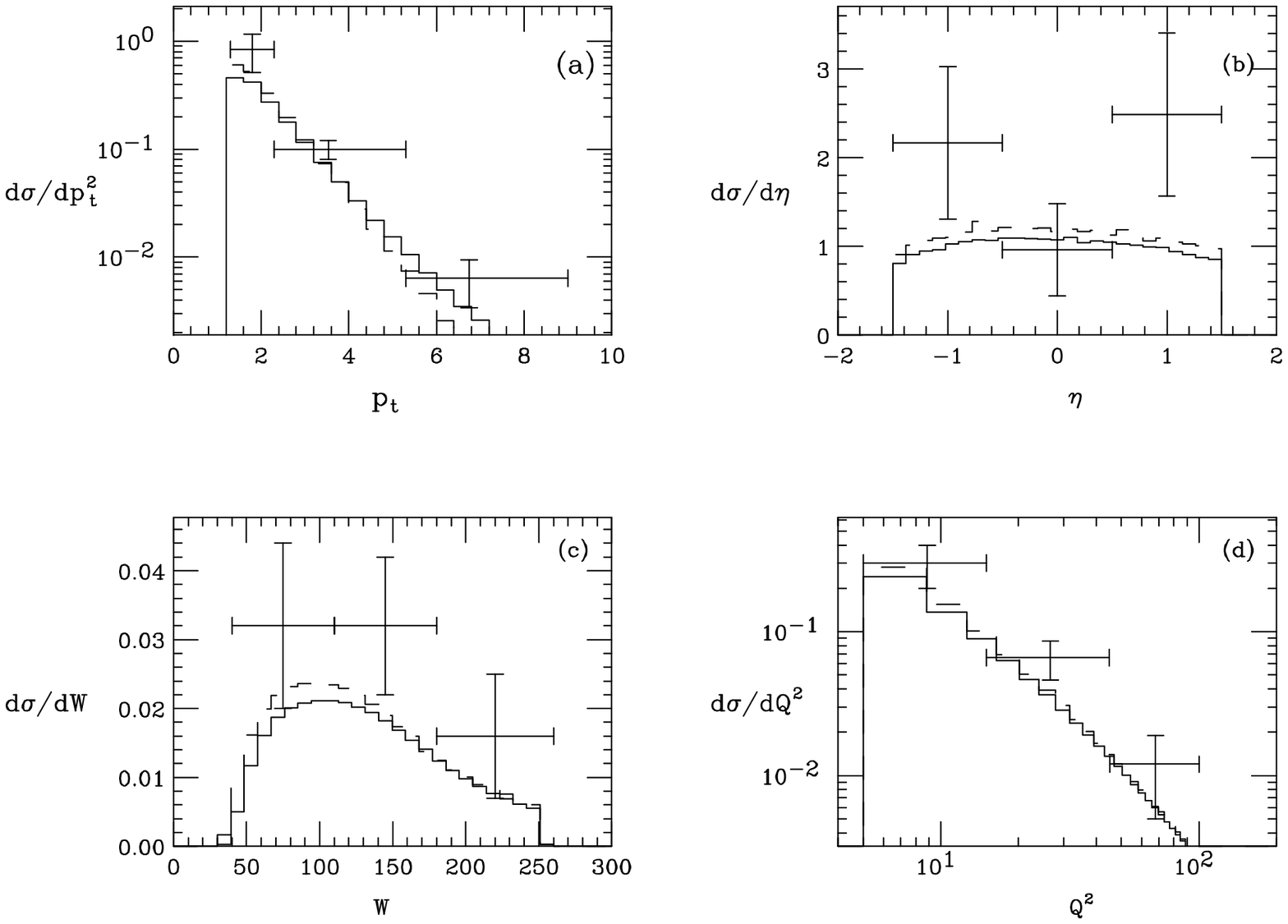,width=0.8\textwidth,clip=}}
\ccaption{}{ \label{f:DISsngl}
Comparison between theoretical predictions and experimental data
for single-inclusive distributions in DIS charm production.}
\end{figure}

In fig.~\ref{f:DISsngl}, taken from ref.~\cite{Laenen96a}, we
show experimental data for some single-inclusive distributions,
superimposed on QCD theoretical predictions at NLO.
The following distributions are shown:
a)~the $D^*$ transverse momentum $p_T^{D^*}$, b)~
its pseudorapidity $\eta^{{D^*}}$, c)~the hadronic
final state invariant mass and d)~the $Q^2$ one,
for the kinematic range 5 GeV$^2 < Q^2 < $ 100 GeV$^2$,
$1.3\,{\rm GeV}<p_T^{D^*}<9\,{\rm GeV}$
and $|\eta^{{D^*}}| < 1.5$. A cut on the DIS $y$ variable $0<y<0.7$
is also applied.
The $D^*$ momenta are obtained by applying a Peterson
fragmentation function to the produced charm quark.
Also shown are experimental data of the ZEUS collaboration \cite{Zeus96}.

The shape of the distributions considered is significantly modified
by the introduction of radiative corrections.
As we can see, the data are in reasonable agreement with theoretical
predictions. More statistics is needed for a meaningful comparison
with the theory.

\subsection{Future physics}
In this subsection we will discuss a few topics on future physics
possibilities in heavy-flavour production at HERA.
With the planned upgrades of the HERA collider, an integrated
luminosity of 100~pb$^{-1}$ or larger will be achieved. This will
probably allow the production of a sizeable sample of double-tagged charm
events \cite{Eichler96},
and therefore to study double-differential cross sections.
The possibility of having polarized beams has also been considered
\cite{Ingelman96}.
Furthermore, the HERA-B fixed-target program will become operational.
\subsubsection{Determination of $f_g^{(p)}$}
An interesting application of double-differential measurements
is the determination of the proton gluon density $f_g^{(p)}$
\cite{Buchmueller91,Riemersma92,Frixione93a}. At present,
this quantity is not directly measured: DIS data allow its indirect
extraction from the $Q^2$ evolution of the structure functions,
and direct photon and inclusive jet measurements constrain it
in complementary regions of $x$ and $Q^2$.

In photon or electron--hadron collisions, the advantage is that the
gluon density enters the cross section in a simpler way than in
the case of hadronic collisions. It is easy to show that this in turn
implies that, if one could determine the invariant mass and rapidity
of the produced system, the cross section would be directly proportional
to $f_g^{(p)}$ and to other calculable factors.
For example, the LO heavy-lavour electroproduction cross section
can be written, in the Weizs\"acker--Williams approximation, and
neglecting the hadronic component, in the following way:
\beq
\frac{d \sigma^{(0)}}{d y_{\qq} \, dM_{\qq}^2 }=
x_g\frac{d \sigma^{(0)}}{d x_g \, dM_{\qq}^2 }= \frac{1}{E^2}\;
f^{(e)}_{\gamma}(x_\gamma,\mu_0^2)
f^{(p)}_{g}(x_g,\muf^2)
\hat{\sigma}^{(0)}_{\gamma g}(M_{\qq}^2),
\label{ptoqq}
\eeq
where $M_{\qq}$ is the invariant mass of the heavy-quark pair, and
$y_{\qq}$ is the rapidity of the pair in the electron--proton
centre-of-mass frame (we choose positive rapidities in the direction
of the incoming proton); $E=\sqrt{S}$ is the electron--proton
centre-of-mass energy, and
\beqn
x_\gamma&=&\frac{M_{\qq} }{E} \exp(-y_{\qq}),
\\
x_g&=&\frac{M_{\qq} }{E} \exp(y_{\qq}).
\label{xgdef}
\eeqn
The function $f^{(e)}_{\gamma}$ is the Weizs\"acker--Williams
function we already discussed in subsection~\ref{subsec:phxsec}.
We observe that
all quantities on the right-hand side of eq.~(\ref{ptoqq}) are calculable, except for $f_g^{(p)}$, which
can therefore be measured.
\begin{figure}[htb]
\centerline{\epsfig{figure=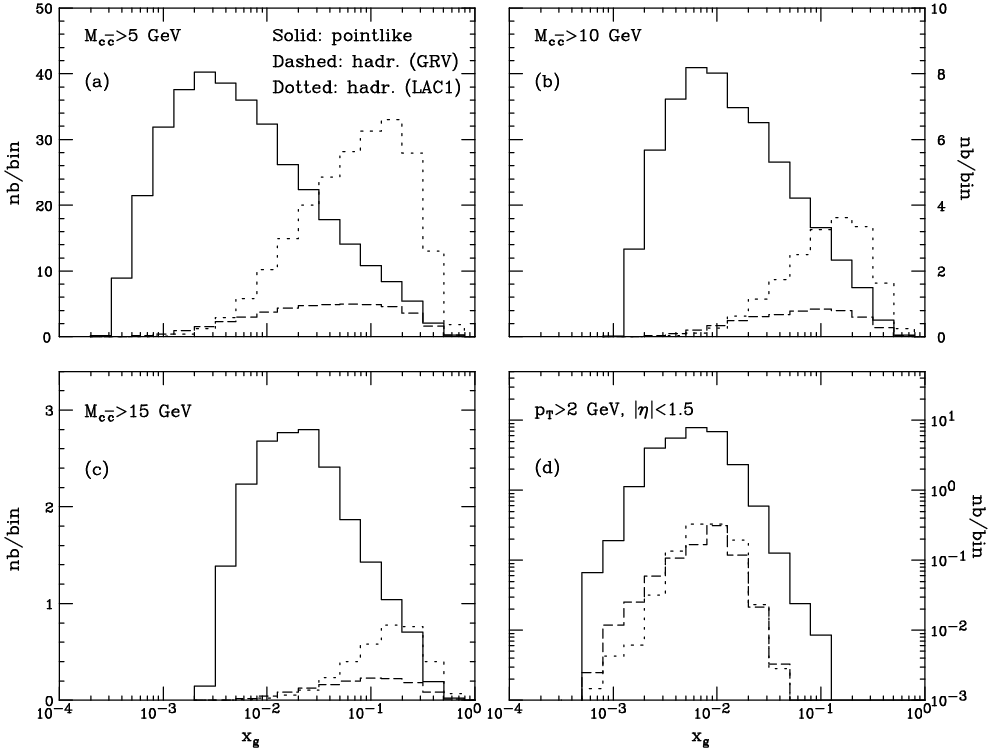,width=0.8\textwidth,clip=}}
\ccaption{}{ \label{f:xgdistr}
$x_g$ distribution in $ep$ collisions (Weizs\"acker-Williams approximation)
at HERA, for $m_c=1.5$~GeV and the MRSG proton parton-densities.}
\end{figure}
The inclusion of radiative corrections does not pose any problem.
This issue is discussed in detail in ref.~\cite{Frixione93a},
where it is shown that $f_g^{(p)}$ can be measured up to NLO
accuracy by using double-differential charm data.

As discussed previously, the heavy-flavour cross section receives a large
contribution from the hadronic component. In order to extract the
gluon density of the proton with the method outlined above,
we have to consider only those kinematical regions where
the hadronic component is suppressed.
We study this possibility in fig.~\ref{f:xgdistr},
where we present the NLO QCD
distribution in the variable $x_g$,
in electron-proton collisions at $\sqrt{S}=314$~GeV.
The partonic densities in the proton are given by the MRSG set, while we
considered both the LAC1 and GRV-HO sets for the photon.
Figures~\ref{f:xgdistr}a)--\ref{f:xgdistr}c) show the effect of
applying a cut on the invariant mass of the pair. Even in the
case of the smallest invariant-mass cut, there is a region of small
$x_g$ where the hadronic component is negligible with respect to the
point-like one. When we increase the invariant-mass cut, we notice that
the hadronic component decreases faster than the point-like one.
This is due to the fact that, for large invariant masses,
the production process of the hadronic component is suppressed by the small
value of the gluon density of the photon at large $x$. By pushing
the invariant-mass cut to 20 GeV, it turns out that the point-like
component is dominant over the hadronic one for $x_g$ values
as large as $10^{-1}$. We then conclude that the
theoretical uncertainties affecting the charm cross section,
in the range of $10^{-3}<x_g<10^{-1}$, are small
enough to allow for a determination of the gluon density in the proton
by using invariant-mass cuts to suppress the hadronic component.
In a more realistic configuration, as the present one of
the detectors at HERA, additional cuts are applied to the data.
Figure~\ref{f:xgdistr}d) shows the effect, on the $x_g$ distribution,
of a small-$p_{\sss T}$ and a pseudorapidity cut applied
to both the charm and the anticharm. In this
case, even without applying an invariant-mass cut, the point-like
component is dominant in the whole kinematically accessible range.
\subsubsection{Polarization asymmetries}
It is conceivable that, in the future, the HERA collider will
be operated in a polarized mode. At present, a good result for
the polarization of the positron beam has been obtained, and
feasibility studies for the polarization of the proton beam
are under way~\cite{Ingelman96}.
At leading order in QCD, the heavy-flavour production
cross section in
polarized $ep$ collisions is proportional
to the polarized gluon density in the proton, $\Delta g$. Therefore,
data on charm production could be
used to measure $\Delta g$ directly, as previously
shown for the unpolarized case. In practice, the situation
for the polarized scattering is much more complicated.
First of all, it is not possible to perform a complete NLO
analysis, because NLO cross sections for
polarized charm photoproduction have never been computed.
Furthermore, there is no experimental
information on parton densities in the polarized {\it photon}.
It is reasonable, however, that charm production
at the HERA collider in the polarized mode can help in constraining
the polarized gluon density in the proton. This possibility
was first suggested in refs.~\cite{Gluck88,Gluck91,Vogelsang91},
and recently reconsidered in refs.~\cite{Frixione96b,Stratmann96}.

In order to reduce the impact of the uncertainties induced by
radiative corrections, it is useful to present
predictions~\cite{Frixione96b} for the ratio $\Delta\sigma/\sigma$
(asymmetry), where $\sigma$ is the unpolarized cross section and
\beq
\Delta\sigma=\frac{1}{2}
\left(\sigma^{\uparrow\uparrow}-\sigma^{\uparrow\downarrow}\right).
\eeq
Here $\sigma^{\uparrow\uparrow}$ and $\sigma^{\uparrow\downarrow}$
are the cross sections for $c\bar{c}$ production with parallel and
antiparallel polarizations of the incoming particles respectively.
One might expect that the effect of the radiative corrections
approximately cancels in the ratio. It must be stressed that, for consistency,
the unpolarized cross section $\sigma$ appearing in the asymmetry
must be calculated at the leading order, as the polarized one.
\begin{figure}[htb]
  \begin{center}
    \mbox{
      \epsfig{file=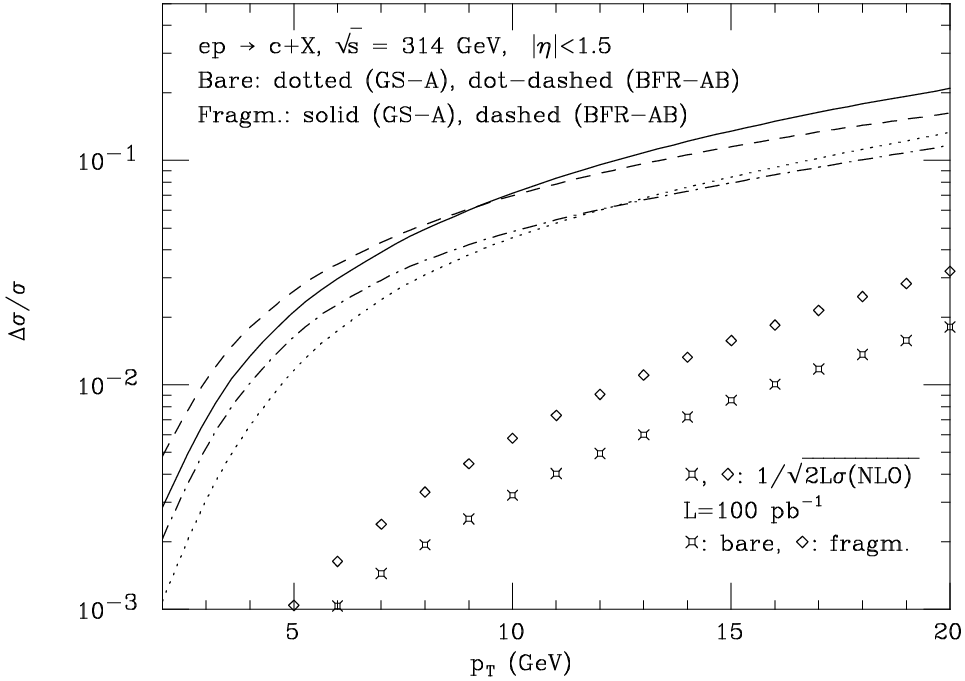,width=0.70\textwidth}
      }
  \ccaption{}{\label{f:ptel}
Asymmetry cross section versus transverse momentum in polarized $ep$
collisions (Weizs\"acker-Williams approximation)
at $\sqrt{s}=314$~GeV. The minimum observable asymmetry,
computed at next-to-leading order, is also displayed.
}
  \end{center}
\end{figure}
The NLO value of $\sigma$ can then be
used to estimate the sensitivity of the experiment.

A rough estimate of the
minimum value of the asymmetry observable at HERA can be obtained by requiring
the difference between the numbers of events with parallel and antiparallel
polarizations of the initial-state particles to be larger than the
statistical error on the total number of observed events. This gives
\beq
\left[\frac{\Delta\sigma}{\sigma}\right]_{min}\simeq
\frac{1}{\sqrt{2\sigma{\cal L}\epsilon}},
\label{minassmtr}
\eeq
where $\cal L$ is the integrated luminosity and the factor $\epsilon$
accounts for the experimental efficiency for charm identification
and for the fact that the colliding beams are not completely polarized.
This procedure can be applied to total cross sections, as well as
to differential distributions; in this case, the values of
$\sigma$ and $\Delta\sigma$ have to be interpreted as {\it cross
sections per bin} in the relevant kinematical variable.

In ref.~\cite{Frixione96b} it was shown that total cross section
asymmetries for the point-like component are quite small in absolute
value, and can be measured only if $\epsilon$ is equal to or
larger than 1\% (0.1\%), assuming ${\cal L}=$~100~pb$^{-1}$
(1000~pb$^{-1}$). Therefore, even with a vertex detector (see
ref.~\cite{Eichler96}), it appears unlikely that this
kind of measurements will be performed at HERA. Furthermore,
in ref.~\cite{Stratmann96}
a rough estimate of the hadronic contribution to the polarized
cross section has been given, assuming polarized parton densities
in the photon to vary between zero and the corresponding unpolarized densities
in order to get a lower and an upper bound on the cross section.
It was found that a non-negligible contamination of the point-like
result might indeed come from the hadronic process.
The situation clearly improves when considering more exclusive
quantities; in ref.~\cite{Frixione96b} it was found that at moderate
$p_{\sss T}$ values the asymmetry for the point-like component
can be rather large, well above the minimum observable
value (in this region, the experimental efficiency is
sizeable~\cite{Eichler96}); this is shown in fig.~\ref{f:ptel}. In
ref.~\cite{Stratmann96} it was argued that the hadronic component
should have a negligible impact in this case.

We conclude that charm data in high-energy polarized $ep$ collisions
will help in the determination of the polarized gluon density
of the proton, provided that the integrated luminosity and the
experimental efficiencies will allow a study of at least
single-inclusive distributions.
\subsubsection{HERA-B}
The HERA-B program will allow a detailed study of $b$
production in proton--copper collisions, at $\sqrt{S}=39.2\;$GeV.
Since this energy is relatively close to threshold,
one may worry about the presence of large soft gluon
resummation effects.
Large resummation effects were found in ref.~\cite{Kidonakis95}.
In ref.~\cite{Catani96a} it was instead argued that these large
effects have a spurious origin, and a properly performed resummation
gives corrections of the order of 10\%.
This issue will be discussed in more detail in section~\ref{sec:resum}.
For example, from the NLO calculation
with the MRSA$^\prime$ \cite{Martin95} parton
densities and $m_b=4.75\,$GeV, we get
$\sigma_{b\bar{b}}=10.5{ +8.2  \atop -4.7}\,$nb,
a range obtained by varying the renormalization
and factorization scales from $m_b/2$ to $2m_b$.
Thus the upper band is 80\% higher than the central value, to be compared with
a 10\% increase from the resummation effects of ref.~\cite{Catani96a}.
This result is much less dramatic than the results of ref.~\cite{Kidonakis95}.
For completeness we report in table~\ref{tab:herabbot} cross section
values obtained using different sets of structure functions, corresponding
to different values of $\lambdamsb$, and to different values
of the bottom quark mass.
Observe that measurements of $b$ cross sections in $pN$
collisions at a beam energy of 800~GeV, close to the HERA-B regime,
have been presented in refs.~\cite{Jansen95,Spiegel96}.
These measurements (see fig.~\ref{bcproton} in section~\ref{sec:ft})
are consistent with the range reported in table~\ref{tab:herabbot}.
\begin{table}[htb]
\begin{center}
\begin{tabular}{|l||c|c|c|} \hline
$m_b$ (GeV)
& MRS, $\as(M_{\rm Z})=0.105$
& MRSA$^\prime$
& MRS, $\as(M_{\rm Z})=0.125$
\\ \hline\hline
4.5  & $ 12.9{+7.7 \atop -5.1} $ & $ 16.5{+12.8 \atop -\phantom{1}7.3} $
    & \addsp{$ 30.2{+43.3 \atop -16.6} $}
\\ \hline
4.75 & $ 8.2{+5.0 \atop -3.3} $ & $ 10.5{+8.2 \atop -4.7} $
    & \addsp{$ 18.6{+27.0 \atop -10.2} $}
\\ \hline
5.0  & $ 5.2{+3.2 \atop -2.1} $ & $ 6.6{+5.4 \atop -3.0} $
    & \addsp{$ 11.5{+16.9 \atop -\phantom{1}6.3} $}
\\ \hline
\end{tabular}
\ccaption{}{\label{tab:herabbot}
Total cross sections in nb for bottom production at HERA-B.
The central values correspond to $\mu=m_b$, upper and lower values
to $\mu=m_b/2$ and  $\mu=2m_b$ respectively.
The structure functions are those of refs.~\cite{Martin95,Martin95a}.
}
\end{center}
\end{table}
%
\section{Heavy-quark production at hadron colliders}\label{sec:colliders}
\sethead{Hadron colliders}
Hadron collisions are by far the largest source of heavy quarks available
today. While the environment of high-energy hadronic interactions does not
allow to trigger on the largest fraction of charm and bottom produced, the
production rates are so large that the number of recorded events
allows today $b$-physics studies that are competitive with those of $e^+e^-$
experiments. The introduction of new experimental
techniques, such as the use of silicon vertex detectors, which enable the
tagging of events containing bottom quarks, led in the recent
years to high-statistics and low-background measurements of the $b$-production
properties over a large range of transverse momenta. The high energies
and luminosities available at the Tevatron $p \bar p$ collider, in addition to
the enhanced $b$-tagging capability, have furthermore led to the discovery of
the top quark \cite{Abe94,Abe95,Abachi95}, opening the way to a new set of
tests of the heavy-quark production dynamics.

From the point of view of QCD studies, heavy-flavour production
in high-energy hadronic collisions has better potentials than in fixed-target
experiments. The $b$ quarks produced at large $p_{\sss T}$ can be studied in
perturbative
QCD with smaller contamination from non-perturbative effects.
For example, the impact of the initial-parton transverse momentum
is largely reduced with respect to fixed-target charm production.
Furthermore, the fragmentation properties of heavy flavours in high
transverse momentum jets can be directly studied, since the transverse
momenta involved are typically perturbative.
Finally, top production should be fully perturbative, and therefore
the ultimate testing ground for the theory of heavy-flavour hadroproduction.

In this section we review the current status of these measurements and
their implications for the study of QCD.
We discuss here first the phenomenology of
charm and bottom production. We cover the inclusive production of $b$ quarks,
$b$ hadrons, jets containing $b$ and $c$ quarks, $b\bar b$
correlations, and we finally discuss the associate production
of heavy flavours and electroweak gauge bosons.
We then discuss the phenomenology of top-quark production.
The large mass of the top quark implies that theoretical
predictions based on NLO calculations should provide a rather accurate
description of its inclusive production properties, with uncertainties in the
range of 10\%. The experimental statistics currently available cannot
probe the theory at this level of accuracy, and significant improvements will
only occur with the next round of data taking at the Tevatron and with the LHC.
Nevertheless, the current data are an important step forward, and it is
encouraging that they basically confirm the theoretical expectations.

\subsection{Inclusive bottom production}\label{Inclusive-Bottom-Production}
\subsubsection{Preliminaries}
The distribution most commonly studied by the hadron-collider experiments is
the $b$-quark differential \pt\ spectrum, integrated within a fixed
rapidity range and above a given \pt\ threshold (\ptmin):
\be \label{eq:ptmin}
        \sigma(\ptmin) = \int_{\vert y \vert < y_{max}} d \, y
        \; \int_{\pt>\ptmin} d\,\pt \frac{d^2\sigma}{dy\, d \pt} \;.
\ee
The UA1 experiment at the CERN $p\bar{p}$
collider used $y_{max}=1.5$, while
the CDF and D0 experiments at the Tevatron use $y_{max}=1$.
QCD calculations for this distribution are available at
NLO~\cite{Nason89,Beenakker91}.

\begin{figure}[htb]
\centerline{\epsfig{figure=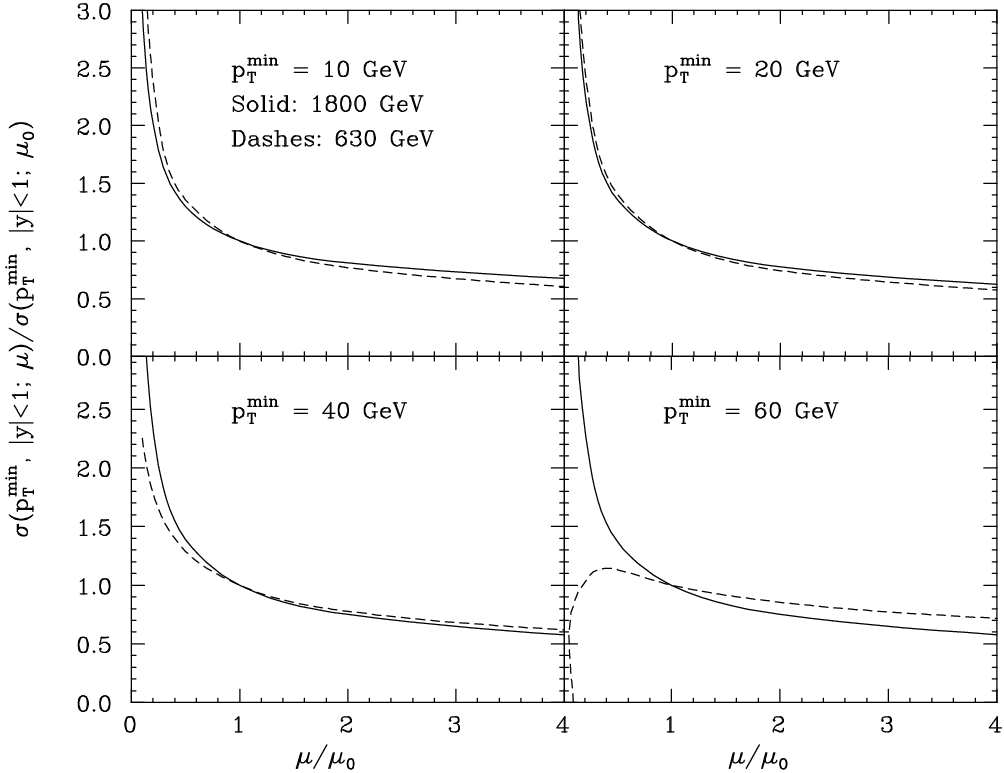,width=0.7\textwidth,clip=}}
\ccaption{}{ \label{fig:bscaledep}
Scale dependence of the integrated $b$-quark \pt\ distribution at 630~GeV (dashed
lines) and at 1800~GeV (solid lines), for different values of $\ptmin$.}
\end{figure}

In analogy to the case of fixed-target
charm production, the study of NLO $b$ cross sections
indicates a strong dependence on
the choice of factorization and renormalization scales. We show in
fig.~\ref{fig:bscaledep} this dependence
at various values of \pt\ for the two energies $\sqrt{S}=630$ and 1800~GeV.
Notice the growth of the cross sections at small scale values. Notice
furthermore that for values of $\ptmin<40$~GeV the scale dependence at 630~GeV
and 1800~GeV are approximately the same. The large scale dependence indicates
that corrections of higher orders are significant both at 630 and at 1800~GeV.
Above \ptmin=~40~GeV, on the other
hand, the scale dependence at 630~GeV is significantly reduced, suggesting a
more reliable perturbative expansion.

In the following we will present results obtained by varying simultaneously
$\muR$ and $\muF$ within the range\footnote{Unless
otherwise specified we will always assume the renormalization and factorization
scales to be equal, and will simply denote them by $\mu$.}
$\muo/2 < \mu < 2\muo$ ($\muo^2=\pt^2+m_b^2$),
which represents an acceptable window within which to explore the scale
dependence.

The uncertainty induced by the choice
of different sets of parton density functions (PDF) is rather marginal,
given the tight
constraints set by the DIS and HERA data.
One should, however, remember that the gluon density, which is the most important
ingredient in $b$ and $c$ production at colliders, has not yet been
measured directly for
the values of $x$ probed by the $b$
measurements at CERN and at the Tevatron, approximately limited to
the range $0.01<x<0.1$. As a standard set of parton densities we
will choose here the MRSA$^\prime$ set~\cite{Martin95}, for which
$\lambdamsb = 152$~MeV.  It should be pointed out, however, that
this value of $\lambdamsb$ yields a value of $\as$ significantly lower
than that extracted from different observables \cite{Barnett96}. Parton
distribution fits have therefore also been performed
with fixed values of $\lambdamsb$~\cite{Martin95a,Lai96}.
To explore the dependence
of our results on the choice of $\as$, we will also present results obtained
using the PDF set MRS125~\cite{Martin95a}, which
was extracted by forcing $\as(M_{\rm Z})=0.125$.
Notice that this procedure is not necessarily consistent, since
the inconsistency in the extraction of $\as$ is present in the data themselves.
Should the future DIS measurements of $\as$ become closer to the LEP
value, as suggested by recent results \cite{Harris96,Seligman97,Schmelling96},
it is very likely that also the PDF fits, and with them
our results, will change.

Given the large values
of \pt\ probed by the collider experiments, the $b$-mass dependence of the
theoretical result is small. We will choose as a default the value
$\mb = 4.75$~GeV, and allow for variations in the range
$4.5~{\rm GeV}<\mb<5$~GeV.

\subsubsection{The effect of higher-order corrections}
The significant scale dependence of the NLO results for bottom production is a
symptom of large contributions from higher orders. We briefly discuss here
two possible sources of higher-order corrections: small-$x$ effects, possibly
relevant in the low-\pt\ domain, and large logarithms, which appear at
high \pt.

The possibility that higher-order corrections at the Tevatron may be very
large had been pointed out early on in
ref.~\cite{Nason88}.
One can show that, in the high-energy limit ($\rho=4m_Q^2/S\; \to\; 0$),
large corrections arise to all orders in perturbation theory.
The source of these large corrections is the presence,
starting from \oacube, of diagrams where a gluon exchanged in the $t$ channel
becomes soft. In this limit, it is easy to show that the dominant
behaviour of the Born and
of the {\oacube} term are
\beqn  \label{eq:smallx1}
   \hat \sigma_{Born}(gg \to Q\bar Q) &\stackrel{\hat\rho \to 0}{\propto}&
   \frac{\as^2}{m_Q^2}\;\hat\rho\;, \\
   \hat \sigma_{(3)}(gg \to Q\bar Q) &\stackrel{\hat\rho \to 0}{\propto}&
   \frac{\as^3}{m_Q^2}\;,
\eeqn
where $\hat\rho=4m_Q^2/\hat s$. A crude but revealing estimate of the
relative size of leading-order and higher-order terms can be obtained by
assuming a simple form of the gluon densities at small $x$, e.g. $f(x) =
A/x^{1+\delta}$, with $\delta < 1$. It is easy to show that the total cross
sections scale as follows:
\begin{equation}\label{eq:smallx3}
 {{\sigma_{(3)}(gg \to Q\bar Q)} \over {\sigma_{Born}(gg \to Q\bar Q)}}
 \sim \left\{ { { {\displaystyle \as\,\log\frac{1}{\rho}\quad
\mbox{(if $\delta\log 1/\rho <1$)} } \atop
{\displaystyle \as\,\frac{1+\delta}{\delta}\quad
\mbox{(if $\delta\log 1/\rho >1$)} } } } \right.
\end{equation}
Notice that in the Feynman scaling approximation (e.g. $x\,g(x)$ constant at
small $x$),
the \oacube\ correction is enhanced by a large logarithm of the ratio between
the total hadronic CM energy and the heavy-quark mass. Logarithms of this type,
also known as {\em small}-$x$ logarithms,
arise at all orders of perturbation theory, and need to be resummed. The
resummation theory for this class of corrections has been applied
to the problem of heavy-flavour production in
refs.~\cite{Ellis90,Collins91,Catani90,Catani91,Levin91}. The results contained
in ref.~\cite{Collins91} indicate an enhancement of the NLO total  bottom cross
section at the Tevatron by no more than 30\%  due to the resummation of
small-$x$ effects, the highest values being obtained with small-$\delta$
gluon densities.
Naive reasoning based upon eq.~(\ref{eq:smallx3}) would lead to a similar
pattern.

These results suggest that small-$x$ effects at the current collider energies
are not large enough to justify discrepancies as large as those initially found
at the Tevatron. Use of a gluon density more singular than $1/x$, such as for
example recent parametrizations obtained from the HERA data, should reduce
these effects even more.

Another class of potentially large higher-order corrections appears when $b$
quarks are produced at high transverse momentum. At large momentum, in fact,
the $b$ quark behaves more and more like a massless particle, radiating an
increasingly large amount of its energy in the form of hard, collinear
gluons. This physical phenomenon is associated with the presence of logarithms of
the form $\log(\pt/m)$, which appear at all orders in perturbation theory.
This problem
was already examined in ref.~\cite{Nason89}. Some logarithmically enhanced
higher-order terms were estimated, and found to be negative and small
up to transverse momenta of the order of 20 GeV.

Techniques are available to resum this class of logarithms.
Two different groups have approached this problem in the recent past. Cacciari
and Greco \cite{Cacciari94} have folded the NLO cross section for production of
a massless parton $i$ ($i=g,q$) \cite{Aversa89}
with the NLO fragmentation function for the
transition $i\to b$ \cite{Mele90,Mele91}.
The evolution of the fragmentation functions resums all
terms of order $\as^n \log^n (\pt/m)$ and $\as^{n+1} \log^n (\pt/m)$.
All the dependence on the $b$-quark mass lies in the
boundary conditions for the fragmentation
functions. This approach ensures a full NLO accuracy, up to corrections of
order $m^2/(m^2+p_{\scriptscriptstyle T}^2)$. In particular, this formalism
describes NLO corrections to the gluon
splitting process, which in the \oacube\ calculation is only included
at the leading order.
One can verify, by looking at the Born-level production process as a
function of the quark mass, that,
in order for the mass corrections not to exceed the 10\% level, it is
necessary to limit the applications of this formalism to the region of
$\pt \gsim 20$~GeV.

\begin{figure}[htb]
\centerline{\epsfig{figure=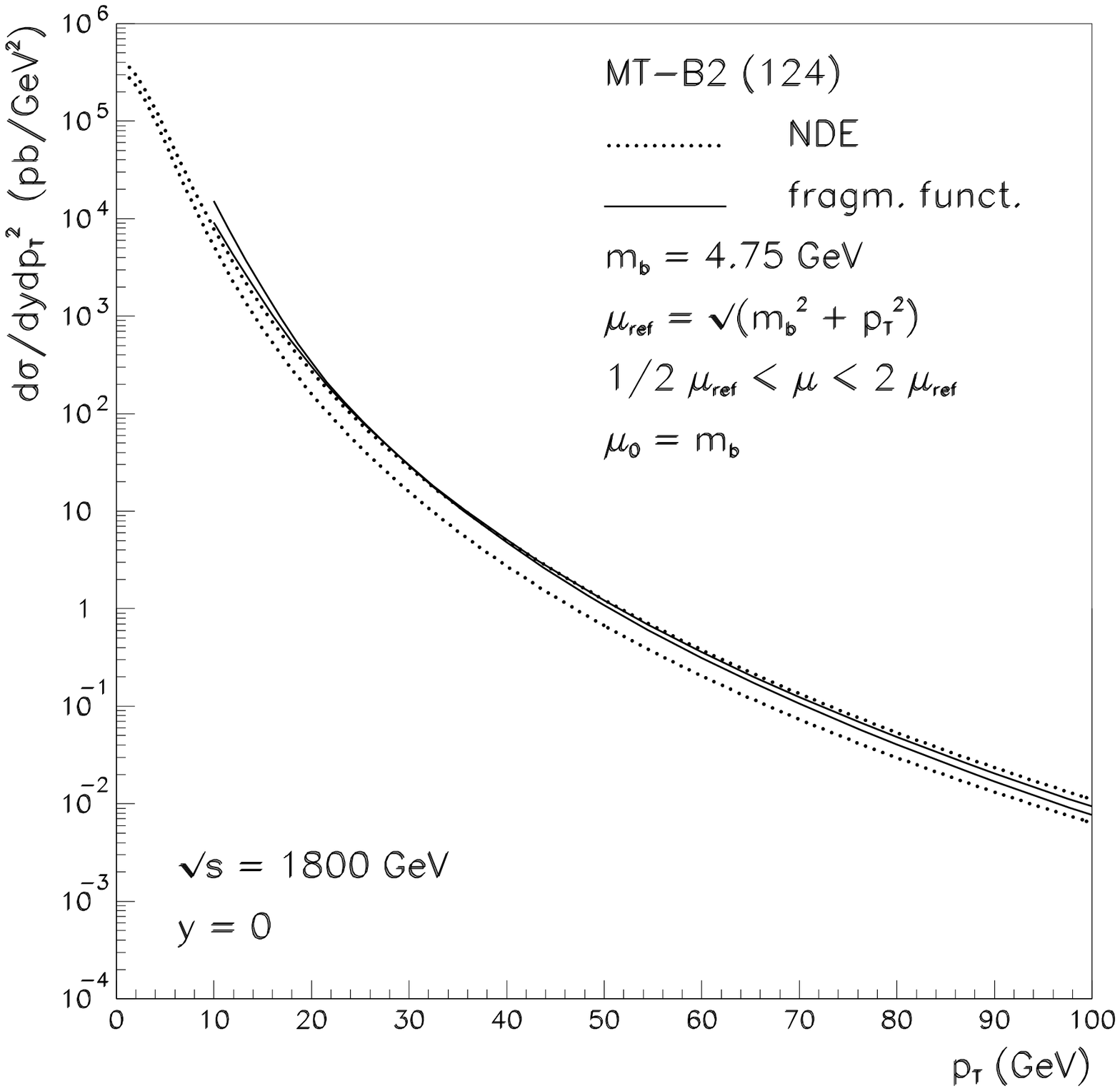,width=0.7\textwidth,clip=}}
\ccaption{}{ \label{fig:greco}
Differential $b$-quark \pt\ distribution: comparison of the fixed-order NLO
calculation (NDE stands for \cite{Nason88})
with the fragmentation-function approach \cite{Cacciari94};
$\mu_0$ is the scale at which the boundary conditions for the fragmentation
functions are set.}
\end{figure}

Figure~\ref{fig:greco} shows the differential $b$-quark \pt\ distribution
obtained in the fragmentation-function approach, compared to the standard
fixed-order NLO result. Several features of this figure should be
noticed. To start with, the scale
dependence is significantly reduced with respect to the fixed-order calculation.
Furthermore, in the range of
applicability of this calculation (i.e. $\pt \gsim 20$~GeV) the result of the
fragmentation-function approach lies on the upper side of the fixed-order NLO
calculation. The resummed calculation is, however, always
within the uncertainty band of the fixed-order one. Finally,
notice that the overall effect of the
inclusion of higher-order logarithms is a steepening of the \pt\ spectrum,
as is natural to expect.

Another calculation has recently appeared, by Scalise, Olness and Tung
\cite{Scalise96}. In this approach the authors employ a strategy developed in
the case of DIS in refs. \cite{Aivazis94,Aivazis94a}.
Here initial- and final-state mass
singularities are resummed as in the fragmentation-function approach, and
the result is then matched in the
low-\pt\ region to the fixed-order NLO massive calculation.
At large \pt\ this calculation does not include as yet, however, the full set
of NLO corrections to the hard-process matrix elements.

The preliminary numerical results of this study \cite{Scalise96}
are consistent with those of
the approach by Cacciari and Greco; in particular, they support the conclusion
that in the \pt\ range explored by the Tevatron experiments the resummed
results are consistent with the fixed-order ones, provided a scale
$\mu$ of the order of $\muo/2$ or smaller is selected.

The fragmentation function formalism discussed so far
can also be applied to high transverse momentum charm-meson
production \cite{Cacciari96b}.

\subsubsection{Comparison with experimental results}
\begin{figure}[htb]
\centerline{\epsfig{figure=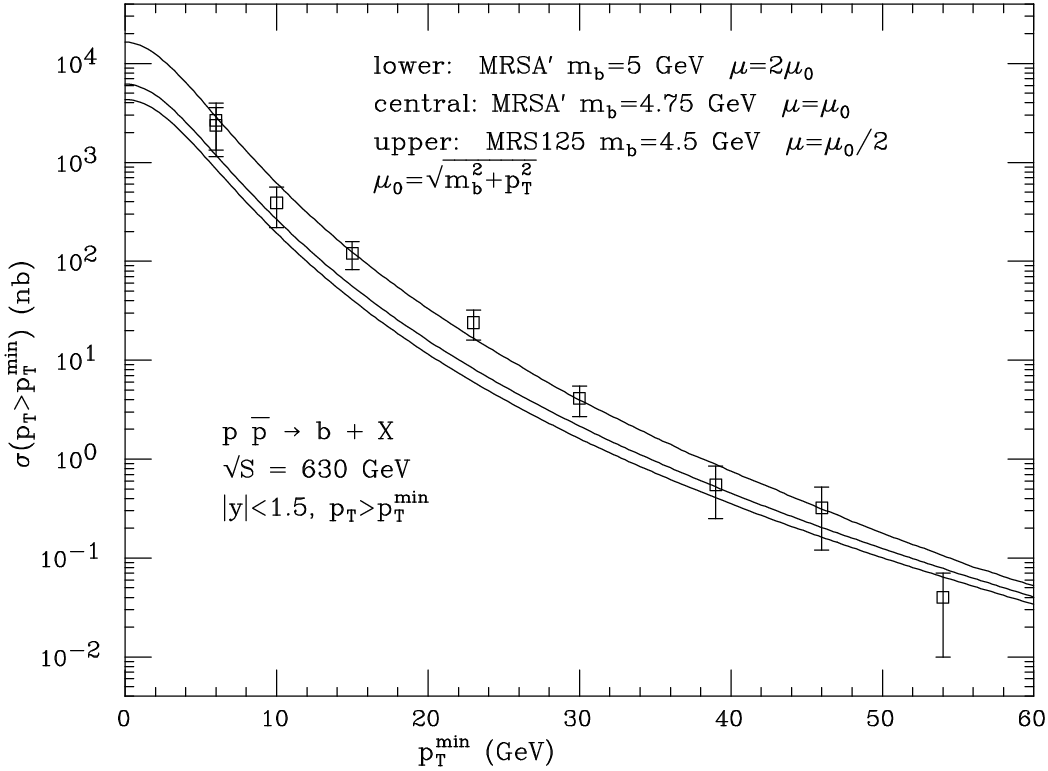,width=0.7\textwidth,clip=}}
\ccaption{}{ \label{fig:ua1bpt}
UA1 data on the integrated $b$-quark \pt\ distribution, compared to the results
of NLO QCD.}
\end{figure}

The comparison of the final experimental data from UA1~\cite{Albajar91} with
theory is shown in fig.~\ref{fig:ua1bpt}.
The data are (approximately) a factor of
2 higher than the central prediction. Good agreement can be obtained by
selecting the largest theoretical prediction.

Initially, the comparison with the Tevatron data led  to pessimistic
conclusions on the capability of NLO QCD to describe the data. Limited
statistics in the channels with fully reconstructed $b$-meson
decays~\cite{Abe92a}, and the insufficient theoretical understanding of the
J$/\psi$ production mechanisms, which was required at that time for the extraction of the
$b$ cross section from the measurement of inclusive
J$/\psi$'s~\cite{Abe92b,Abe93b}, led to evidence that the production rate of
$b$ quarks exceeded by several times the one predicted by NLO QCD.
The disagreement with theory was less pronounced in the case of
higher-statistics measurements using semileptonic $b$ decays~\cite{Abe93,Abe93a},
which however probe production at \pt\ values higher than those based on
exclusive and J$/\psi$ decays.

The analysis of the large data samples accumulated during the Tevatron data
taking of 1992--93 has recently been completed, and the first results
from the data of the 1994--96 run are being released. This wealth of data
and the improved detection capabilities offer today a rather precise picture of
inclusive $b$ production at 1800~GeV. The CDF experiment primarily used its
silicon vertex detector to improve the background rejection in the
reconstruction of fully exclusive $b$-hadron decays and to separate the
J$/\psi$'s coming from $b$ decays from those promptly produced, thus
eliminating an important source of systematic error. The measurement of fully
reconstructed $b$-hadron decays also allows a precise determination of the
differential \pt\ spectrum of $b$ hadrons, free of the uncertainties related to
the modelling of fragmentation and decay. The experiment D0 exploited its low
muon backgrounds to extend the measurement of the $b$ spectrum to high
values of \pt~\cite{Abachi95a}, as well as using dimuon
pairs~\cite{Abachi96,Abachi96a} and forward muons for a measurement of
small-angle $b$ production~\cite{Abachi96b}.

\begin{figure}[htb]
\centerline{\epsfig{figure=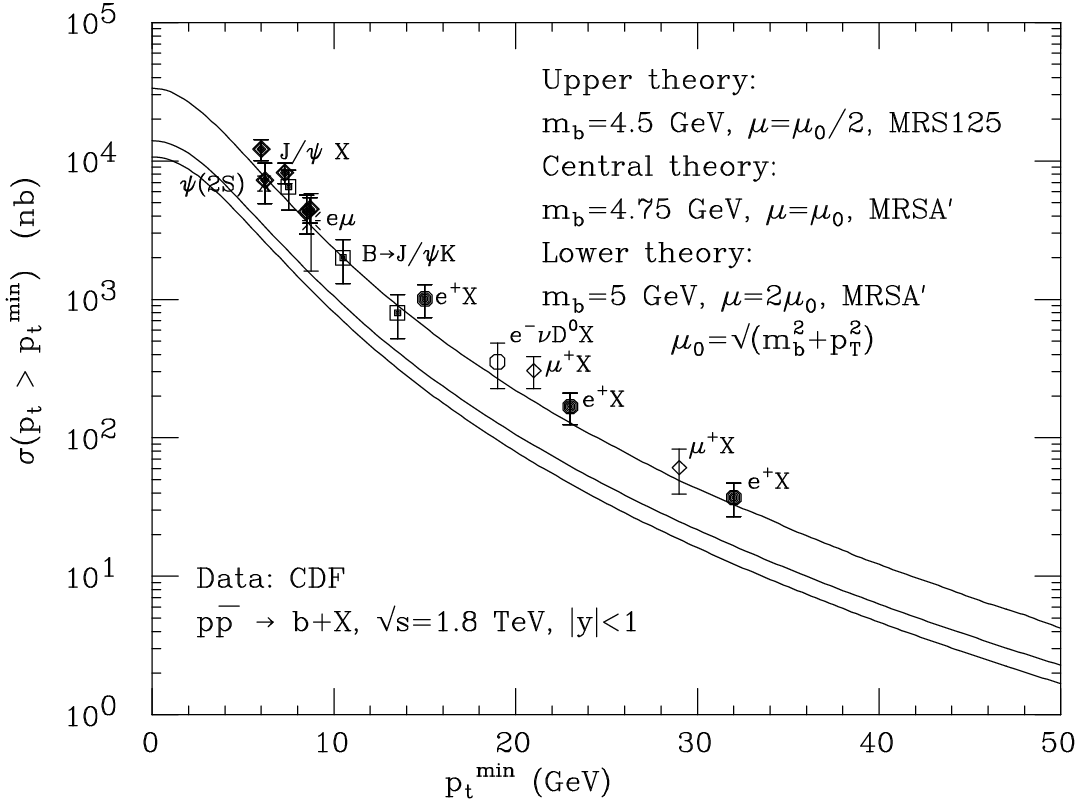,width=0.7\textwidth,clip=}}
\ccaption{}{ \label{fig:cdfbpt}
CDF data on the integrated $b$-quark \pt\ distribution, compared to the results
of NLO QCD.}
\end{figure}

\begin{figure}
\centerline{\epsfig{figure=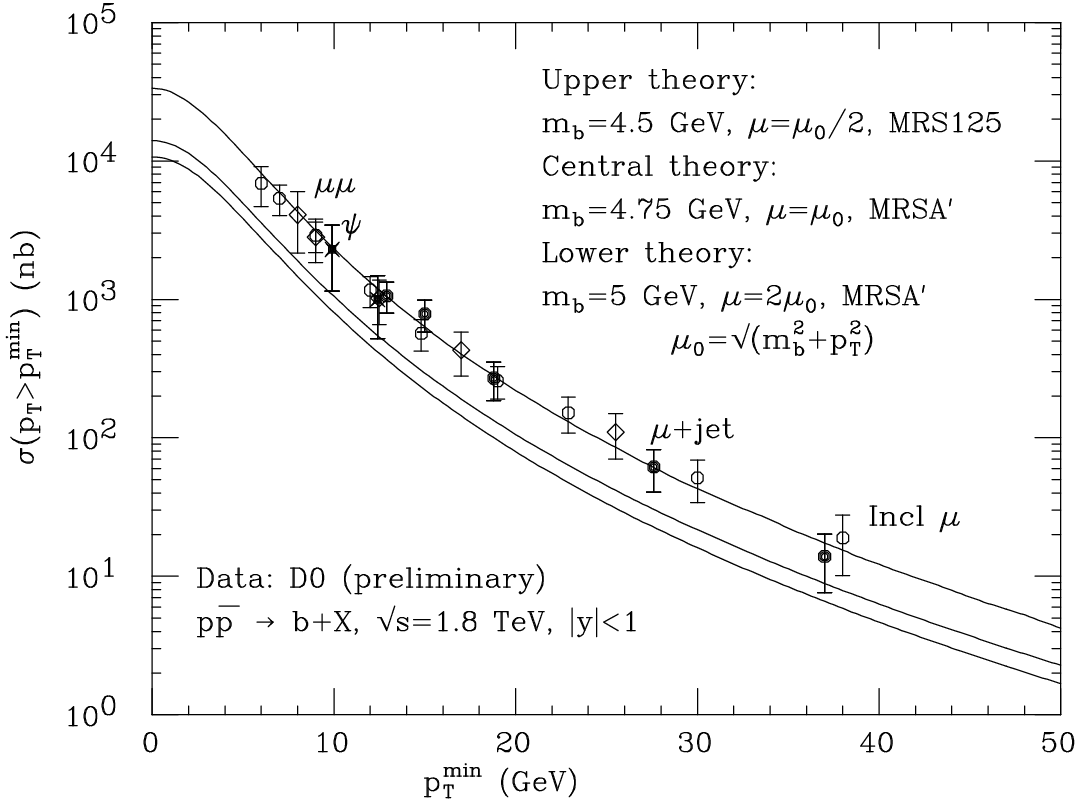,width=0.7\textwidth,clip=}}
\ccaption{}{ \label{fig:d0bpt}
D0 data on the integrated $b$-quark \pt\ distribution, compared to the results
of NLO QCD.}
\end{figure}

The results obtained by the two collaborations are shown in
figs.~\ref{fig:cdfbpt} and ~\ref{fig:d0bpt}, compared with the theoretical
predictions.


\begin{figure}[htb]
\centerline{\epsfig{figure=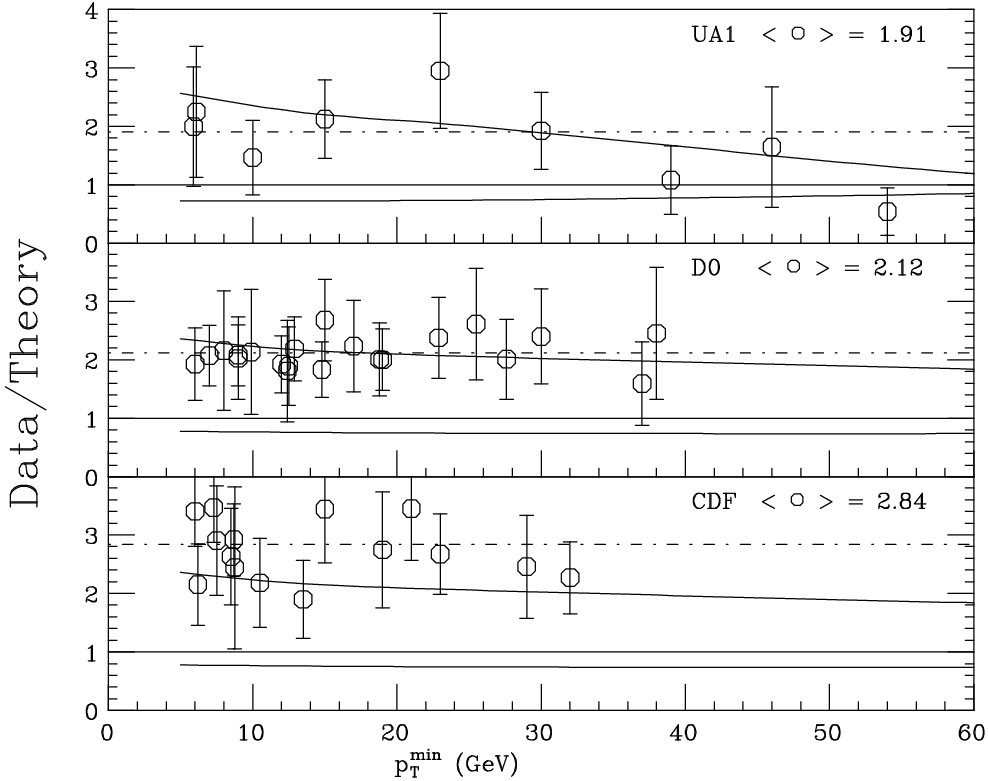,width=0.7\textwidth,clip=}}
\ccaption{}{ \label{fig:mrscomp}
Linear comparison between experimental data and theory for the
integrated $b$-quark \pt\ distribution. See the text for details.}
\end{figure}

For an easier evaluation of the results, we present them on a linear scale
in the form Data/Theory in fig.~\ref{fig:mrscomp}, where we include the UA1
data as well.
We divided the data by our central theoretical prediction. The dot-dashed
straight lines are constant fits to the ratios, weighed by the inverse of
the experimental uncertainties.
For comparison,
we also include the upper and lower theoretical curves divided by
the central one (solid lines).

The first important thing to notice is that,
independently of the chosen input parameters, the shape of the theoretical
curves agrees very
well with the data. Secondly, it must be pointed out that the results at
$\sqrt{S}=630$~GeV are consistent with those at $\sqrt{S}=1800$~GeV. The
average ratio Data/Theory measured by UA1 differs by less than 10\% from the
ratio measured by D0, independently of the input parameters chosen.
The difference is between 40 and 50\% if one uses the CDF data. If we
naively average the Tevatron data, we conclude that the relative
discrepancy between theory and data when changing the value of $\sqrt{S}$ from
630 to 1800~GeV is about 30\%, a number of the same order as
the estimate of small-$x$ effects.

Independently of the beam energy, the data are higher by a factor
of about 2 than the default prediction based on $\mu=\muo$.
They are, however, well
reproduced by the upper theoretical curve. Therefore, while the
overall uncertainty of the theoretical prediction due to the scale choice
is large, there is currently no inconsistency between theory and data for the
inclusive \pt\ distribution of $b$ quarks at the Tevatron.
The 30\% discrepancy between the results of CDF and D0 is
comparable to the discrepancy between the extrapolation of the UA1 data to CDF,
while UA1 and D0 data agree at the level of 10\%.

Recently, new data on $b$ production at 630~GeV have been presented by
CDF \cite{Abe96a}.
The measurement was performed during a special Tevatron run at
$\sqrt{S}=630\;$GeV, using large-impact-parameter muons from $b$-meson
decays. The measurement was compared with a similar one carried out at
$\sqrt{S}=1800\;$GeV, to reduce common systematic errors. The preliminary
result of this study is:
\be
\frac{\sigma_b(\pt>9.5\,{\rm GeV},\vert y \vert <1, \sqrt{S}=630\,{\rm GeV})}
     {\sigma_b(\pt>9.5\,{\rm GeV},\vert y \vert <1, \sqrt{S}=1800\,{\rm GeV})}
\; = \;
     0.193{\pm}0.025\,\mbox{stat}\pm0.023\,\mbox{syst} \; .
\ee
To compare this result with a theoretical expectation, we have to choose
the scales to be used at the two different energies. While there is no theorem
on how to do this,
it is however reasonable to choose the scales to be the same.
This can be partly justified by the similarity of the scale
dependence at the two energies, shown in fig.~\ref{fig:bscaledep}.
Assuming $\mu(630)=\mu(1800)=\muo$ and using MRSA$^\prime$ parton densities,
we find a theoretical ratio of $0.189\pm0.012$.
The theoretical error is obtained
by varying $\mu$ in the range $\muo/2<\mu<2\muo$ and $m_b$ in the range
$4.5~{\rm GeV}<m_b<5~{\rm GeV}$.
The ratio (Data/Theory)(630)/(Data/Theory)(1800) measured by CDF is therefore
equal to $1.0\pm 0.2$. This is a very important measurement, because a large
fraction of the systematic uncertainties present in the individual measurements
cancel out in the ratio. This result indicates that possible sources of
corrections to the theoretical calculations (e.g. higher-order corrections,
small-$x$ effects) should have the same impact at 630 and at 1800~GeV.
One would reach the same conclusion by comparing the results of UA1 and D0,
although the comparison of the systematic errors is in this case more complex.

\begin{figure}[htb]
\centerline{\epsfig{figure=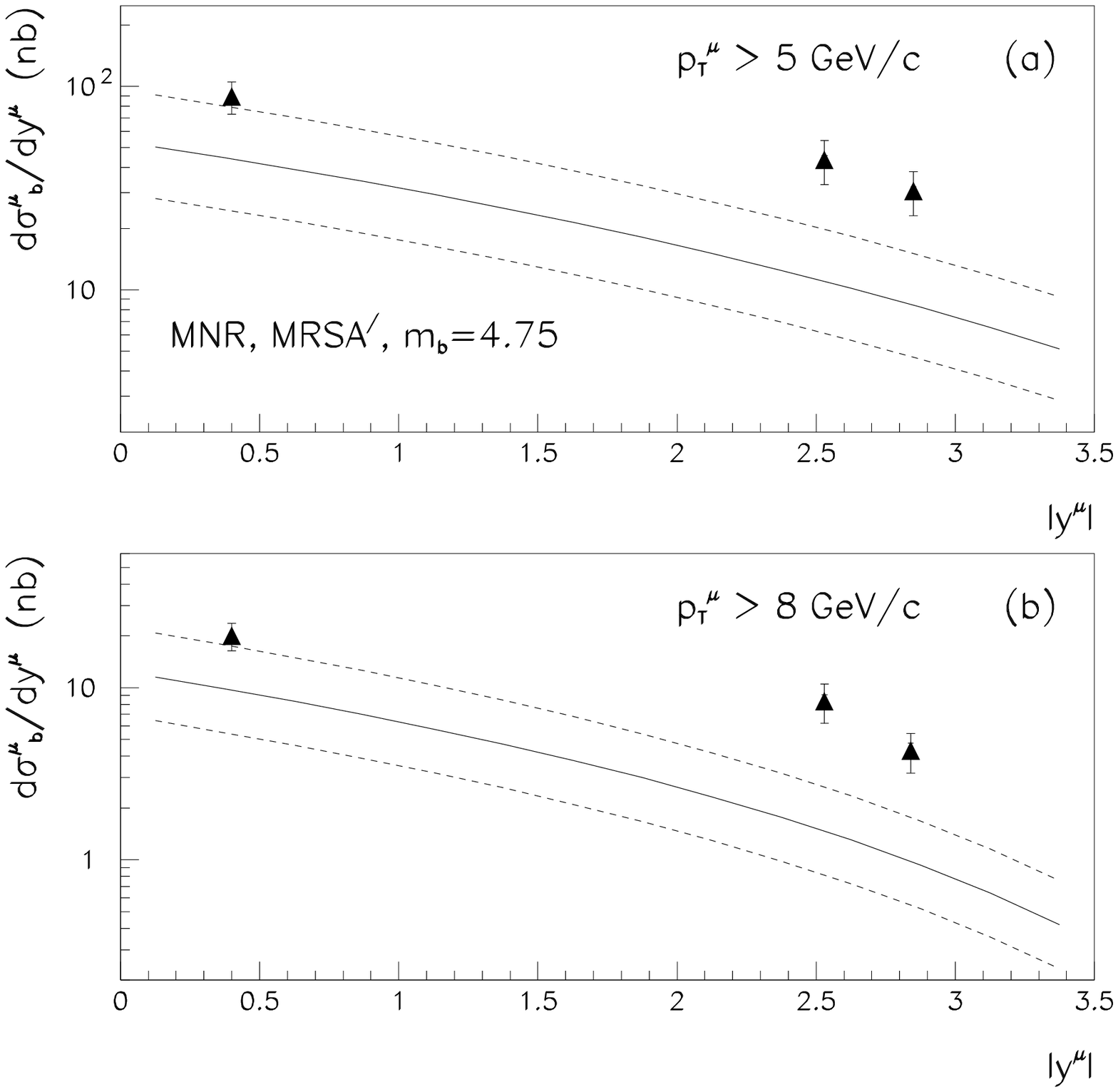,width=0.7\textwidth,clip=}}
\ccaption{}{ \label{fig:d0fmdy}
Comparison between D0 data and theory for the $p\bar{p}\to (b\to\mu)+X$
cross section at large rapidity.}
\end{figure}

A recent measurement by D0~\cite{Abachi96b},  using muonic semileptonic decays
of $b$ quarks, explored the production of $b$ quarks in the forward region
($2.4<\vert y^{\mu} \vert<3.2$). The results, shown in fig.~\ref{fig:d0fmdy},
indicate a rate in excess by a factor of approximately 4 over the
central theoretical prediction.
The large samples collected at 1800~GeV have also allowed a measurement of the
differential~\pt\ spectrum of fully reconstructed $B$ mesons. CDF performed
this measurement by detecting several hundreds of events in the final states $B\to
{\rm J}/\psi K^{(*)}$ \cite{Abe92a,Abe95a,Abe96f}. The full reconstruction of the
final state allows a precise measurement of the hadron momentum, therefore
providing a direct measurement of the differential \pt\ spectrum. The
systematic errors due to the modelling of the $b$-hadron decays are
significantly reduced relative to the measurements in the inclusive channels.
The comparison with the NLO QCD calculations, however, still requires a
description of the $b$-quark to $b$-hadron fragmentation function. The NLO
calculation of the $b$-quark \pt\ distribution contains the effects of
final-state gluon radiation, although limited to the emission of one gluon. As was
done in the fixed-target section, we then
fold the NLO quark \pt\ spectrum with a
phenomenological parametrization of the non-perturbative quark-to-hadron
transition. We use the so-called Peterson fragmentation function~\cite{Peterson83}, already described in
section~\ref{sec:ft}.  For the parameter $\epsilon$
that defines the shape of the
distribution we choose the value $\epsilon_b=0.006\pm0.002$~\cite{Chrin87}.
We furthermore assume that the fraction of $b$ quarks that
fragments into the hadron species identified in the experimental analyses
($B_{u,d}$) is 75\%, a number supported by recent direct
measurements~\cite{Abe96c}.

\begin{figure}[htb]
\centerline{\epsfig{figure=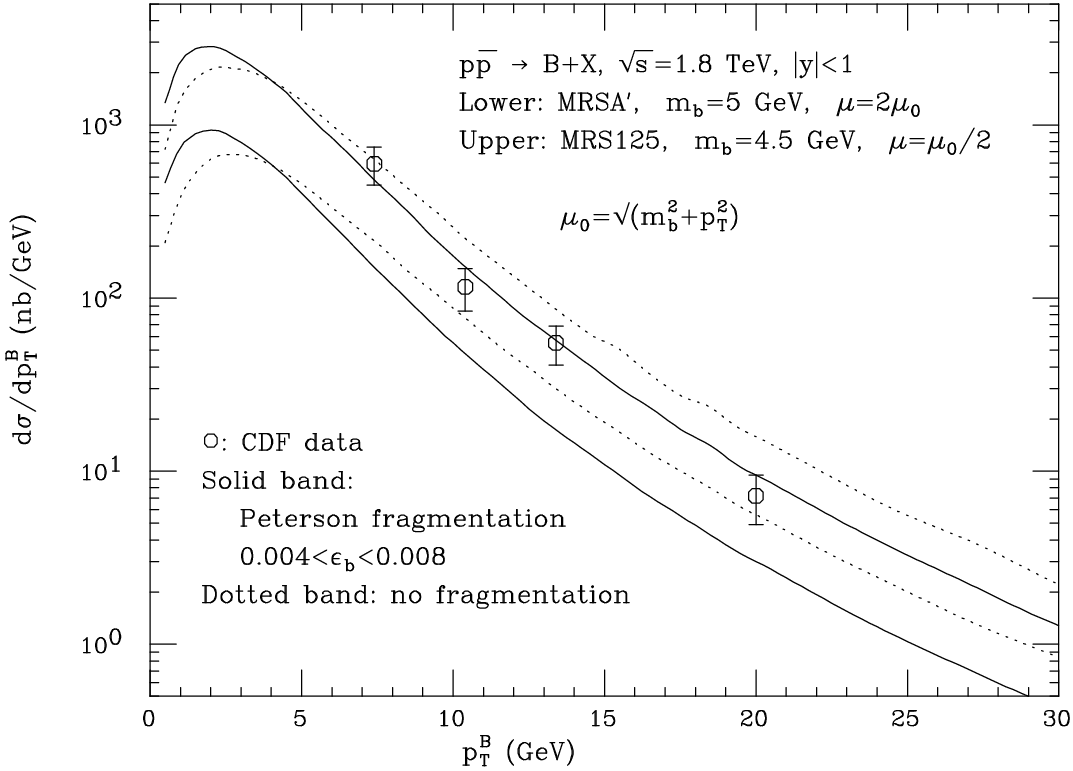,width=0.7\textwidth,clip=}}
\ccaption{}{ \label{fig:bptmes}
Comparison between CDF data and theory for the differential
$B$-meson $p_{\sss T}$ distribution.}
\end{figure}

Figure~\ref{fig:bptmes} shows the latest CDF data~\cite{Abe96f}, compared to the
NLO QCD predictions. The ratio Data/Theory is consistent with that
found in the inclusive measurements, with the possible exception of the point
at the lowest \pt, which is higher than expected. It is premature to conclude
whether this is just a statistical fluctuation,
or whether this is a sign of small-$x$ effects.

As in the case of charm production at fixed target, we should comment here
that the use
of the Peterson fragmentation function might not be justified in the
context of hadroproduction of heavy quarks.
As a simple observation, we point out here that while the measurement of
heavy-quark spectra in $e^+e^-$ data is mostly sensitive to the first moment
of the fragmentation function, corresponding to the average jet
energy carried by the heavy hadron, the \pt\ distributions in hadronic
collisions are sensitive to higher moments of the non-perturbative
fragmentation function. In
fact, assuming for simplicity a perturbative \pt\ spectrum of the form:
\be
        \frac{d\sigma}{d\pt} = \frac{A}{\pt^n} \; ,
\ee
it is easy to prove that the resulting hadron spectrum, after convolution with
a given fragmentation function $f(z)$, will be given by:
\be
        \frac{d\sigma}{d\pt} = \frac{A}{\pt^n} \;
        \times \; \int \, dz \, z^{n-1} \, f(z) \; .
\ee
In the case of the Tevatron, $n$ turns out to be approximately 5.
It is not unlikely, therefore, that  alternative models for the
non-perturbative fragmentation of heavy quarks, which could equally well fit
the $e^+e^-$ data, could give rise to significant differences when applied to
production in hadronic collisions. We also remark that
the gluon component of the fragmentation function is not important
in $e^+e^-$ physics, while it may be crucial in hadroproduction.

To display the size of the hadronization corrections induced by the Peterson
fragmentation, we also include in fig.~\ref{fig:bptmes} the $b$-quark \pt\
distribution.
Notice that, neglecting the lowest-\pt\ point,  both shape and
normalization of the data are well described by the central theoretical
prediction without fragmentation.

We summarize the main conclusions of the studies presented in this
section:
\begin{enumerate}
\item There is good agreement between the shape of the $b$-quark \pt\
distribution predicted by NLO QCD and that observed in the data for central
rapidities.
\item Although the data are higher by a factor of approximately 2
with respect to the theoretical prediction with the default choice
of parameters, extreme (although acceptable)
choices of $\lambdamsb$ and of renormalization and factorization scales
bring the theory in perfect agreement with the data of UA1
and D0, and within 30\% of the CDF measurements.
\item The choice of low values of $\muR$ and $\muF$ is favoured by studies of
higher-order logarithmic corrections.
\item There is a difference at the level of 30\%
between the measurements of CDF and D0 at 1800~GeV,
and between CDF and UA1 at 630~GeV.
\item The CDF measurements at 630 and 1800~GeV indicate that theory correctly
predicts the scaling of the differential \pt\ distribution
between 630 and 1800~GeV, a fact that had often been questioned in the past
and now finds strong support.
\item Forward production of $b$ quarks indicates a larger discrepancy between
theory and data.
\item More theoretical studies should be devoted to the understanding of the
non-perturbative fragmentation function for heavy quarks. As already stressed
in the section on fixed target, there is no firm
evidence, either theoretically or experimentally, that the standard Peterson
parametrization is well suited for the description of the hadroproduction data.
\end{enumerate}

\subsection{$ b \bar b$ correlations}
Studies of the one-particle inclusive distributions are a benchmark test of
QCD. However, these tests do not probe all of the important features of the
production mechanism. In several cases the
correlations between the quark and the antiquark can provide additional
important information. We already discussed how, in the
case of fixed-target charm production, the shape of the azimuthal
correlations between the $c$ and $\bar c$ quarks is sensitive to the amount of
intrinsic transverse momentum carried by the partons in the hadron. Given the
large mass of the $b$ quark and the large \pt\ values at which $b$'s are probed
in hadronic collisions, these effects should be negligible. However, one could
observe similar behaviours induced by the build-up of large transverse momentum
due to the perturbative evolution of the initial state (i.e.\ multi-gluon
emission). This phenomenon would be even more pronounced at small \pt\ and in
the presence of important small-$x$ effects.

The knowledge of heavy-quark correlations is also an important element in the
study of detector acceptances and detection systematics. For example, in
studies of $b$ tagging in samples where both $b$'s have been tagged, it is
important to know what the correlations between the $b$ and $\bar b$ momenta
are, since the tagging efficiency depends on the $b$ momentum. These studies
have implications on the determination of tagging efficiencies for a multitude
of important measurements, from the study of top production~\cite{Abe94}
to the study of $R_b$~\cite{Nason96}.

The study of $b\bar b$ correlations in hadronic collisions was pioneered by
UA1~\cite{Geiser92,Albajar94,Albajar96}. For these measurements they used
their sample of high-mass and non-isolated dimuon events.
The shapes of the \dphi\ and $\Delta R$
distribution for the data are both in good agreement with the
theoretical expectations~\cite{Albajar94}. The rate is approximately 30\%
higher than the central value of the theoretical predictions, a result
consistent with the UA1 determination of the inclusive cross section.
They also separated the sample of
dimuon events into two classes, defined by two-body and three-body topologies.
Contributions to the three-body topology come from events with hard-gluon
bremstrahlung, and are therefore proportional to $\as^3$.
Two-body final states, with the $b$ and the $\bar b$ almost back-to-back, are
dominated by Born-like processes, and their rate is of $\oatwo$. A comparison
between the two samples, carried out as a function of the \pt\ of the $b$
quark, allowed UA1 to study the dynamical evolution of $b\bar b$ correlations
and to extract the value of $\as(M_{\sss Z})=0.113{{+0.007}\atop{-0.006}}_{\rm
exp}{{+0.008}\atop{-0.009}}_{\rm th}$~\cite{Albajar96}. The same study led to a
3$\sigma$ evidence for the running of $\as$.

\begin{figure}[htb]
\centerline{\epsfig{figure=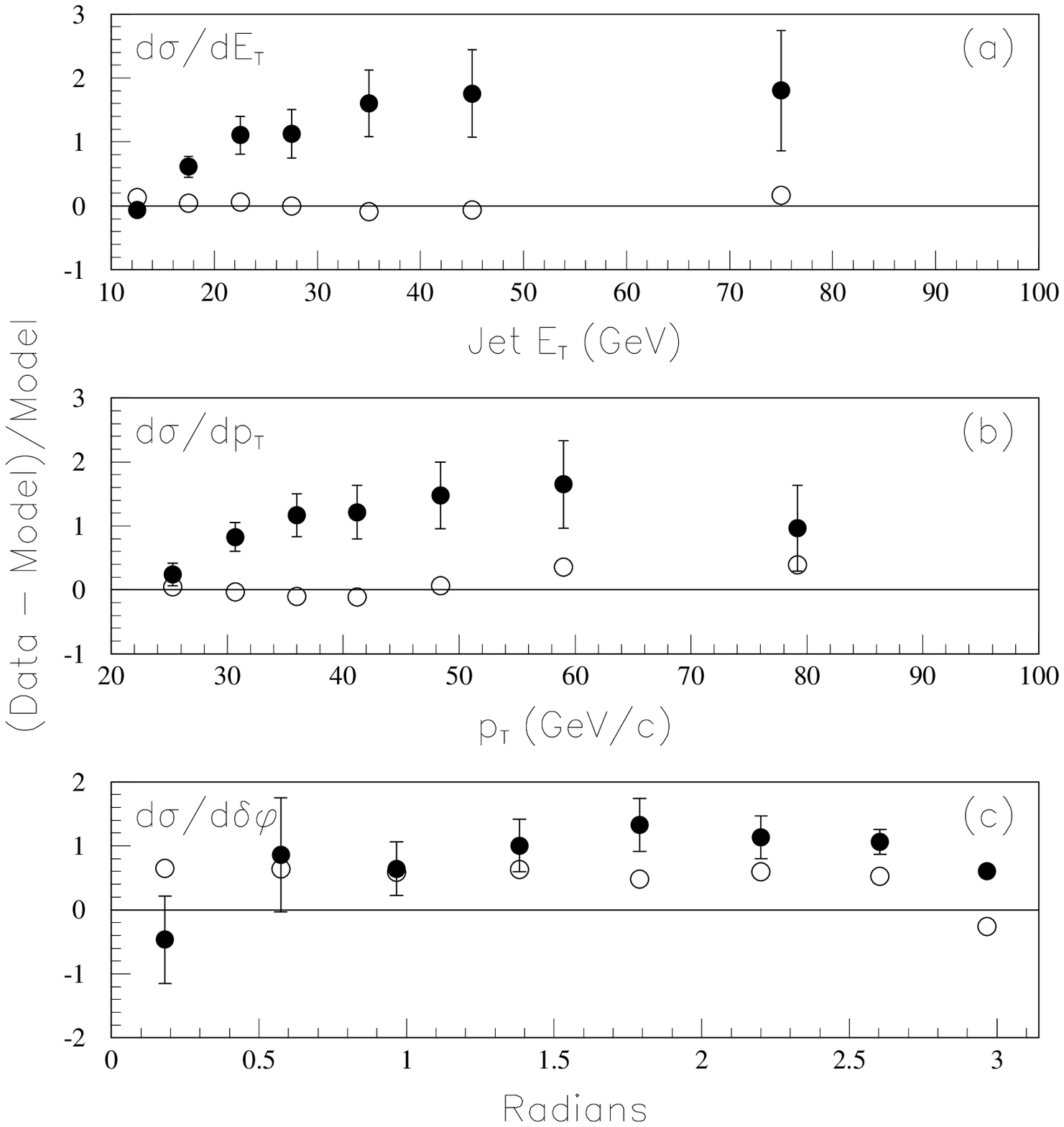,width=0.6\textwidth,clip=}}
\ccaption{}{ \label{fig:cdfmub}
CDF results on the $b\bar b$ correlations using $\mu$+$b$-tagged jet final
states~\cite{Abe96b}, compared to NLO QCD. Solid points correspond to the
default theoretical prediction, with scales $\mu=\muo$, empty circles
correspond to the difference between the choice $\mu=\muo/2$ and $\mu=\muo$.
Top figure: tagged-jet \et\ distribution.
Central: muon \pt\ spectrum. Bottom: azimuthal correlations.}
\end{figure}

CDF presented their first study of $b\bar b$ correlations using a sample of
$e\mu$ pairs \cite{Abe94a}. Once again, good agreement was found between the
shape of the data and the QCD expectations.
A second measurement has been recently reported by CDF,
using muons plus tagged $b$ jets~\cite{Abe96b}. The shape of the \dphi\
distribution is in reasonable agreement with QCD, while the distributions of
the jet \et\ and of the muon \pt\ are not (see fig.~\ref{fig:cdfmub}).

Contrary to the case of $c\bar c$ correlations measured in fixed-target
experiments,
the measurement at hadron colliders is
sensitive to the modelling of the heavy-quark fragmentation, because of
possible trigger biases. A harder (softer) fragmentation function would enhance
(decrease) the efficiency for the detection of the softest of the two $b$
quarks. These effects could have an impact on the distributions reported in
this study by CDF. This collaboration explored the effects of changes in the $\epsilon_b$
parameter within the Peterson fragmentation model, finding them negligible.
As we argued earlier, it cannot be excluded that a systematic study of other
possible parametrizations for the fragmentation modelling could lead to
significant effects.
Also the possible effects of the $k_{\sss T}$ kick have been studied by
CDF~\cite{Abe96b}, with the conclusion that not even an average $k_{\sss T}$
as large as 4~GeV could improve the agreement between theory and data for the
measurements considered.

\begin{figure}[htb]
\centerline{
   \epsfig{figure=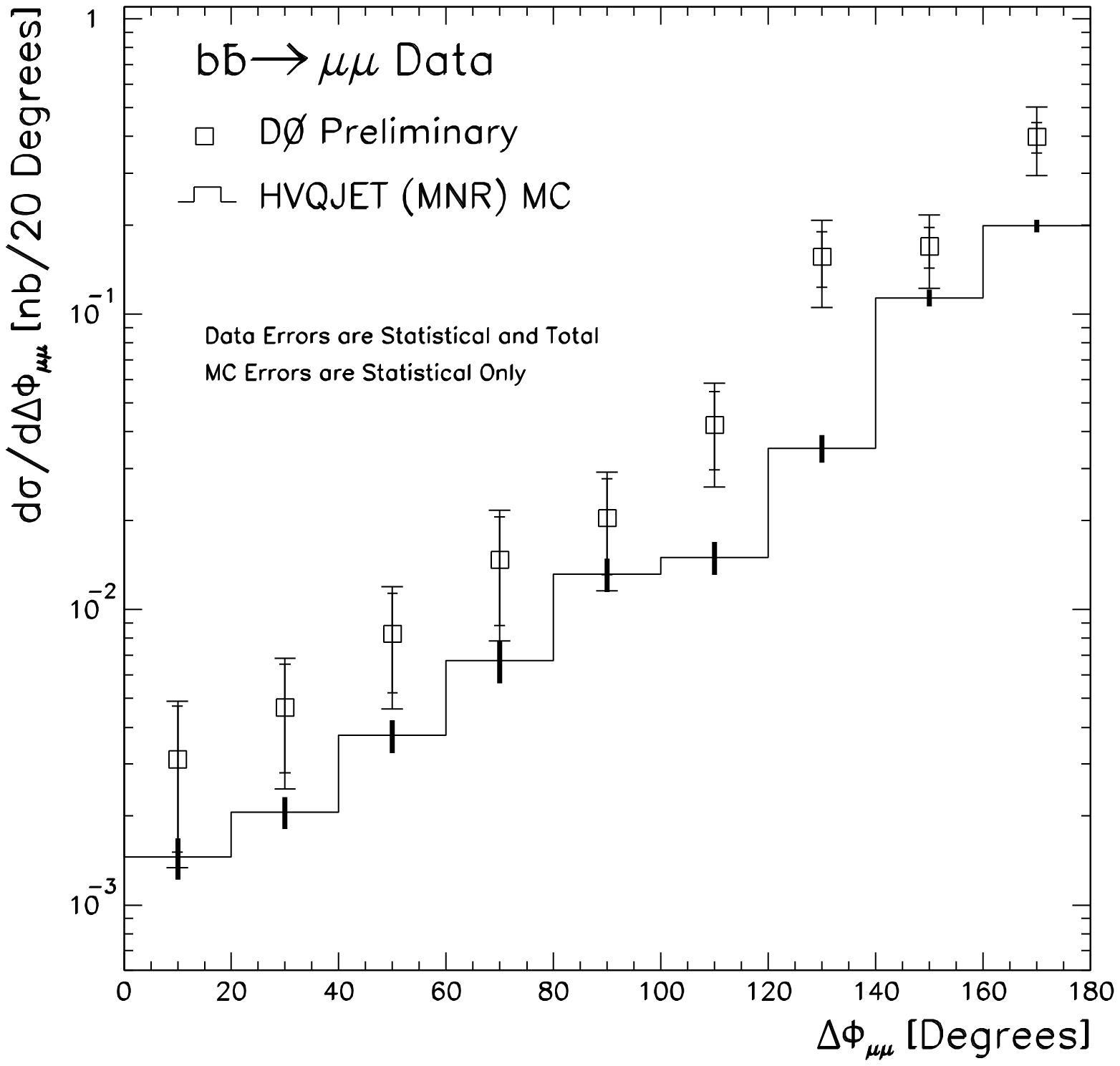,width=0.4\textwidth,clip=}
   \hfill
   \epsfig{figure=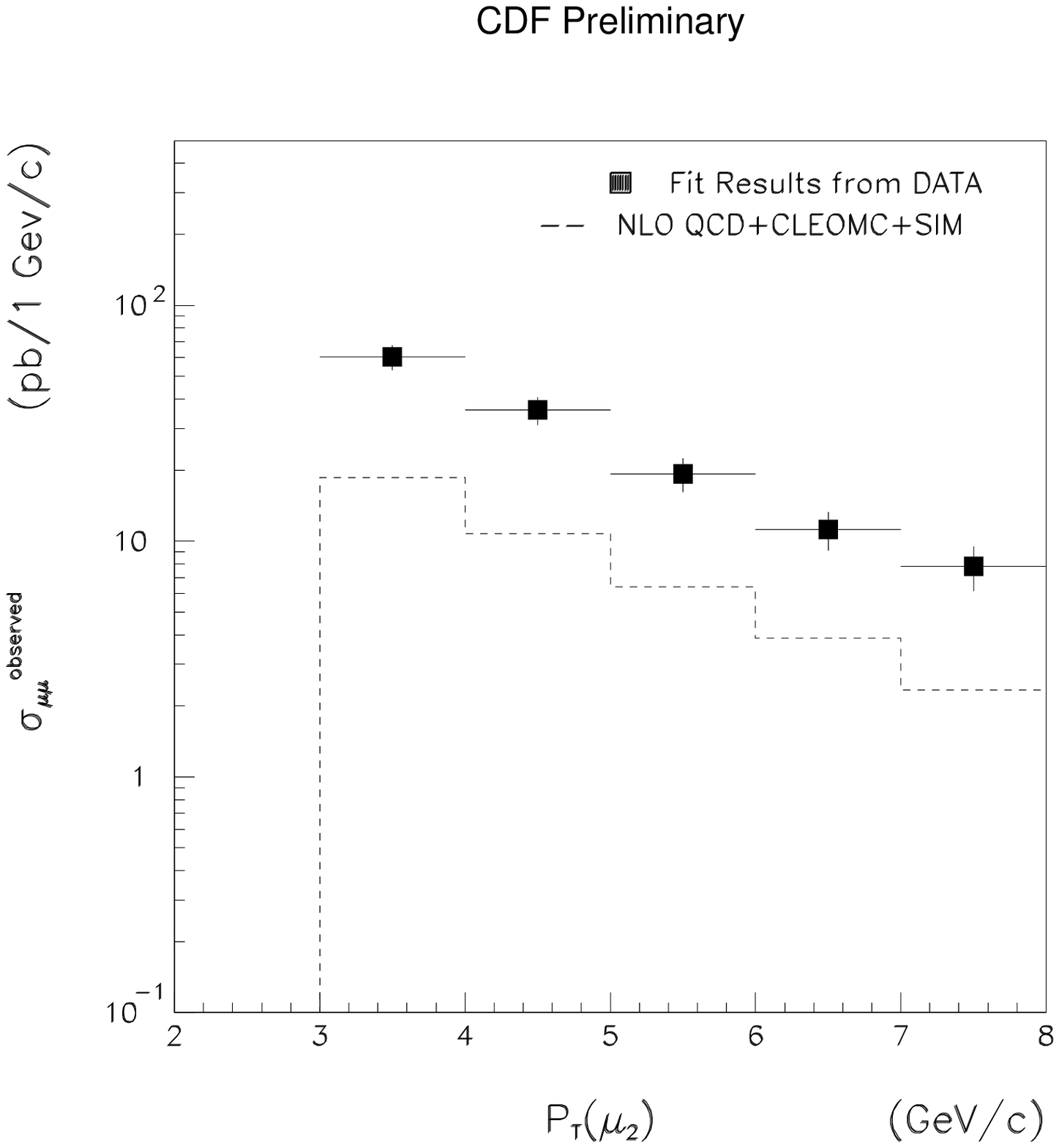,width=0.4\textwidth,clip=} }
\ccaption{}{ \label{fig:mmcorr}
Results on $b\bar b$ correlations using dimuon final states,
compared to NLO QCD. Left: azimuthal-correlation results from D0.
Right: \pt\ distribution of the slowest muon from CDF.}
\end{figure}

Both CDF and D0 recently presented studies based on samples of
high-mass dimuons~\cite{Abe96d,Abachi96a}. The shapes for
both $\Delta\phi$
and $p_{\sss T}^{b}$ for a given $p_{\sss T}^{\bar b}$ are well reproduced
by theory, within the large uncertainties, while there is a
normalization discrepancy relative to the central predictions
(fig.~\ref{fig:mmcorr}).

In conclusion we can say that, overall, there are good indications that NLO QCD
correctly describes the correlations between $b$ pairs produced in
hadronic collisions. Most of the large theoretical systematics present in the
case of fixed-target charm production are not relevant here. The discrepancies
observed in the CDF measurement using the muon+tagged-jet final states,
however, indicate that there still is a possible interplay between
the theoretical and experimental systematics, which is not entirely understood
as yet.
\subsection{Heavy-quark jets in perturbative QCD}
\subsubsection{Preliminaries}
An interesting way of understanding the production
mechanism of heavy flavours is to consider the cross section of
jets that contain a heavy quark (in short:
heavy-quark jets) \cite{Frixione96}.  The main difference between the
study of a heavy quark and of a heavy-quark jet is that in the former we  are
interested in the momentum of the quark itself, regardless of the properties of
the event in which the quark in embedded, while in the latter we are
interested in the properties of a jet containing one or more heavy quarks,
regardless of the momentum fraction of the jet carried by the quark. A priori
it is expected that variables such as the \et\ distribution
of a heavy-quark jet should be described by a finite-order QCD calculation
more precisely than the \pt\ distribution of open quarks.
This is because the large logarithms $\log(\pt/m_b)$ arising from
the emission of collinear gluons from the heavy quark cancel in this case.
Furthermore, the experimental measurement of the \et\ distribution of
heavy-quark jets does not depend on the knowledge
of the heavy-quark fragmentation functions, contrary to the case
of the \pt\ distribution of open heavy quarks.
Experimental systematics, such as
the knowledge of decay branching ratios for heavy hadrons or
of their decay spectra, are also largely reduced.

The calculation of the heavy-quark jet rate is very similar to
the one of the generic jet cross section.
Two important differences have nevertheless to be stressed:
by its very definition, a heavy-quark jet is not flavour-blind; one
has to look for those jets containing a heavy flavour. Furthermore, the
mass of the heavy flavour is acting as a cutoff against
final-state collinear divergences.
This in turn implies that the structure of the
singularities of the heavy-quark jet cross section is identical
to that of the open-heavy-quark cross section. The
heavy-quark jet cross section at NLO can therefore be written in
the following way, which implicitly defines $d\Delta/d\et$
(this term will be denoted in the following as the ``jet-like'' component
of the cross section)
\beq
\frac{d\sigma}{d\et}=\frac{d\sigma^{(open)}}{d\et}+\frac{d\Delta}{d\et}\,,
\label{xsecsplit}
\eeq
where $d\sigma^{(open)}/d\et$ is the open-heavy-quark cross
section.
Observe that  $d\Delta/d\et$ has no contributions at the Born level.
Gluon-radiation effects start contributing to it at order $\as^3$.
Normally the heavy-quark jet is defined by a cone algorithm.
In our phenomenological applications, we will use the Snowmass
convention~\cite{Huth91}, whereby particles are clustered in
cones of radius $R$ in the pseudorapidity--azimuthal angle plane.
One then requires that the $b$ quark be inside the jet.
At the NLO, the heavy-quark jet can be the heavy quark,
or it can contain the heavy quark and the light parton, or
the heavy quark and the heavy antiquark. In other words, there is
the possibility to cluster more than one parton to get the heavy-quark jet,
which will be eventually observed.
It turns out that, if the jet definition is appropriately infrared-safe,
all the subtractions needed to get an infrared-safe result
are contained in the term $d\sigma^{(open)}/d\et$.
This is shown in details in ref.~\cite{Frixione96}. This result
is however quite intuitive. In fact, singularities arise because
of radiation collinear to the incoming partons, which cannot enter
the jet cone, or to soft radiation, which does not sensibly alter the
jet $\et$.
\subsubsection{The structure of heavy-quark jets at the
Tevatron}\label{sub:hqjet}
We present in this section some results of interest for the Tevatron
collider~\cite{Koehn95}.
We consider jets produced within $|\eta|<1$,
in order to simulate a realistic geometrical acceptance of the Tevatron
detectors. We will use the parton distribution set
MRSA$^\prime$~\cite{Martin95a}.
Our default values of the parameters entering the calculations are
\mc~=~1.5~GeV, \mb~=~4.75~GeV
and $\mur=\muf=\muo\equiv\sqrt{E_{\sss T}^2+m_Q^2}$.
In practice, since we will consider jets of energy much larger than the
heavy-quark mass, the dependence of our results upon the mass value
is almost negligible. We will therefore study only the uncertainties
associated with the choice of renormalization and factorization scales.

\begin{figure}
\centerline{\epsfig{figure=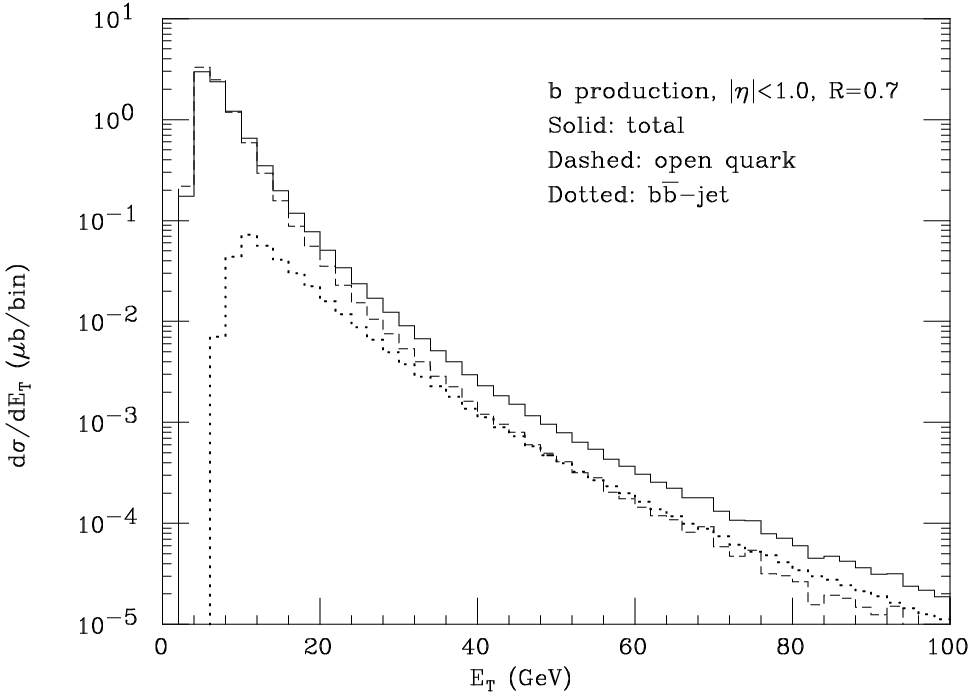,width=0.7\textwidth,clip=}}
\ccaption{}{ \label{fbinc}
$b$ jet inclusive \et\ distribution in $p\bar p$ collisions at 1.8 TeV, for
$|\eta|<1$, $R=0.7$ and $\muf=\mur=\muo$ (solid line).
For comparison, we also show the open-quark inclusive
\et\ distribution (dashed line). The component of the jet-like
contribution due to jets containing both $b$ and $\bar b$ is represented by the
dotted line.}
\end{figure}

Figure~\ref{fbinc} shows the prediction for the \et\ distribution
of $b$ jets at the Tevatron for $R=0.7$. For the
purpose of illustration, the open-quark component is presented separately.
It is apparent that the jet-like component
becomes dominant as soon as \et\ becomes
larger than 50~GeV. It can be shown~\cite{Frixione96} that this value
depends significantly on the cone size,
being equal to 25 and 100~GeV for $R=1$ and 0.4 respectively.
Also, for a fixed $R$, the value of \et\ at which the jet-like
component becomes dominant is smaller for $c$ jets than
for $b$ jets. This is in agreement with naive expectations.
The relative probability of finding a heavy-quark pair inside
a high-\et\ gluon scales in fact like
$\log (\et/m_Q)$~\cite{Mueller85,Mueller86,Mangano92,Seymour95}.

We also show the part of the jet-like component due to
jets including the \bbbar\ pair (we will call these $\bbbar$ jets).
The figure suggests that for this \et\
range and with $R=0.7$ this is the dominant part of the jet-like component.
This is consistent with the expectation that, for large enough \et, and
provided  that the majority of the final-state generic jets are composed of
primary gluons, heavy-quark jets are dominated by the process of gluon
splitting, with the jet formed by the heavy-quark pair.
As we will show later, the situation changes at higher values of \et\, where
heavy quarks are mostly produced via the $s$-channel annihilation of light
quarks.

\begin{figure}
\centerline{\epsfig{figure=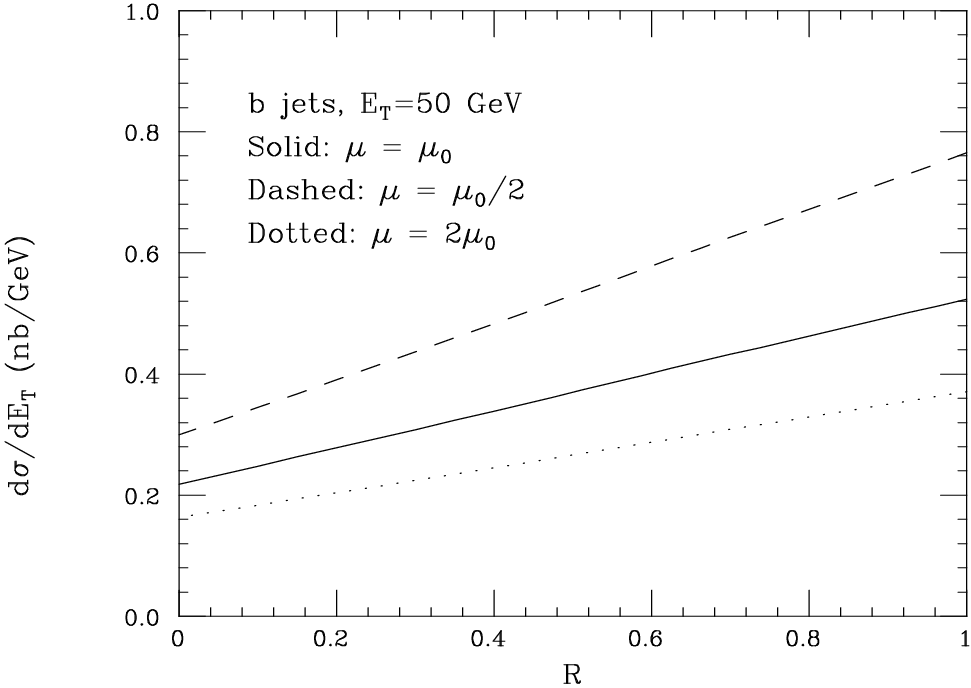,width=0.5\textwidth,clip=}
            \hspace{0.3cm}
            \epsfig{figure=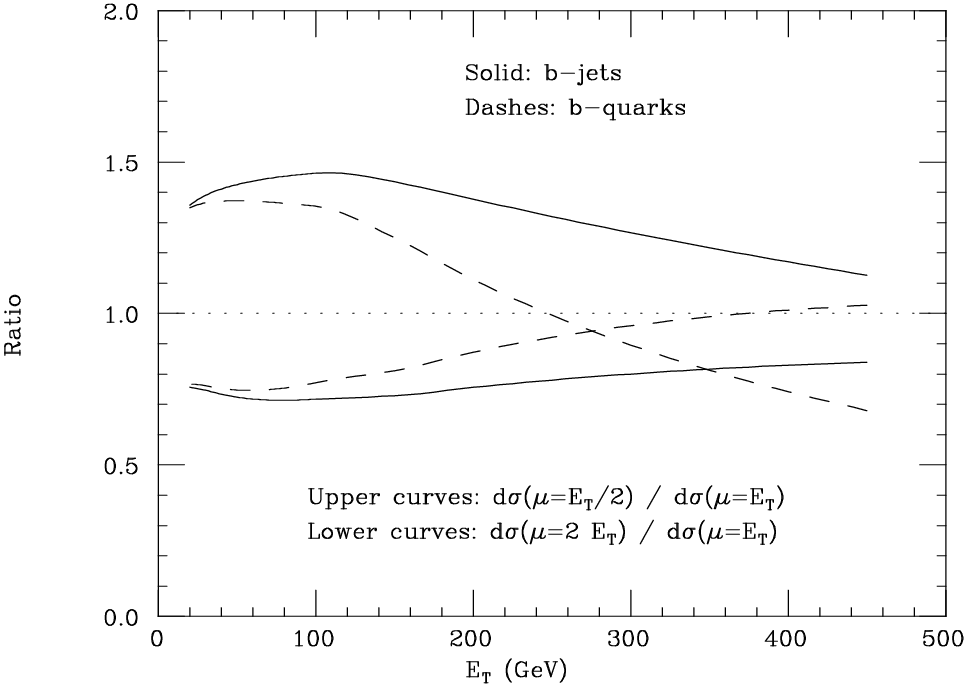,width=0.5\textwidth,clip=}}
\ccaption{}{ \label{fbscale}
Left: $b$ jet inclusive \et\ rate, as a function of the cone size
$R$, at \et~=~50~GeV and for various scale choices ($\mur=\muf\equiv\mu$).
Right: Scale dependence of the $b$-jet \et\ distribution ($R=0.4$,
solid lines) and of the open-quark inclusive \et\ distribution
(dashed lines).}
\end{figure}

The left-hand side of fig.~\ref{fbscale} presents the theoretical prediction
for the absolute $b$-jet rate at $\et=50$~GeV versus the
cone size, for different choices of the factorization and
renormalization scales.
The cross section at $R=0$
is well defined, and it is equal
to the open-bottom cross section. This should be contrasted with the case
of generic jets, in which the cross section at $R=0$ is not well defined,
being negative at any fixed order in perturbation
theory~\cite{Aversa89,Aversa90a,Ellis89S,Ellis90S}.
Completely analogous results hold in the case of $c$-jet production.
The right-hand side of fig.~\ref{fbscale} shows the scale dependence of the
$b$-jet cross section ($R=0.4$) as a function of \et,
for values up to 450~GeV.

The strong scale dependence exhibited by the absolute rate at low and
moderate \et\ values is of the same size as the one present in the
inclusive \pt\ distribution of open bottom quarks.
This scale dependence is usually attributed to the
importance of the gluon splitting contribution.
One therefore expects that, in a regime where the
gluon-splitting contribution is suppressed by the dynamics,
the scale dependence
should be milder. We will show later that this suppression does indeed
take place for high transverse energies. This explains why,
in the high-$\et$ region, the scale dependence is indeed reduced to the
value of $20$\% when the scale is varied in the range $\muo/2<\mu<2\muo$, a
result consistent with the limited scale dependence of the NLO
inclusive-jet cross sections~\cite{Aversa89,Aversa90a,Ellis89S,Ellis90S}.
In spite of the strong scale sensitivity at the smaller values of \et,
it is reasonable to expect that the ratio of the
$b$- and $c$-jet rates be a stable quantity under scale variations.
It can be shown~\cite{Frixione96} that this is indeed the case: by taking
$\muo/2<\mu<2\muo$ the ratio changes at most by 10\%
with respect to the default prediction; furthermore, the ratio of
$b$  to $c$ jets is more stable than the one of $b$ to $c$ quarks.

\begin{figure}
\centerline{\epsfig{figure=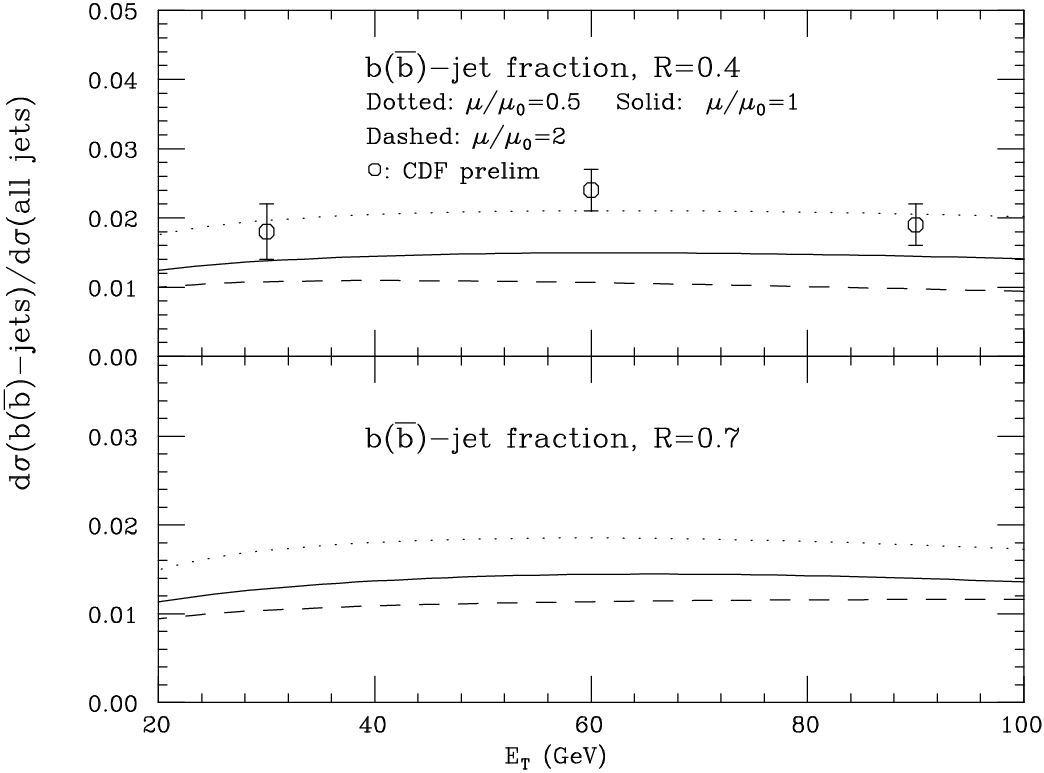,width=0.5\textwidth,clip=}
            \hspace{0.3cm}
            \epsfig{figure=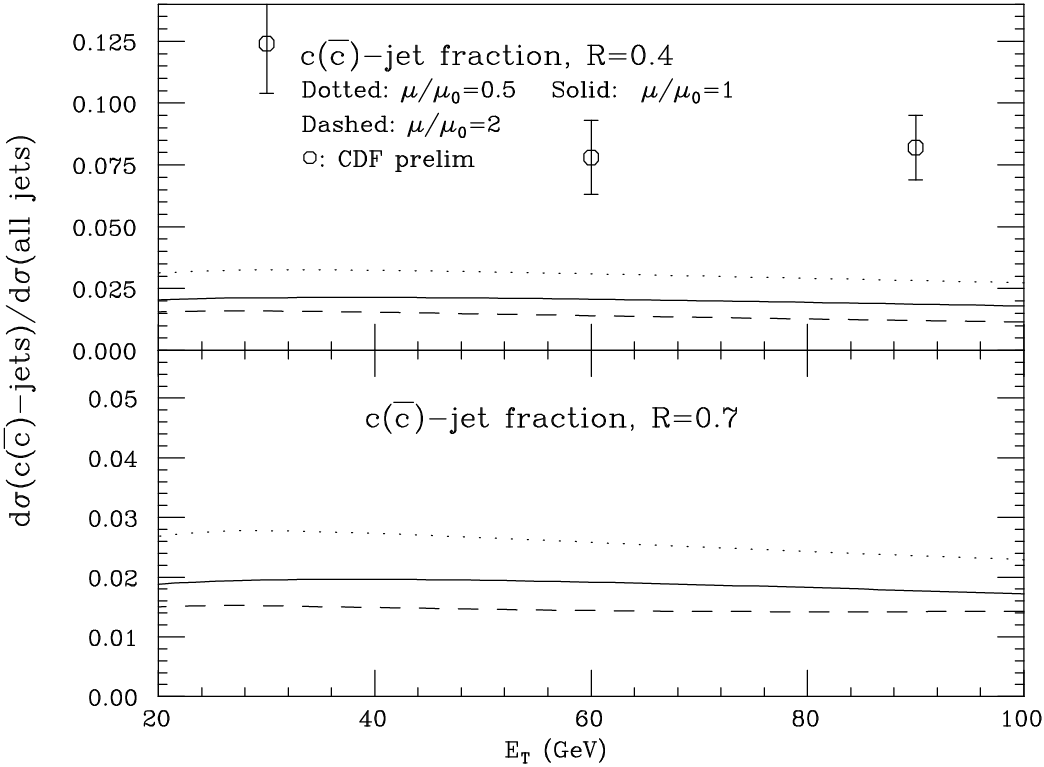,width=0.5\textwidth,clip=}}
\ccaption{}{ \label{fjetfrac}
Ratio of the heavy-quark jet (for bottom, left, and charm, right)
to inclusive-jet \et\ distributions, for different choices of
renormalization and factorization scales ($\mur=\muf\equiv\mu$),
for $R=0.4$ (top) and $R=0.7$ (bottom).
The data points for $R=0.4$ represent preliminary results
from the CDF experiment, for which only the statistical
uncertainty is shown.}
\end{figure}

Of direct interest for studies of heavy-flavour tagging and for searches of
possible new physics is the fraction of heavy-quark jets relative to generic
jets. This is also in principle the most straightforward measurement from the
experimental point of view.
We present in fig.~\ref{fjetfrac} the ratio of the heavy-quark jet
to the inclusive-jet \et\
distributions~\cite{Aversa89,Aversa90a,Ellis89S,Ellis90S} for
bottom (left) and charm (right). The inclusive-jet \et\ cross section
used here was calculated with the JETRAD
program~\cite{Giele94}, using the same choices of parton densities and
$(\mur,\muf)$ that were adopted for the $b$-jet and $c$-jet calculations.
Contrary to the figures presented so far, which showed results for the heavy
quark only (i.e. no antiquark contribution), we adopt for this figure the
prescription used in the definition of the data presented by
CDF~\cite{Koehn95}.
The $Q$-jets are defined there as jets containing either a $Q$ or a
$\overline{Q}$ quark, jets containing both being counted only once. We will
call these $Q(\overline{Q})$ jets. This distribution can be obtained by
subtracting the contribution of $Q\overline{Q}$ jets from twice the total
$Q$-jet rate.

It is interesting to notice that, as far as $b$ production
is concerned, there is a good agreement between
the CDF data and the theoretical prediction obtained with
$\mur=\muf=\muo/2$; this choice of scale is also supported
by the inclusive-jet \et-spectrum data~\cite{Abe92,Abe96,Blazey96}.
This is particularly significant since the choice of scale for the
heavy-quark jet cross section is not independent from the scale chosen to
predict the open-heavy-quark one (see eq.~(\ref{xsecsplit})).
On the other hand, to get a satisfactory description of
the data for the open-bottom \pt\ spectrum, smaller scale
values (of the order of $\muo/4$) or larger $\Lambda_{\sss QCD}$
values (compatible with LEP measurements) have to be
chosen~\cite{Frixione94,Frixione96a}. Should this situation persist when
additional data on $b$ jets will become available, it would
indicate an inconsistency in describing two phenomena due to
the same underlying physics. The poor understanding of the
fragmentation mechanism is very likely a source of this inconsistency.
As for the large disagreement with the charm data, we have no significant
comment to make. Hopefully, additional data will soon be available, as well as
estimates of the experimental systematics.
Notice that the largest contribution expected from higher-order perturbative
corrections is given by the production of $c\bar c$ pairs from soft gluons
emitted during the gluon-shower evolution. However, these effects have been
estimated in refs.~\cite{Mueller85,Mueller86,Mangano92,Seymour95}, and
have been shown to be negligible at the
energies of interest for the current measurements.

To conclude this section, we discuss the behaviour of the $b$-jet production
cross section at high \et.
The interest of this item stems from the
discrepancy reported by CDF~\cite{Abe92,Abe96} in the
tail of the jet distribution. If this discrepancy could not be accommodated by
new theoretical developments in QCD~\cite{Catani96a}
or in the fitting of parton densities~\cite{Huston95,Glover96},
a study of the flavour composition of these
high-energy jets could help in understanding the nature of the phenomenon.

\begin{figure}
\centerline{\epsfig{figure=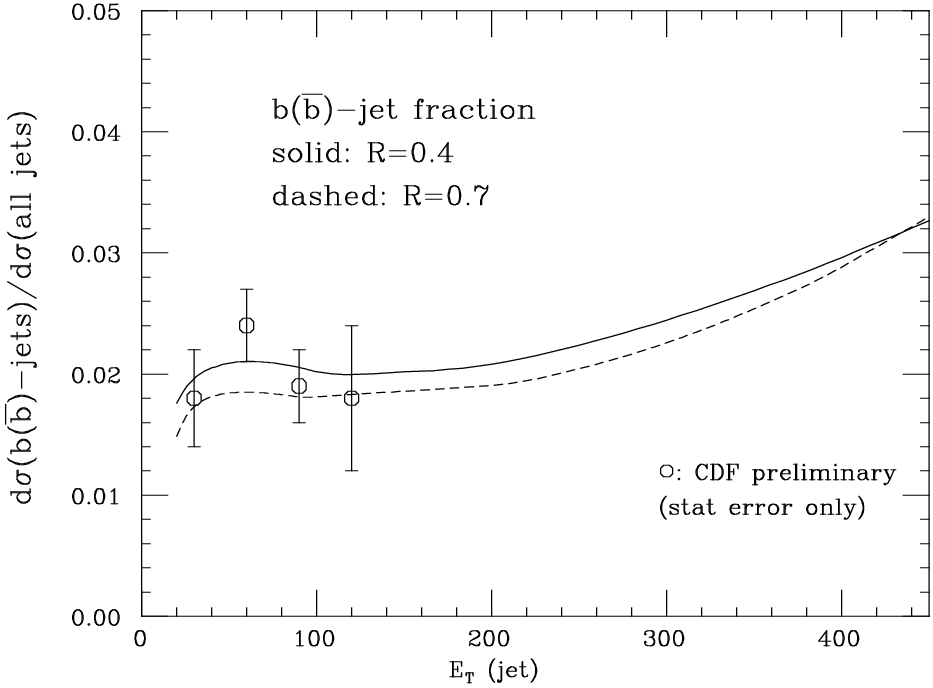,width=0.7\textwidth,clip=}}
\ccaption{}{ \label{highet}
Ratio of the $b(\bar b)$ jet to inclusive-jet \et\
distributions for $\mur=\muf=\mu/2$ and with $R=0.4$ (solid) or
$R=0.7$ (dashes). The data points
are preliminary CDF data~\cite{Koehn95}, obtained with $R=0.4$.}
\end{figure}

\begin{figure}
\centerline{\epsfig{figure=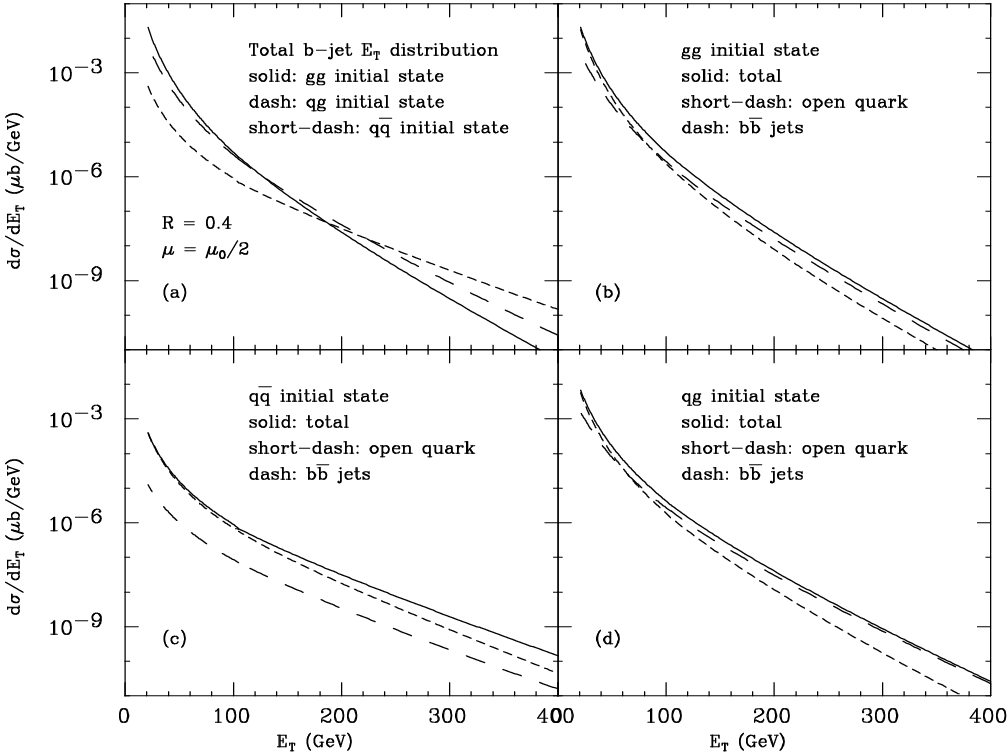,width=\textwidth,clip=}}
\ccaption{}{ \label{bhtot}
Initial-state composition of the $b$-jet production processes,
calculated for $\mur=\muf=\muo/2$ and $R=0.4$ (upper left).
Different components of the
production processes: $gg\;\to\; b$-jet (upper right), $q{\bar q} \;\to\;
b$ jet (lower left) and $qg\;\to\; b$ jet (lower right).}
\end{figure}

Figure~\ref{highet} shows the $b(\bar b)$-jet fraction for \et\ values up
to 450 GeV, for two different values of the cone size ($R=0.4$ and $R=0.7$)
and at $\mur=\muf=\muo/2$.
Notice that while the fraction remains constant
through most of the \et\ range, a rise is observed above 300~GeV. To better
understand the origin of this rise, we present in fig.~\ref{bhtot}a the
separate contribution to the $b$-jet cross section
of the three possible initial states, $gg$, $q\bar q$ and $qg$. Notice that the
$q\bar q$ contribution becomes dominant for $\et>250$~GeV.
Figures~\ref{bhtot}b--d show, for each individual channel,  the separate
contribution of the open-quark and \bbbar-jet components. For \et\ large
enough, the dominant component of the $gg$ and $qg$ channels is given by the
\bbbar-jet contribution, because of the gluon-splitting dominance. In the case
of the $q\bar q$ channel, on the contrary, the \bbbar-jet term is always
suppressed, and most of the $b$ jets are composed of a single $b$
quark, often accompanied by a nearby gluon.
We conclude that at high \et\ the dominant mechanism for the production of a
$b$ jet is the $s$-channel annihilation of light quarks. Since
mass effects are negligible at high \et,
1/5 of the jets produced in $s$-channel annihilation are $b$ jets. A simple
LO calculation shows that the fraction of the two-jet rate due to $s$-channel
light-quark annihilation is about 20\% at \et~=~450~GeV, giving an overall
$b$-jet over inclusive-jet fraction of approximately 4\%. This explains the
rise of the $b$ fraction at high \et, and provides a nice consistency check of
our results. Notice also that while the probability that a gluon jet
will split into a \bbbar\ pair grows at large \et\
faster than the \oacube\
result~\cite{Mueller85,Mueller86,Mangano92,
Seymour95}\footnote{This happens because  of pairs emitted at
higher orders from the gluons of the shower.},  the fraction of primary gluons
in the final state is so small that the overall effect on our predictions is
negligible.

\subsection{Associated production of heavy quarks with $W$ or $\gamma$.}
In addition to the standard measurements of inclusive heavy-quark production,
the high energies and luminosities available at the Tevatron collider allow the
detection of the associated production of heavy quarks and electroweak vector
bosons ($\gamma$, $W$ and $Z$).
The study of these processes goes beyond the goal
of testing perturbative QCD, and has a wide range of applications, which will
be briefly described in this section.

\subsubsection{Photon plus heavy quarks}
The associate production of a photon recoiling against a jet
is dominated by the Compton scattering of a gluon and a quark in the
proton sea. This process, in the most common experimental configurations,
dominates over the quark--antiquark annihilation channel.
In the case of a heavy-flavoured jet we would then be sensitive
to the heavy-flavour component of the proton sea.
The measurement of the \pt\ spectrum of photons produced in
association with charm jets, and the knowledge of the gluon parton density,
could therefore provide in principle a direct measurement of the charm
density inside the proton~\cite{Fletcher89}.
The similar process $g b\to \gamma b$ would allow the study of the $b$ density.
In addition, these processes
provide a useful control sample for the study of heavy-quark tagging.

\begin{figure}
\centerline{\psfig{figure=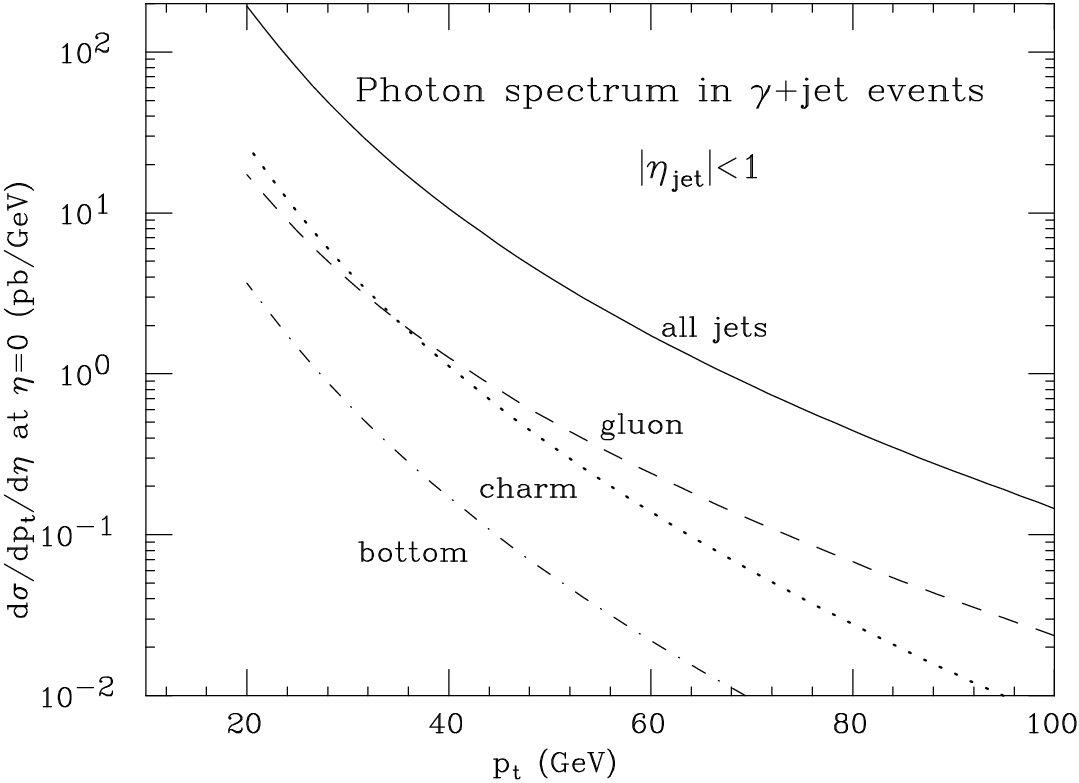,width=0.7\textwidth,clip=}}
\ccaption{}{\label{fig:phojet}
Jet-type composition in $\gamma$+jet events.}
\end{figure}

Figure~\ref{fig:phojet}
contains the distributions of jets of various flavours produced with photons,
as estimated with a LO QCD calculation and CTEQ1M parton
densities~\cite{Botts93}.
Because of the difference in charge and partonic density, associate production
with a bottom quark is suppressed by a factor of 8--10 relative to charm.
Notice also the curious fact that, because of the suppression of the light-quark
annihilation channel, it is more likely for a jet produced in association with
a photon to be a charm jet rather than a gluon jet, at least for transverse
momenta up to 30 GeV.

There is a theoretical interest in these processes because the density
of a heavy quark inside the proton is in principle calculable perturbatively.
Neglecting higher-order logarithmic corrections, which can be resummed using
the Altarelli--Parisi evolution, the inclusive process $p\bar p \to \gamma Q +X$
can be calculated by evaluating the partonic process $gg\to Q \overline{Q}
\gamma$ and
integrating over the phase space of the $\overline{Q}
$~\cite{Mangano94,Stratmann95}. This process is dominated by
configurations where the quark being integrated over is produced at large
rapidity and small \pt. No divergence will appear, because of the heavy-quark
mass. A consistent definition of the heavy-quark density, including threshold
effects \cite{Collins86},
should then reproduce the result of this calculation,
at least at moderate $\pt$. Comparison
against the experimental data is however important, to verify that no
additional non-perturbative effect is at work.

\begin{figure}
\centerline{\psfig{figure=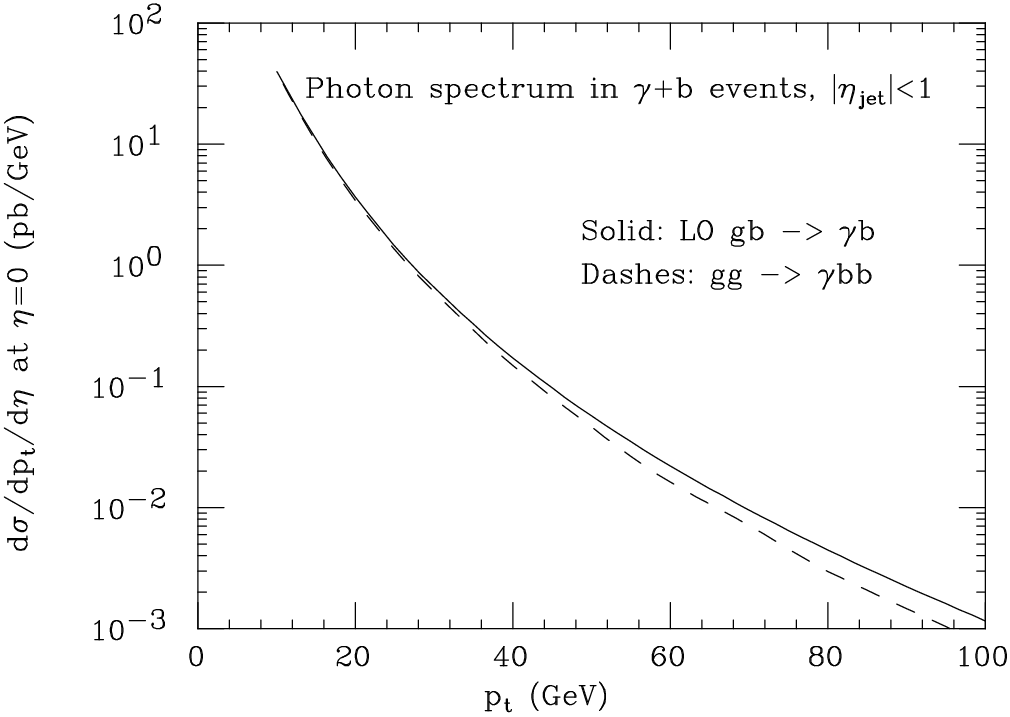,width=0.7\textwidth,clip=}}
\ccaption{}{\label{fig:phobot}
Photon \pt\ distribution in events with a central $b$ jet
in direct photon events,
showing the results of the LO structure-function approach and of the $\oatwo$
calculation. }
\end{figure}
We show in fig.~\ref{fig:phobot}\ the photon \pt\ distribution in events with a
central $b$, calculated~\cite{Mangano94} using the structure function
approach ($bg\to b\gamma$, solid line, CTEQ1M PDF's \cite{Botts93}) and
using the exact matrix elements for the $gg+q\bar q\to\gamma Q\overline{Q}$
processes~\cite{Ellis88} (dashed line).
The $q\bar q$ channel produces a heavy-quark pair via gluon splitting, and
cannot be accounted for by the structure-function calculation. As the plot
shows, there is perfect agreement between the two approaches, at least in the
region where gluon splitting is small. This indicates that the effects of
initial-state evolution for the $b$ quark at these values of $x$ and $Q^2$ are
not important. Notice that this conclusion is important for the consistency of
the NLO evaluation of the inclusive $b$ cross sections,  which contains
only the  flavour excitation diagrams present in the $\oacube$
matrix elements.

Similar results are obtained in the case of charm production, where
there is, however, some larger sensitivity to the choice of the charm mass.
A more comprehensive study of the charm case can be found in
ref.~\cite{Stratmann95}.

A NLO QCD calculation of this process has also been recently
performed~\cite{Bailey96}. This
calculation uses massless charm quarks;  it is therefore valid in the
limit of large transverse momentum. The inclusion of photon isolation and
of a $c\to\gamma$ fragmentation function is then necessary to remove the
divergences due to the collinear emission of a photon from the
final-state charm quark.
The first experimental results on the associated production of photons and
charm have recently appeared from CDF~\cite{Abe96e}.
The quoted result is the cross section for the process $p\bar p \to \gamma
D^{*+}+X$,
for $\pt(\gamma)>16$~GeV, $\pt(D^{*+})>6$~GeV, $\vert \eta(\gamma)\vert < 0.9$
and $\vert \eta(D^{*+})\vert < 1.2$.
The reported measurement is $\sigma=0.38\pm0.15\pm0.11$~nb.
While the statistics of this study (based on approximately 19~pb$^{-1}$, i.e. a
20\% fraction of the full data set currently available) are not sufficient to
produce a distribution in \pt, the agreement
of theory and data is satisfactory. Estimates performed using the LO PYTHIA
Monte Carlo program~\cite{Sjostrand87}, using as renormalization scale
$\mu=\pt(\gamma)$, give rates in the range
0.18--0.22~nb, depending on the set of parton densities used. Given the large
K-factor values found in ref.~\cite{Bailey96} between the LO and NLO
predictions, the CDF results disfavour
a significant presence of a non-perturbative component of charm
quarks inside the proton in the regions of $x$ explored by this measurement.

\subsubsection{$W$ bosons plus heavy quarks}
The associated production of $W^{\pm}$ bosons and heavy quarks is interesting
for several reasons. Production of $W$ plus charm is, in analogy to
the $\gamma$ plus charm case, a potential probe of the partonic densities of
the proton. In this case, the Cabibbo-allowed channel $g s\to Wc$ is the
leading production mechanism, and provides a direct probe of the strange
content of the proton at large $Q^2$~\cite{Baur93}.

The $Wc$ final state is also one of the dominant sources of
events with a $W$ plus one jet with a reconstructed secondary vertex.
As such, these
events enter in the analysis of the low jet multiplicity $W+$jet events, which
is performed as ancillary work to the isolation of the $t\bar t$
signal (see the extensive discussion in ref.~\cite{Abe94}).
Although no cross sections for the $W+c$ process have been explicitly quoted
by the experiments,
the agreement of the data with the LO theoretical predictions, modelled for
the presence of several analysis cuts, is good~\cite{Abe94,Abe95,Abachi95}.
The full set of NLO ($\oatwo$) corrections to this process has recently been
evaluated~\cite{Giele96}, resulting in an improved stability
with respect to
renormalization and factorization scale variations.

Because of the small value of $V_{cb}$ and of the
charm density in the proton, which suppress the LO process $g c\to W b$,
associate production of $W$ and bottom quarks has
its dominant contribution from the higher-order process $q\bar q \to W b\bar
b$~\cite{Kunszt84,Mangano93a}. Interest in this process is once again
related to its relevance for the isolation of the $t \bar t$ final state, which
has always among its decay products a $W$ boson and a $b\bar b$ pair. Direct
evidence for the prompt production of $Wb\bar b$ events comes from the study of
tagged jets in the $W+2J$ sample, performed by the CDF and D0
collaborations in the top analyses (for a detailed discussion, see
ref.~\cite{Abe94}). Because of the small tagging efficiencies for charm
quarks, and because of the limited statistics, there is currently no
experimental result on the production of $Wc\bar c$ events.

Additional interest in the $Wb\bar b$ process comes from the studies of direct
backgrounds to the detection of an intermediate-mass Higgs. This
could be observed with possible future luminosity upgrades of the
Tevatron collider in the mass range $65<M_{\sss H}<110$~GeV using the $p \bar p
\to WH \to W b\bar b$ channel~\cite{Stange94,Stange94a,Amidei96}.
The D0 collaboration has recently presented the first study of a direct search
for Higgs bosons in the $Wb\bar b$ channel~\cite{Abachi96d}. The findings
are consistent with the QCD background estimates, providing preliminary
evidence for the correctness of the QCD calculations of the $Wb\bar b$ rate,
although no explicit cross section has as yet been quoted for this process.
\subsection{Production of top quarks}
\def \mur  {\mbox{$\mu_{\rm \scriptscriptstyle{R}}$}}
\def \muf  {\mbox{$\mu_{\rm \scriptscriptstyle{F}}$}}
\def \muo  {\mbox{$\mu_0$}}
\def \mt   {\mbox{$m_t$}}
\def \ttbar {\mbox{$t \bar t$}}
\def \ptpair {\mbox{$p^{\ttbar}_{\scriptscriptstyle T}$}}
\subsubsection{Total $t\bar t$ production cross sections}
The top quark having been found~\cite{Abe94,Abe95,Abachi95},
the comparison between its observed
production properties and those expected from the Standard Model will be an
important probe of the possible existence of new phenomena.
One of the most important tests to be performed concerns the total production
cross section. This is the most inclusive quantity available, and is a
priori the least sensitive to a detailed understanding of the higher-order
corrections that influence the evolution of the initial and final states.
Within QCD, one expects the perturbative expansion in powers of
$\as(\mt)$ to be well behaved and to provide an accurate estimate of the
total cross section already at low orders.
In particular, the first estimates of the total production cross section using
the full NLO matrix elements~\cite{Nason88,Beenakker89,Ellis91}
gave an increase (relative to the Born result) of the order
of 30\% for masses above 100 GeV. The residual perturbative QCD uncertainty,
evaluated by
varying the renormalization and factorization scales, was estimated
to be no larger
than 10\%. The choice of parametrization for the input parton densities was
also shown to give effects of this order of magnitude,
by using the available sets.

It was later pointed out in refs.~\cite{Laenen92,Laenen94} that logarithmic
contributions associated with the emission of soft gluons from the initial
state could significantly enhance the NLO result.
Similar conclusions were reached in refs.~\cite{Berger95,Berger96}.
More recent
studies~\cite{Catani96,Catani96a},  which will be described in more detail in
Section~\ref{sec:resum},
have proved that the effect of soft-gluon resummation is actually
very small, of the order of a 1\% correction to the total cross section.

Beside the soft-gluon emission effects, also Coulomb
effects~\cite{Sommerfeld39}
may enhance or deplete the cross section near threshold, as originally
discussed in ref. \cite{Fadin90}.
These effects can be evaluated in the following way.
We separate the partonic Born cross-section formulae into their colour
singlet and octet components
\beqn
\hat{\sigma}^{(gg)}&=&\hat{\sigma}_{(8)}^{(gg)}+\hat{\sigma}_{(1)}^{(gg)}
\\
\hat{\sigma}^{(q\bar{q})}&=&\hat{\sigma}_{(8)}^{(q\bar{q})}
\eeqn
where $\hat{\sigma}^{(q\bar{q},gg)}$ can be found in ref.~\cite{Nason88}, and
\beq
\hat{\sigma}_{(1)}^{(gg)}(\rho)=\frac{\as^2}{m^2}\;
\frac{\beta\rho\pi}{384}\left[\frac{1}{\beta}\log\frac{1+\beta}{1-\beta}
(4+4\rho-2\rho^2)-4-4\rho\right]\;.
\eeq
Here $\beta=\sqrt{1-\rho}$ and $\rho=4m^2/\hat s$.
The Coulomb-resummed cross section is given as
\beq
\hat\sigma^{\rm Coul}(\rho)
=\hat\sigma_{(8)}(\rho)\frac{\pi\as/(6\beta)}{\exp(\pi\as/(6\beta))-1}
+\hat\sigma_{(1)}(\rho)
\frac{4\pi\as/(3\beta)}{1-\exp(-4\pi\as/(3\beta))}\;.
\eeq
We do not include bound-state effects, which, as shown in ref.~\cite{Fadin90},
are much smaller.
As shown in ref.~\cite{Catani96}, the
relative correction due to the resummation
of Coulomb effects not already included into the NLO results is an effect
smaller than 1\% of the NLO cross section.
Together with soft-gluon resummation effects, higher-order Coulomb corrections
can therefore be entirely neglected, given the theoretical NLO uncertainties
and the current experimental accuracy.

Electroweak processes can give rise to single top
production. This happens through the so-called $Wg$ fusion
process~\cite{Dawson85,Willenbrock86,Yuan90,Ellis92,Bordes95},
whereby a virtual $W$, emitted from an initial-state light quark,
interacts with
an initial-state gluon and produces a $t\bar b$ pair, and through the decay
of an off-shell $W$ boson into a $t\bar b$ pair~\cite{Cortese91,Stelzer95}.
While in principle these processes do not contribute to the observation of
$t\bar t$ pairs, and therefore should not interfere with the measurement and
study of the QCD production properties of top quarks, in practice they
constitute an irreducible background to the detection of $t\bar t $ events.
From the point of view of QCD studies, it is therefore a welcome fact that
their total contribution to the inclusive top-production cross section is only
a fraction of the order of 25\%.
Their study is however important for electroweak measurements, as it probes
the $Wtb$ vertex and the value of $V_{tb}$ directly.
A complete review of the phenomenological applications of single top
production, in addition to a complete list of the relevant references, can be
found in ref.~\cite{Amidei96}.

Electroweak radiative corrections to top production in hadronic collisions have
been considered in  refs.~\cite{Kao93,Stange93,Beenakker94,Harlander95}.
They range from $-0.97\%$ to $-1.74\%$ of the Born cross section, for a Higgs
mass of 60 and 1000~GeV respectively\footnote{W.~Hollik and D.~Wackeroth,
private communication.}.

\begin{figure}
\centerline{\epsfig{figure=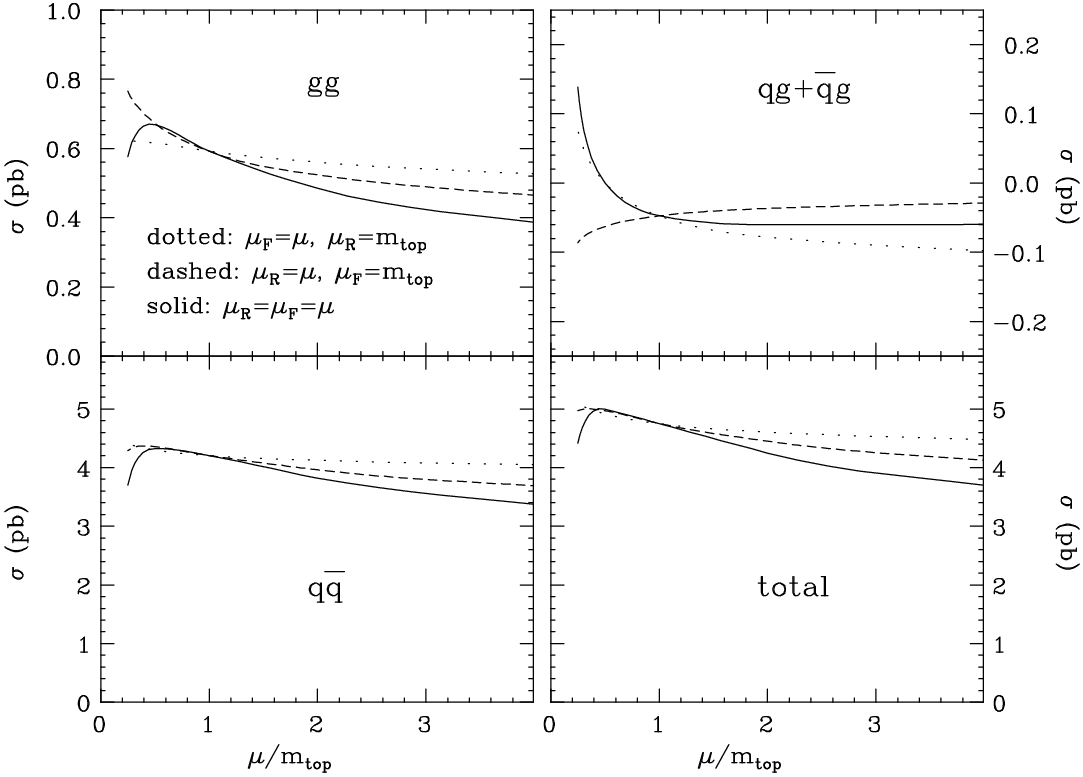,width=0.7\textwidth,clip=}}
\ccaption{}{ \label{fig:topscales}
Scale dependence of the top cross section at the Tevatron, for
$m_t=175\,$GeV.
Dotted lines are the $\muf$ dependence at fixed  $\mur$,
dashed lines are the $\mur$ dependence at fixed  $\muf$,
and solid lines are obtained by the simultaneous variation of $\mur$
and $\muf$.
}
\end{figure}
In this section we will confine our theoretical analysis to results
obtained within the QCD $\oacube$ approximation.
It should be pointed out that this study is performed within the strict domain
of the Standard Model. Corrections to the top cross section much larger than
the Standard Model QCD uncertainties can be obtained in the presence of new
phenomena. Virtual corrections due to loops of supersymmetric particles have
been studied, for example, in refs.~\cite{Li95,Li96,Alam96,Kim96,Sullivan96}.  For a
partial list of examples of alternative production mechanisms for top quarks,
arising in theories beyond the SM such as Supersymmetry or Technicolor,  see
for instance refs.~\cite{Eichten94,Hill94,Holdom95,Casalbuoni96,Kane96}.

We now proceed to a review of the theoretical uncertainties on the total
cross section for $t\bar t$ production, which arise at NLO. Uncertainties due to
unknown higher-order effects are usually accounted for by varying the
renormalization ($\mur$) and factorization ($\muf$) scales. In
principle, independent variations of the two scales should be considered.  In
fig.~\ref{fig:topscales} we show the scale dependence of the top cross section.
Notice that there is a compensation of the renormalization-scale dependence
when the different subprocesses are added up.
In fact, the renormalization-scale
dependence in the $gg$ and $qq$ channels has a behaviour opposite to that
in the $qg$ channel. It is only the combined
scale dependence that can be considered an estimate of the neglected
subleading corrections. As a second point, we observe that
the maximum of the cross section is reached around $m_t/2$. Thus, the usual
choice of the range $m_t/2<\mu<2m_t$
appears to be particularly justified in this case. As a third point,
we observe that the scale dependence in the cross section goes in the same
direction for the two scales, so that, for the purpose of estimating
the associated uncertainty, it is sufficient to consider the simultaneous
variation of $\mur$ and $\muf$.

\begin{figure}
\centerline{\epsfig{figure=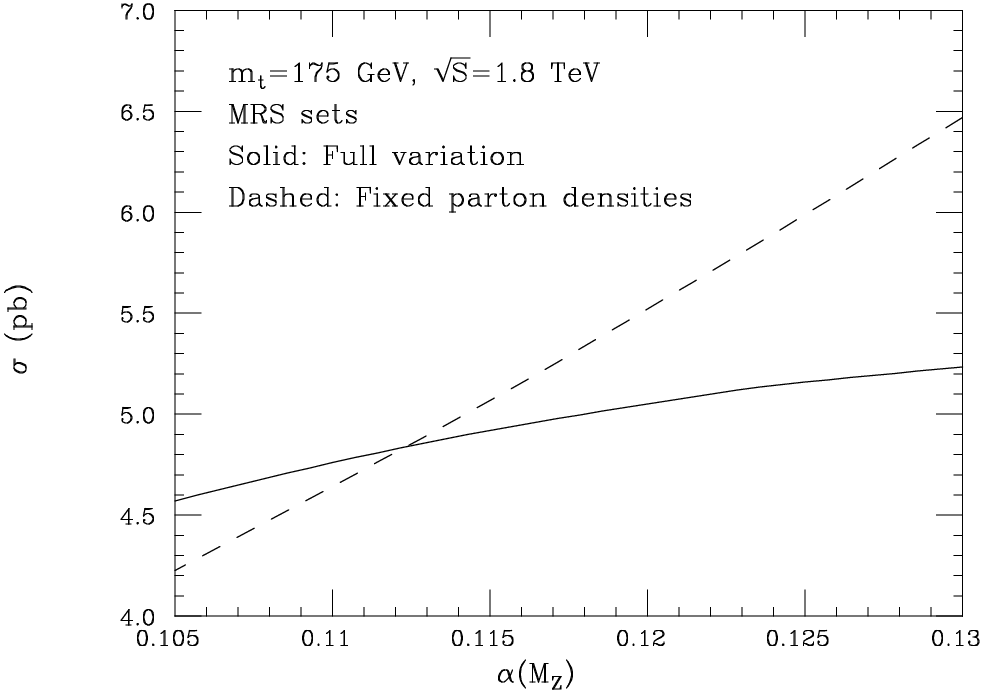,width=0.7\textwidth,clip=}}
\ccaption{}{ \label{fig:lambdavar}
Top cross section as a function of $\as(M_{\rm Z})$.
The dashed line (obtained with the MRSA$^\prime$ set)
does not include the variation of the parton
densities due to the change in $\as$. }
\end{figure}
Aside from the scale uncertainties, which reflect the limitations of
the perturbative QCD calculation,
there are uncertainties associated with our imprecise
knowledge of the physical parameters involved. In particular,
the strong coupling constant is determined within a certain accuracy.

In the case of top production at the Tevatron, there is
fortunately a compensating mechanism that reduces the dependence
of the cross section upon $\as$. In fact, the top cross section is dominated
by the $q\bar{q}$ annihilation process. Quark densities in the proton are
directly measured in DIS at a scale around $10\;$GeV$^2$, and the QCD
evolution to scales of the order of the top mass makes them softer.
Therefore, for larger coupling, the partonic cross section increases, but the
quark luminosity decreases.

In order to perform a fair estimate of the uncertainty due to
$\Lambda_{\rm QCD}$, we need sets of parton densities
fitted with different values
of the strong coupling. The sets of ref.~\cite{Martin95a} meet our
purpose\footnote{The values of $\Lambda_4$ that accompany the FORTRAN
program for the structure function sets of ref.~\cite{Martin95a}
are not consistent with the values of $\as$ quoted there,
the differences being of the order of 1\%. In the present work,
we extract the values of $\Lambda_5$ from their quoted values of $\as$ using
the standard two-loop formula~\protect{\cite{Barnett96}}.}.
In fig.~\ref{fig:lambdavar} we show the cross section as a function of the
strong coupling.

For comparison we also show the $\as$ dependence
if the parton densities are kept fixed to the MRSA$^\prime$ set.
We see the remarkable reduction
in the  $\as$ dependence, due to the compensation of the rise
of the partonic cross section and the decrease of the quark parton densities.

Our results for the top cross section for $m_t = 175$~GeV
are collected in table \ref{tab:top175}.
We also show results obtained with the recent parametrization derived in
ref.~\cite{Huston95} by including the CDF jet data. We stress that the
numbers in the table are obtained with a standard one-loop calculation. No
resummation effects are included.
{\renewcommand{\arraystretch}{1.8}
\begin{table}
\begin{center}
\begin{tabular}{|c||c|c|c|c|c|c|c|c|} \hline
& CTEQ1M & CTEQ$^\prime$ & MRSA$^\prime$ & \multicolumn{5}{c|}{MRS,
variable $\Lambda$,  $\as(M_{\rm Z})=$} \\ \cline{5-9}
 $\mur=\muf$   &   &       &       &$0.105$ &
$0.110$ & $0.115$ & $0.120$ &
$0.125$
\\ \hline\hline
\mt/2 & 5.24 & 5.07 & 5.00 & 4.78 & 4.99 & 5.18 & 5.34 & 5.48
\\ \hline
\mt   & 4.96 & 4.86 & 4.75 & 4.57 & 4.76 & 4.92 & 5.05 & 5.16
\\ \hline
2\mt  & 4.38 & 4.38 & 4.25 & 4.13 & 4.27 & 4.38 & 4.47 & 4.52
\\ \hline
\end{tabular}
\ccaption{}{\label{tab:top175} Total cross sections (in pb) for $\mt=175$~GeV
at NLO. The set CTEQ$^\prime$ is taken from ref.~\protect{\cite{Huston95}}.}
\end{center}
\end{table} }

Our range for the top cross section at $m_t=175$~GeV ($m_t=170$~GeV)
is given by $4.75{+0.73\atop-0.62}$~pb ($5.57{+0.86\atop-0.73}$~pb).
This should be compared with the current experimental results:
\ba
     \sigma_{t\bar t}(m_t=175\;{\rm GeV}) &=& 7.5 \pm 1.6 \; {\rm pb} \quad
     \mbox{(CDF \cite{Caner96})} \;  ;  \\
     \sigma_{t\bar t}(m_t=170\;{\rm GeV}) &=& 5.2 \pm 1.8 \; {\rm pb} \quad
     \mbox{(D0 \cite{Narain96})} \;  .
\ea
As a central value for our determination we have chosen the
MRSA$^\prime$~\cite{Martin95} result with $\mu_{\rm R}=\mu_{\rm F}=m_{\rm t}$,
in association with a value of $\Lambda_5=0.152\,$GeV (which corresponds to
$\as(M_Z)=0.1113$,
according to the standard two-loop formula \cite{Barnett96}).
For the MRS sets with variable $\Lambda$, we have used
$\Lambda_5=0.0994$, 0.140, 0.190, 0.253, 0.328~GeV (which correspond to
$\as(M_Z)=0.105,\,0.110,\,0.115,\,0.120,\,0.125$).

\begin{figure}
\centerline{\epsfig{figure=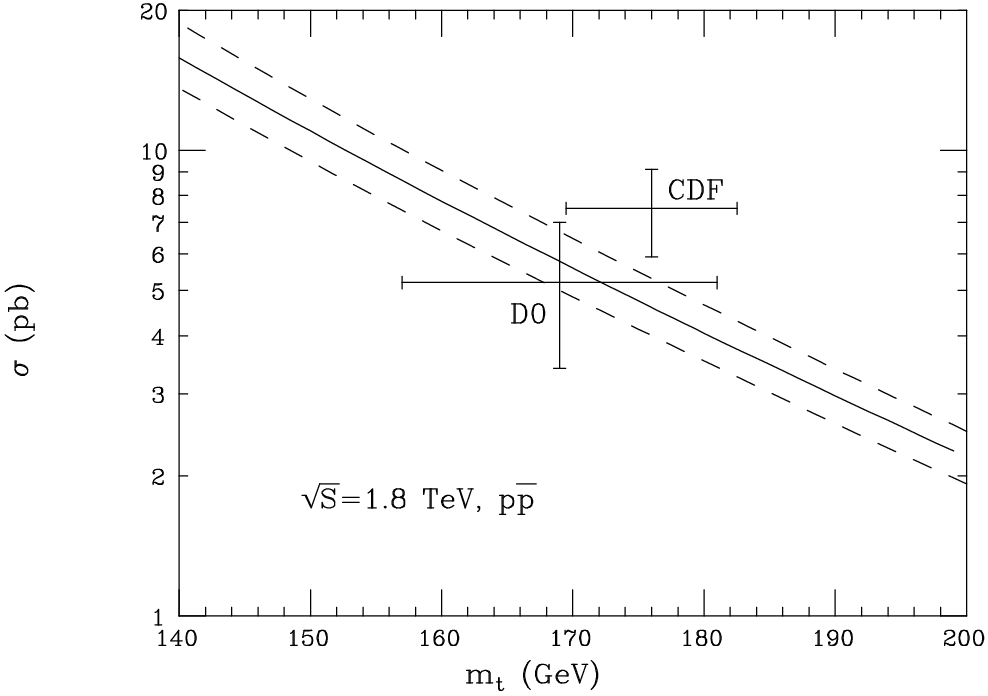,width=0.7\textwidth,clip=}}
\ccaption{}{ \label{fig:topxsc}
Top cross-section at the Tevatron at $\sqrt{S}=1.8\;$TeV.
The solid line is obtained with MRSA$^\prime$
parton density, and the dashed lines correspond to the upper and
lower values obtained in table~\protect{\ref{tab:top175}}. The experimental
data are taken from refs.~\protect{\cite{Caner96,Narain96}}.}
\end{figure}

The cross section bands are also shown in fig.~\ref{fig:topxsc}.
The agreement of theory and data is good, but it is clear that higher
statistics should be collected before a significant test is achieved.

For reference, we also quote the cross section for top production at the
LHC. We get $\sigma(t\bar{t})=0.77{+0.25\atop -0.12}\;$nb,
for $m_t=175\;$GeV
and $\sqrt{S}=14\;$TeV. The error is obtained with the same scale and $\as$
variations that we used for the Tevatron cross section. Notice that
the relative error is much
larger than at the Tevatron, because of the greater importance
of the $gg$ initial-state contribution at the LHC energy.

\subsubsection{Top kinematical distributions}
Because of the high precision of the theoretical prediction,
the measurement of the top cross section could become a sensitive probe
of new physics.
The only way to unambiguously disentangle the standard sources of top production
from possible new physics will nevertheless rest on the detection of
significant discrepancies in the kinematical features of top production from
what is expected from QCD.  For example, the detection of mass peaks in the
invariant mass distribution of top pairs would indicate the possible presence
of $s$-channel resonances strongly coupled to the top~\cite{Eichten94,Hill94}.

\begin{figure}
\centerline{\psfig{figure=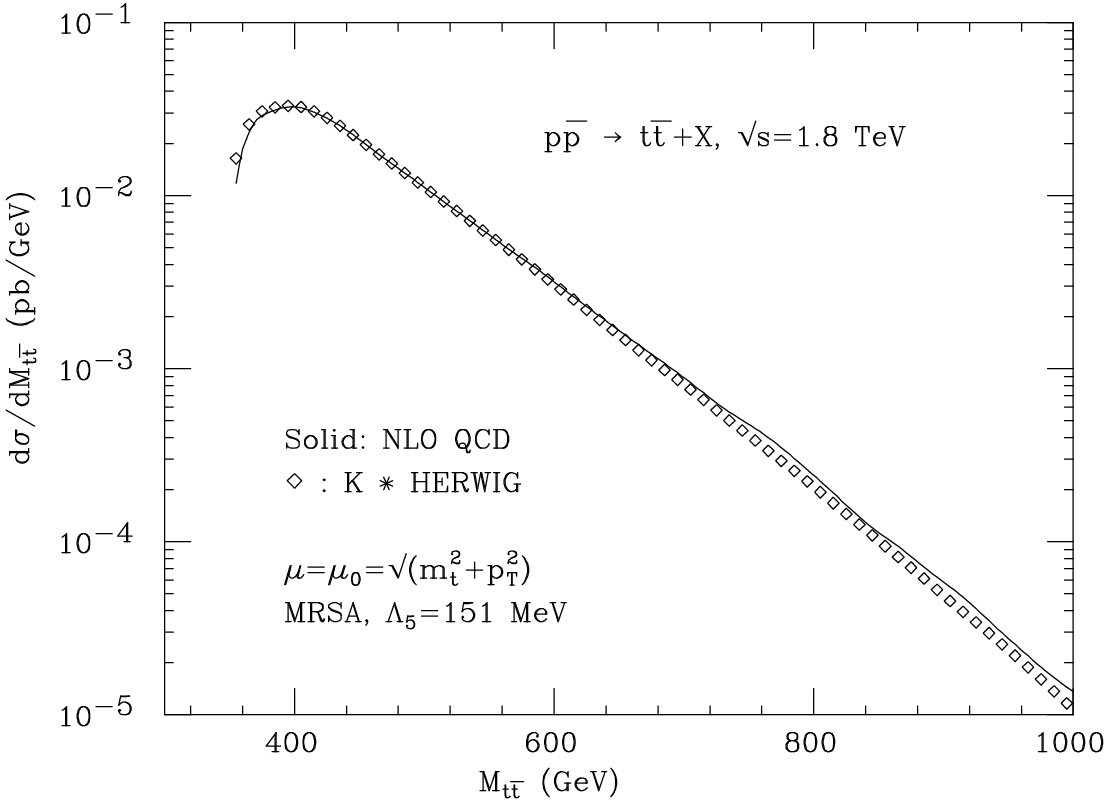,width=0.7\textwidth,clip=}}
\ccaption{}{ \label{fig:mttbar}
Invariant-mass distribution of the $t\bar{t}$ pair.
The HERWIG prediction has been rescaled by a constant factor $K=1.34$.}
\end{figure}

\begin{figure}
\centerline{\psfig{figure=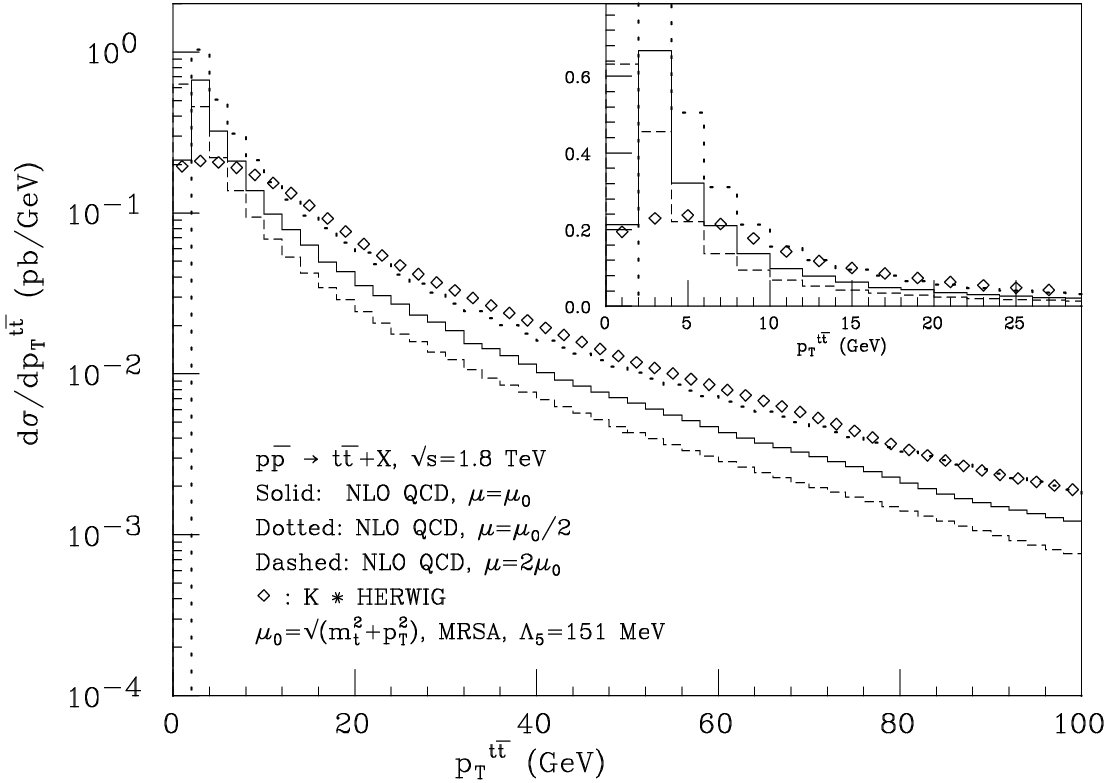,width=0.7\textwidth,clip=}}
\ccaption{}{ \label{fig:ttbpt}
Transverse-momentum distribution of the $t\bar{t}$ pair.
The HERWIG prediction has been rescaled by a constant factor $K=1.34$.}
\end{figure}

Kinematical distributions for top-quark pairs produced at the Tevatron have
been calculated at NLO in ref.~\cite{Frixione95}.
Since top quarks manifest themselves exerimentally in a rather indirect way,
their identification relies on complex series of experimental
cuts~\cite{Abe94,Abe95,Abachi95,Campagnari96,Sinervo96}. The
study of their kinematical properties therefore  strongly relies on the
modelling of the top-quark production mechanism itself. In particular,
experimental analyses are performed using parton shower Monte Carlo (MC)
programs, such as HERWIG~\cite{Marchesini88,Marchesini92},
ISAJET~\cite{Paige86} or PYTHIA~\cite{Sjostrand87}.
It is therefore important to compare the distributions predicted by these
MC calculations with what is expected from NLO QCD. The study performed in
ref.~\cite{Frixione95} indicates an excellent agreement between these
results, at least for distributions that are non-trivial at LO in
perturbation theory, such as the top inclusive \pt\ distribution
or the invariant-mass
distribution of the top-quark pair (fig.~\ref{fig:mttbar}).
The theoretical uncertainty from higher-order corrections and from
hadronization effects is very small for these variables, making them a rather
solid term of comparison in the search for new physics. Distributions that
are non-trivial only at $\oacube$, such as the \pt\ of the top pair,
show instead significant higher-order corrections. This is visible, for the
example of the top-pair \pt\ distribution,
in fig.~\ref{fig:ttbpt}, where a comparison is made between the NLO QCD
predictions and those of the HERWIG MC. The multiple gluon emission
from the initial state of the hard process, included in the MC calculation,
significantly increases the average transverse momentum of the pair, as can
also be seen by the numbers quoted in table~\ref{tab:ttbpt}. Notice the
significant scale dependence of the NLO result.
Similar discrepancies between the results of higher-order exact matrix element
calculations and shower MC predictions have been reported
in~\cite{Orr95,Orr95a}.

It should be pointed out that
the transverse momentum of the top pair, directly related to the
jet activity that accompanies a top production event, has important
consequences on the determination of the experimental top detection efficiency,
as well as a large impact on the reliability with which the mass of the top
quark can be measured~\cite{Abe94,Abe95,Abachi95,Campagnari96}. The
uncertainty related to the modelling of the gluon radiation emitted by initial
and final state, in $t\bar t$ production and decay, is in fact one of the
largest systematics present in the most recent top-mass
determinations~\cite{Galtieri96,Grannis96}.
{\renewcommand{\arraystretch}{1.6}
\begin{table}
\begin{center}
\begin{tabular}{|l||c|c|c|} \hline
  & $\langle \ptpair \rangle$ (GeV) &  F($\ptpair>10$ GeV) (\%)
  &  F($\ptpair>20$ GeV) (\%) \\ \hline\hline
NLO QCD, \mur=\muf=\muo   &  11.9 &    30 & 15 \\ \hline
NLO QCD, \mur=\muf=2\muo  &  9.1  &    23 & 11 \\ \hline
NLO QCD, \mur=\muf=\muo/2 &  16.6 &    44 & 22 \\ \hline
HERWIG,  \mur=\muf=\muo   &  17.5 &    51 & 28 \\ \hline
\end{tabular}
\ccaption{}{\label{tab:ttbpt}
Average value of \ptpair\ and fraction of the cross section with
$\ptpair>10$ and {\rm 20} GeV for different calculations. }
\end{center}
\end{table} }

\begin{figure}
\centerline{\psfig{figure=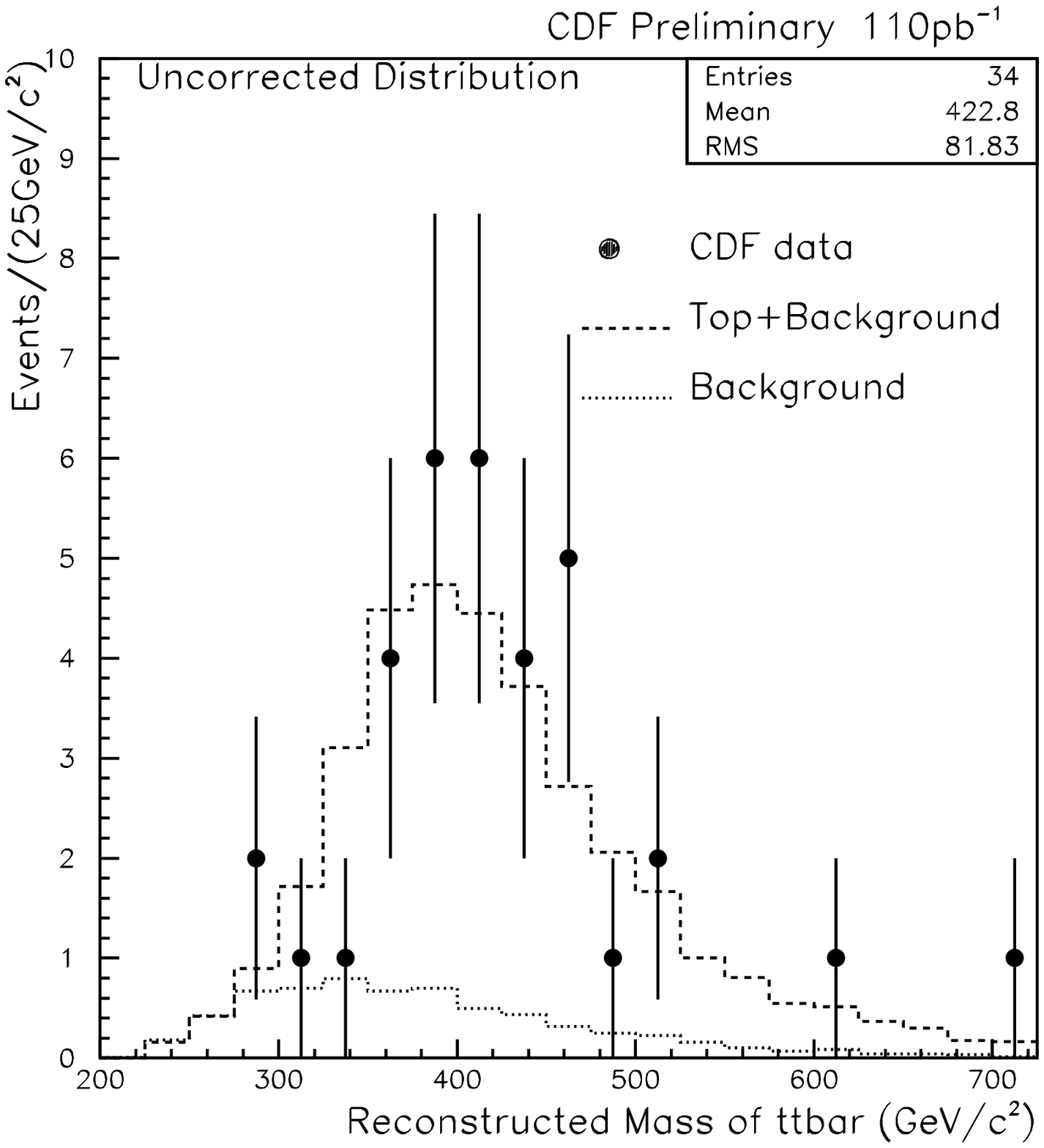,width=0.5\textwidth,clip=}}
\ccaption{}{ \label{fig:cdfmtt}
CDF data on the invariant mass of top pairs, from $W+4$ jets (one of which tagged
as a $b$), compared with the HERWIG MC expectations.}
\end{figure}

\begin{figure}
\centerline{\psfig{figure=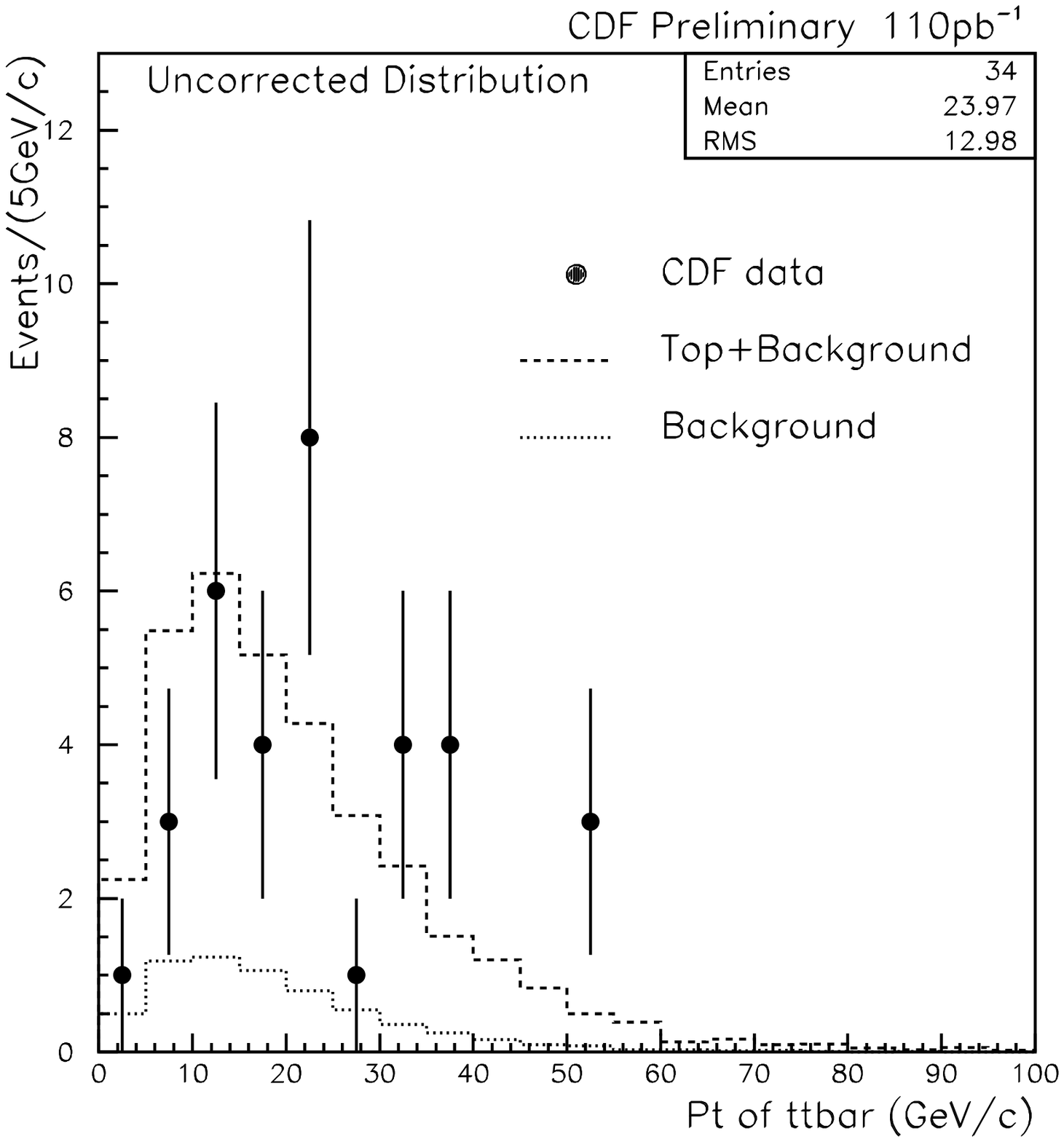,width=0.5\textwidth,clip=}}
\ccaption{}{ \label{fig:cdfptg}
CDF data on the transverse momentum distribution of top pairs, from $W+4$ jets
(one of which tagged as a $b$), compared with the HERWIG MC
expectations.}
\end{figure}

Although the statistics of the data on top production are still low, the first
distributions have been presented by CDF~\cite{Caner96}.
The top-pair invariant-mass spectrum is in good agreement with the theoretical
expectations, and currently shows no evidence of an anomalous production
source in the $s$-channel (fig.~\ref{fig:cdfmtt}). On the contrary, the
pair transverse momentum is harder than the HERWIG MC predicts
(fig.~\ref{fig:cdfptg}).
As was pointed out above, this distribution is not, at the moment, predicted very
accurately. It is therefore premature to conclude that the present
discrepancy is physically significant, especially in view of the limited
statistics.

To summarize this section on top production, we conclude that nothing
surprising has
been discovered so far in the comparison of data with the predictions of NLO
perturbative QCD.  As expected, the agreement between data and theory is
already more solid than in the case of bottom or charm production. Given the
claimed accuracy of the theoretical predictions, at least for the total cross
section and for inclusive quantities, only additional statistics will however
make these comparisons more compelling. Detailed features of the structure of
the final state in top production, such as studies of the jet activity, have so
far been probed only indirectly in the context of the mass measurements.
It is likely that a lot will be learned in the future from these more detailed
analyses, and that significant progress will be made in their MC modelling.
%
\newcommand\bz{b_0}
\newcommand\hrho{\hat{\rho}}
\section{Higher orders and resummation}\label{sec:resum}
\sethead{Higher orders}
In this section we will deal with the problem of the
resummation of logarithmically enhanced effects  in the vicinity of
the threshold region in hard hadroproduction processes.
Drell--Yan lepton-pair production has been in the past the best
studied example of this sort.
The threshold region
is reached when the invariant mass of the lepton pair
approaches the total available energy.
A large amount of theoretical
and phenomenological work has been done on this subject.
The articles in refs. \cite{Sterman87,Catani89} summarize the
theoretical status
of the subject, and also include references to the extensive literature
in this field. For the Drell--Yan process the resummation of soft gluons
has been computed to NLO accuracy. Extension of the
NLO resummation formalism to heavy-flavour production is under
way \cite{Kidonakis96,Contopanagos96}.

Resummation formulae have also been used in estimating heavy-flavour
production \cite{Laenen92,Laenen94,Berger95,Berger96,Kidonakis95}.
These works indicate the presence of very large higher-order corrections
to heavy-flavour production at colliders, in particular in the $gg$
production channel.
Since top production at the Tevatron is dominated by the $q\bar{q}$
channel, the claimed size of the resummation effects is
about 15\%, below the current experimental uncertainties on
the cross section. Nevertheless, it is important to establish whether
threshold resummation effects are really so large, a fact that would
cast doubts on every QCD calculation of hadronic cross sections.
Recently, in ref.~\cite{Catani96,Catani96a}, it was argued that in fact
resummation effects are not as large as found in previous computations.
In particular, it was pointed out there
that large spurious terms are present in certain formulations
of the resummation problem
that are not justified by the threshold approximation.
Furthermore, it was also shown that resummation may be formulated
in such a way that these terms are absent. In the following
we will present a brief review of this problem.

\subsection{What are soft-gluon effects}
\label{sec:softgluons}
\begin{figure}[htbp]
  \begin{center}
    \leavevmode
    \epsfig{file=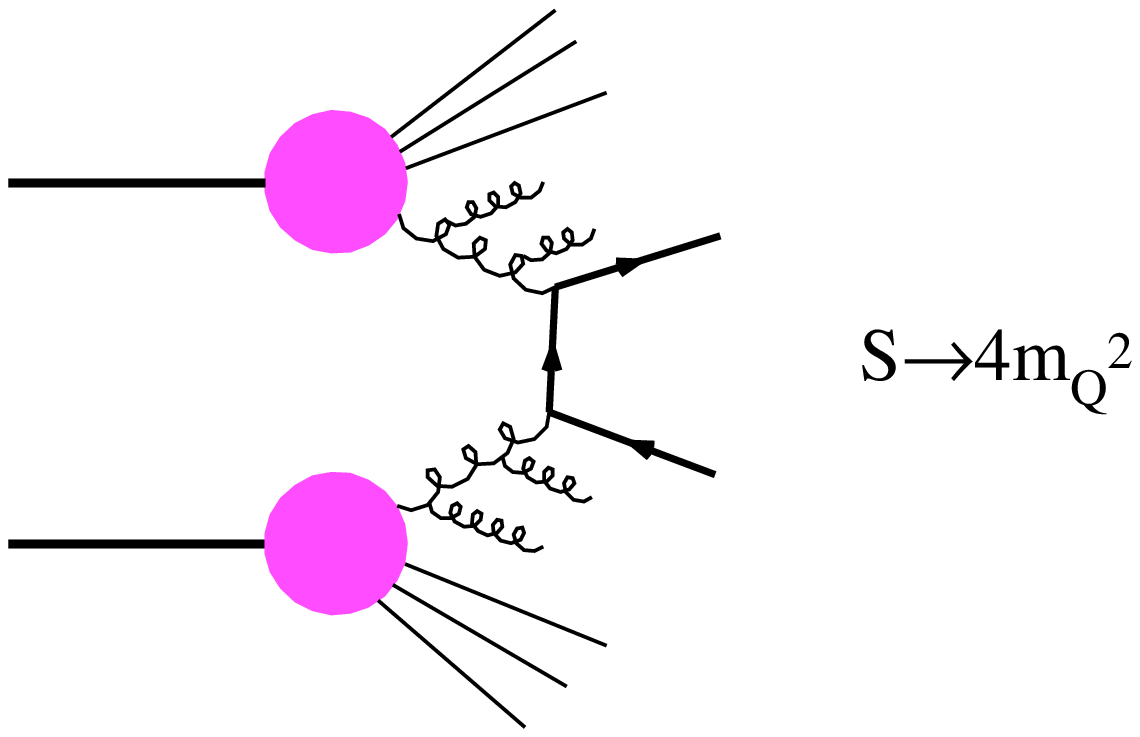,width=6cm}
    \ccaption{}{\label{fig:hvqprod}
      Heavy-flavour production near threshold.}
  \end{center}
\end{figure}

Coloured particles emit soft gluons with high probability.
Normally, the effect of soft-gluon emission is small (at least in
inclusive quantities), since they only slightly affect the kinematics
of a process. However, in a production process of high-mass
objects, when we approach the threshold, soft-gluon emission
becomes important. Let us fix our attention on heavy-flavour
production in hadronic collisions near threshold.
The process is schematically depicted
in fig.~\ref{fig:hvqprod}.
The incoming protons make a big effort in providing partons
with a large fraction of the longitudinal momentum, thus going towards
the very large $x$ region of the structure functions.
Under these circumstances, even the small amount of energy wasted by
soft-gluon radiation yields an important suppression of the cross section.
In the usual application of the factorization theorem, part of the radiative
corrections due to gluon radiation are included in the structure functions.
Thus, depending
upon the factorization scheme, the left-over suppression may be positive
or negative. In the \MSB\ and DIS scheme the left-over is negative,
so that the suppression effect appears instead as an enhancement
of the cross section.

In the case of heavy-flavour production,
the perturbative expansion
for the partonic cross section at ${\cal O}(\as^3)\,$
(omitting obvious indices, such as the incoming parton types)
has the structure
\beq
\hat\sigma(\hat{s})=\sigma_0(\hat{s})\,
\left[1+C\as\log^2(1-\hrho)+{\cal O}(\as\log(1-\hrho),\as^2)\right]\,,
\eeq
where $\hat{s}$ is the square of the partonic centre-of-mass energy,
$\hrho=4m^2/\hat{s}\,$, $\sigma_0(\hat{s})$ is the Born cross section
\beq
\sigma_0(\hat{s})=\frac{\as^2}{m^2}f^{(0)}(\hrho)\;,
\eeq
and $\as=\as(m^2)$.
The reader can find in ref.~\cite{Nason88} explicit formulae for the functions
$f^{(0)}(\hrho)$, as well as for the constant coefficient $C$.
The precise value of $C$ depends upon the type of
incoming partons (i.e. quarks or gluons) and upon the factorization
scheme. In both the \MSB\ and the DIS scheme it is positive,
so that in the following we can think of it as being a positive constant.
Resummation, according to refs.~\cite{Laenen92,Laenen94}, gives
\beq\label{ResLaenen}
\hat\sigma^{\rm (res)}(\hat{s}) = \frac{\as^2}{m^2}f^{(0)}(\hrho)
\exp\left[C\as(s^\prime)\log^2(1-\hrho)\right]~~,
\eeq
where $s^\prime$ is a scheme-dependent function of $\hat{s}$
that goes to zero as $\hat{s}\,\to\, 4m^2$.
We have $s^\prime=(1-\hrho)^\eta m^2$, where $\eta=1$ in
the DIS scheme, and 3/2 in the \MSB\ scheme.
Formula (\ref{ResLaenen}) is supposed
to include all terms of order $\as^m\log^n (1-\hrho)$ with $n>m$
in the exponent.
Remember in fact that
\beq
\as(s^\prime)=\frac{1}{\bz \log\frac{s^\prime}{\Lambda^2}}
=\frac{1}{\bz \log\frac{m^2}{\Lambda^2}+\bz\eta \log(1-\hrho)}
=\as\left(1-\as\bz\eta\log(1-\hrho)+\ldots\right)\;,
\eeq
so that in the exponent there are terms with arbitrary powers
of $\as$, and a power of the logarithm that is always larger than the power
of $\as$. The first subleading terms have the form
$\as^k\log^k (1-\hrho)$.

In order to get a physical cross section, the partonic cross section
given above
should be convoluted with parton luminosities:
\beq\label{ResLaenenhad}
\sigma^{\rm (res)}
=\frac{\as^2}{m^2}\int {\cal L}(\tau)f^{(0)}(\hrho)
\exp\left[C\as\left((1-\hrho)^\eta m^2\right)\log^2(1-\hrho)\right]
d\tau~,
\eeq
where, omitting obvious parton indices,
\beq
{\cal L}(\tau)=\int F(x_1)F(x_2)\delta(x_1 x_2-\tau)\,dx_1 dx_2\,.
\eeq
Here we can spot a problem in the resummation formula.
When performing our integral over $\tau$, as $\hrho\,\to\,1$
the argument of $\as$ in the exponential approaches zero.
Before it actually hits the zero, it will hit
a singularity in the running coupling $\as(s^\prime)$,
causing the integral to diverge. In order to avoid this divergence, a cutoff
$\mu_0$ was introduced in the literature \cite{Laenen92,Laenen94,Appel88}.
Observe that this cutoff has nothing
to do with the standard factorization and renormalization scale $\mu$.
It is essentially a cutoff on soft-gluon radiation, imposed in order to avoid
the blowing up of the running coupling associated with soft-gluon emission.

The use of a cutoff seems an ad hoc procedure in this case.
However, it can be justified to some extent. Suppose, for example,
that we end up in a QCD calculation with a formula like
\beq\label{example}
G=\int_0^{Q}\as(k^2) G(k^2) dk~,
\eeq
where $G(k^2)$ is a smooth function as $k^2\,\to\, 0$.
Integrals of this kind are often found, for example,
in the computation of shape variables in jet physics.
This expression is divergent as $k^2\,\to\,0$, since at some
point $\as$ approaches the Landau pole. The divergence can be handled
by a cutoff $\mu_0$, which has to be large enough for $\as$ to be barely
perturbative. For example, we may choose $\mu_0=5\Lambda$, a value around
2 GeV. We can then argue that
\beq
G=\int_{\mu_0}^{Q}\as(k^2)\, G(k^2)\, dk +C\frac{\mu_0}{Q}\,.
\eeq
In fact, the divergence of the 1-loop expression of $\as$
does not signal a real physical divergence. More likely,
the point at which $\as$ becomes of order 1 signals the breakdown of
perturbation theory. We therefore exclude this region, estimate
its contribution by dimensional analysis, and obtain a power correction.
A slightly more formal justification makes use of the concept of
IR (infrared) renormalons.
We expand eq.~(\ref{example}) in powers of $\as=\as(Q^2)$, using
\begin{equation}
  \label{asexp}
\as(k^2)=\frac{1}{\bz \log\frac{k^2}{\Lambda^2}}
=\frac{1}{\bz \log\frac{Q^2}{\Lambda^2}+\bz \log\frac{k^2}{Q^2}}
=\as\sum_{j=0}^\infty \left(-\as\bz\log\frac{k^2}{Q^2}\right)^j\;,
\end{equation}
and we get
\begin{eqnarray}
  G&\propto&\int_0^{Q} \as\sum_{j=0}^\infty
  \left( -\as\bz\log\frac{k^2}{Q^2} \right)^j \;dk
\nonumber \\ \label{renorms}
 &=& {\as} \sum_{j=0}^\infty (\as\bz)^j\int_0^\infty t^j e^{-t/2} dt~,
\end{eqnarray}
where $t=\log Q^2/k^2$. The integral can be performed, and one gets
\begin{equation}\label{divexp}
  G \propto 2\as\sum_{j=0}^\infty j!\,(2\as b_0)^j\;,
\end{equation}
which is a divergent series, since a factorial grows faster than any power.
This lack of convergence is in fact a general feature of the perturbative
expansion in field theory. The perturbative expansion should be interpreted
as an asymptotic one.
The terms of the expansion (\ref{divexp}) decrease for moderate
values of $j$. As $j$ grows, the factorial takes over, the terms
stop decreasing and begin to increase. This happens at the value of $j$
at which the next term is equal to the current one
\beq
(j+1)!\,(2\as b_0)^{j+1}=j!\,(2\as b_0)^j\;
\eeq
or roughly
\beq
j_{\rm min}\simeq\frac{1}{2\as \bz}\;.
\eeq
Asymptotic expansions are usually handled by summing their terms as
long as they decrease. Of course, in this resummation prescription
there is an ambiguity, which is of the order of the size of the first
neglected term. In our case
\beq
j_{\rm min}!\,(2\as b_0)^{j_{\rm min}}\approx e^{j\log j - j}\frac{1}{j^j}
=e^{-\frac{1}{2\as\bz}}\approx\frac{\Lambda}{Q}\;,
\eeq
which gives a power-suppressed ambiguity with the same power law
that we found using the cutoff procedure.

From the discussion given above, we would expect that the resummation
formulae should include a cutoff of the order of a typical hadronic
scale; varying the cutoff within a factor of order 1
should affect the cross section by terms of order $\Lambda/Q$.
In fact, this is not the case.
The cutoff has a dramatic effect on the cross section,
as can be seen from figs.~2 and 3
of ref.~\cite{Laenen92}. For example, the uncertainty band obtained
by varying the scale $\mu_0$ between $0.2m$ and $0.3m$, for top
production in the $gg$ channel, brings about a change in the cross section
by a factor of 2, for a top mass between 100 and 200~GeV.
These two values of $\mu_0$ correspond to cutting off the gluon
radiation at energies of the order of 20 to 30 GeV, therefore much
larger than a typical hadronic scale.

Other proposals for the resummation procedure have appeared in the
literature. In refs.~\cite{Contopanagos94,Alvero95} a method was
developed in the context of Drell--Yan pair production,
which was applied to the heavy-flavour case in
ref.~\cite{Berger95}. There,
as can be seen from formula (116) and the subsequent discussion,
unphysically large cutoffs are present, much larger than the typical
hadronic scale one would expect.

In the following section, we will show that the presence of
large cutoffs and of large ambiguities in the resummation formula
is not at all related to the blowing up of the coupling constant.
In other words, there are other sources of factorial growth
of the perturbative expansion for the resummation of soft gluons,
and they largely dominate the factorial growth due to the running
coupling. We will also show that these terms are spurious, and that
soft-gluon resummation can be easily formulated in such a way that these
terms are not present.

\subsection{Problems with the $x$-space resummation formula}\label{ProbSection}
For definiteness, let us focus upon the resummation
formula (\ref{ResLaenenhad}). We pointed out earlier that
this formula is divergent when the argument of $\as$ becomes
too small and the coupling constant blows up. In fact,
formula (\ref{ResLaenenhad}) is divergent, even for fixed coupling
constant. At fixed coupling it can be written as
\beqn
\sigma^{\rm (res)}&=& \frac{\as^2}{m^2}
\int d\tau\; {\cal L}(\tau)f^{(0)}(\hrho)
\exp\left[\as C\log^2(1-\hrho)\right]
\nonumber \\ \label{FixedCoupling}
&=& \frac{\as^2}{m^2} \int_\rho^1 d\hrho\;
\frac{\rho}{\hrho^2}{\cal L}\left(\rho/\hrho\right)f^{(0)}(\hrho)
\exp\left[\as C\log^2(1-\hrho)\right]\,,
\eeqn
where $\rho=4m^2/S$,
and the integral diverges as $\hrho\to 1$, since the exponential
\beq
\exp(a\log^2(1-\hrho))=(1-\hrho)^{a\log(1-\hrho)}
\eeq
grows faster than any inverse power of $1-\hrho$ as $\hrho\,\to\,1$.
This divergence can again be related to the factorial growth in the
perturbative expansion.
Expanding formula (\ref{FixedCoupling}) in powers of $\as$,
we get
\beq
\sigma^{\rm (res)}=\frac{\as^2}{m^2}\sum_{k=0}^\infty \frac{1}{k!}(C\as)^k
\int_\rho^1 d\hrho\;
\frac{\rho}{\hrho^2}{\cal L}\left(\rho/\hrho\right)
f^{(0)}(\hrho)\log^{2k}(1-\hrho)\;.
\eeq
It is now easy to see that, because of its singularity for $\hrho\to 1$,
the integral of $\log^{2k}(1-\hrho)$ grows like $(2k)!\,$.
Let us make here the simplifying assumption that
$f^{(0)}(\hrho)\approx\theta(1-\hrho)\,$. In a neighbourhood
of the singularity, the integral behaves like
\beq
\int_{1-\epsilon}^1 d\hrho \, \log^{2k}(1-\hrho)
=\int_{\log 1/\epsilon}^\infty  dt\, e^{-t} t^{2k}\;,
\eeq
where we have performed the substitution $t=|\log(1-\hrho)|$.
The above integral equals
\beq
\int_{\log1/\epsilon}^\infty dt\; e^{-t} t^{2k}
=\int_0^\infty dt \; e^{-t} t^{2k}-\int_0^{\log1/\epsilon}  dt
e^{-t} t^{2k} = (2k)!\;
-\int_0^{\log 1/\epsilon}  dt\,
e^{-t} t^{2k}\; .
\eeq
Since
\beq
\int_0^{\log 1/\epsilon}  dt\,
e^{-t} t^{2k} < (\log 1/\epsilon)^{2k+1}
\eeq
we see that the contribution to the integral near the singularity
is dominated by the term $(2k)!\,$. The power
expansion for the cross section is then
\begin{equation}\label{factgr}
\sigma^{\rm (res)}\approx \sum_{k=0}^\infty \frac{(2k)!}{k!}
\left( C\as\right)^k \approx
\sum_{k=0}^\infty k!
\left(4  C\as\right)^k \;.
\end{equation}
The above formula is in fact appropriate for the case of the heavy-flavour
cross section with a lower cut on the invariant mass of the pair\footnote{
In the case of the total cross section, using the known behaviour of
$f^{(0)}$ near threshold, one would get an extra factor of 4/9 in front of
$C$ in eq.~(\ref{factgr}).}.
As in the previous example, the factorial growth of formula~(\ref{factgr})
will give rise to ambiguities in the resummation of the perturbative
expansion. These ambiguities are not, however, related to renormalons,
since they occur also at fixed coupling constant. In fact, they are in
general fractional powers of $\Lambda/Q$, where $Q$ is the scale
involved in the problem, and $\Lambda$ is a typical hadronic scale.
For example, as shown in ref.~\cite{Catani96a},
in the case of heavy-flavour production through the gluon-fusion mechanism,
at fixed invariant mass of the heavy-quark pair,
the ambiguity would have the form $(\Lambda/Q)^{0.16}$,
and it would thus be extremely relevant, even for very massive
heavy-quark pairs.

These large terms in the perturbative expansion are in fact spurious.
They are an artefact of the $x$-space resummation procedure.
This can be easily understood with the following argument.
Exponentiation of the gluon emission is possible because,
roughly speaking, each soft gluon is emitted independently.
However, this independence is only approximate: the total momentum
must be conserved. Momentum conservation, however, is a subleading
effect in the soft resummation formula. Yet, its violation leads
to factorially growing terms. These terms are subleading from
the point of view of the logarithmic behaviour, but very important
from the point of view of the factorial growth of the perturbative expansion.
The presence of large factorial terms due to momentum non-conservation
can be understood also by simple arguments. The emission of $k$
gluons, where each gluon has a limit on its energy $E_i<\eta$,
leads to a phase space that is larger by a factor of $k!$
than the case when the total energy of emission is bounded
$\sum E_i<\eta$. Thus phase space alone provides a $k!$ factor.
We can see in more detail the origin of the $(2k)!\,$ term
by considering the formula for the partonic cross section
with two emitted soft gluons, implementing momentum conservation.
We have
\beq\label{twosoft}
\sigma^{(2)}(\hrho)=\frac{1}{2}(2C\as)^2\frac{\as^2}{m^2}
\int \left[\frac{\log(1-z_1)}{1-z_1}\right]_+
     \left[\frac{\log(1-z_2)}{1-z_2}\right]_+
f^{(0)}({\hrho}^\prime)\,\delta(\hrho-z_1 z_2{\hrho}^\prime)\,
dz_1 dz_2 d{\hrho}^\prime\,.
\eeq
The leading-logarithmic term of the above integral is given by
\beq
\sigma^{(2)}(\hrho)=\frac{1}{2}(2C\as)^2\frac{\as^2}{m^2}
\log^4(1-\hrho)\,f^{(0)}(\hrho)+\ldots\,,
\eeq
where terms with less than 4 powers of logarithms are neglected.
We now see that the integral of the leading-logarithmic term
of $\sigma^{(2)}(\hrho)$ in $\hrho$ has a large factor $\approx 4!$
due to the integral of $\log^4(1-\hrho)$, while the integral
of the full expression, eq.~(\ref{twosoft}), gives
\begin{equation}
  \label{fullint}
  \int_0^1 d\hrho\, \sigma^{(2)}(\hrho)=
\frac{1}{2}(2C\as)^2\frac{\as^2}{m^2}\left(\int_0^1 dz
\left[\frac{\log(1-z)}{1-z}\right]_+\right)^2 \int_0^1 d{\hrho}
f^{(0)}(\hrho)\;=0\,,
\end{equation}
since the integrals of the soft emission
factors vanish with the $+$ prescription.
In general, we see that if we take generic moments of $\sigma^{(k)}(\hrho)$
(i.e. $\int \hrho^m d\hrho \,\sigma^{(k)}(\hrho)$),
the leading-log expression grows like $(2k)!\,$, while the full expression
grows only geometrically with $k$.

The criticism described so far applies to the calculations
of soft-gluon effects in heavy-flavour production given in refs.~\cite{Laenen92,Laenen94}
and \cite{Berger95}. As one may expect, the large factorial
terms give rise to large corrections to the cross section.
Since the terms of the perturbative expansion grow strongly with the
order, they also give large uncertainties. In refs.~\cite{Laenen92,Laenen94},
the presence of large uncertainties is in fact recognized.
In ref.~\cite{Berger95} it is claimed that the uncertainties are
small. Even there, however, an unphysically large cutoff is needed
in order to make sense out of the resummation formulae.

A second, more subtle problem with resummation prescriptions has to do with
the presence of $1/Q$ corrections that arise from infrared renormalons.
It was shown in ref.~\cite{Beneke95} that soft-gluon resummation
does not yield the correct renormalon structure of the Drell--Yan cross
section.
More precisely, it was shown that in a fully calculable model
the soft-gluon approximation yields a $1/Q$ correction, while
in the exact result $1/Q$ effects are absent.
This result suggests the absence of $1/Q$ corrections in Drell--Yan
cross sections, an issue that is still much debated in the literature
\cite{Korchemsky95,Dokshitzer95,Akhoury95,Akhoury95a,Akhoury96,Beneke96,
Korchemsky96}.
We would like to remark, however, that even if $1/Q$ corrections were present
in Drell--Yan and heavy-flavour production, they would
only be of the order of 1\% for top production at the Tevatron,
and also that the result of ref.~\cite{Beneke95} shows that
the soft-gluon approximation cannot describe them adequately.

It is possible to formulate the resummation
of soft gluons in such a way that the kinematic constraints are
explicitly satisfied, and no factorial growth arises
in the perturbative expansion. It is enough to formulate
the resummation problem in the Mellin transform space\footnote{In fact,
resummation formulae are usually derived in Mellin space.}.

In ref.~\cite{Catani96a} a resummation
prescription using the Mellin space formula is given.
It is shown there that there are, with this prescription,
no factorially growing terms
in the resummed perturbative expansion. Yet, soft effects are
consistently included. This prescription is called
``Minimal Prescription'' (MP hereafter), because it does not
introduce large terms that are not justified by the soft-gluon
approximation.

\subsection{Phenomenological applications}
\begin{figure}[htb]
\centerline{\psfig{figure=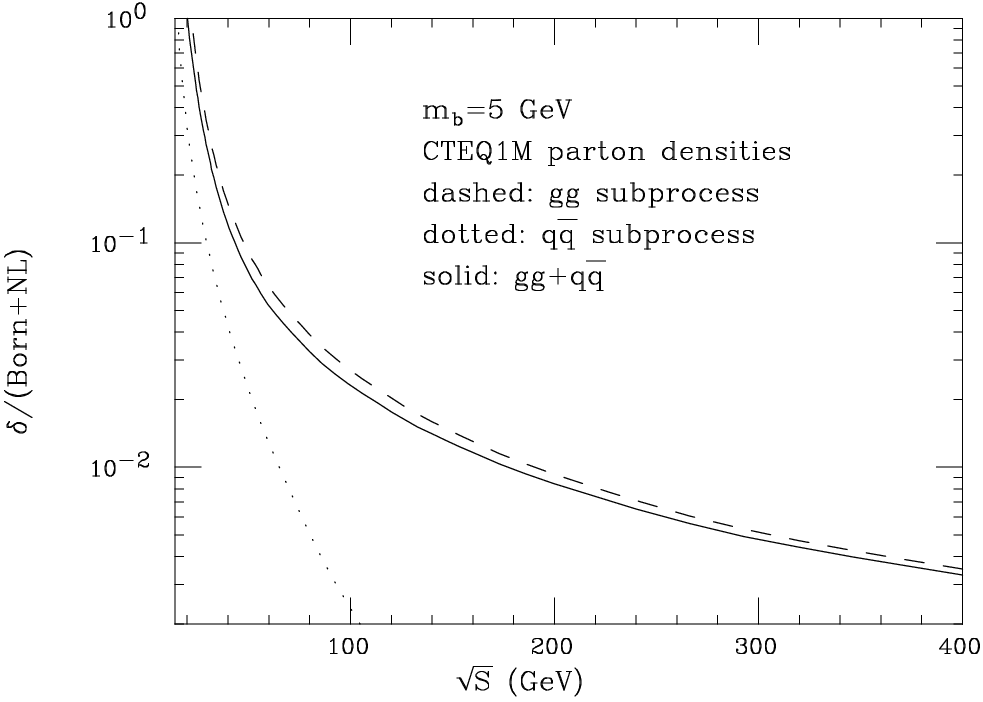,width=0.7\textwidth,clip=}}
\ccaption{}{ \label{bot-kfac}
Contribution of gluon resummation at order $\as^4$ and higher, relative to the
NLO result, for the individual channels and for the total,
for bottom production
as a function of the CM energy in $pp$ collisions.}
\end{figure}

\begin{figure}
\centerline{\epsfig{figure=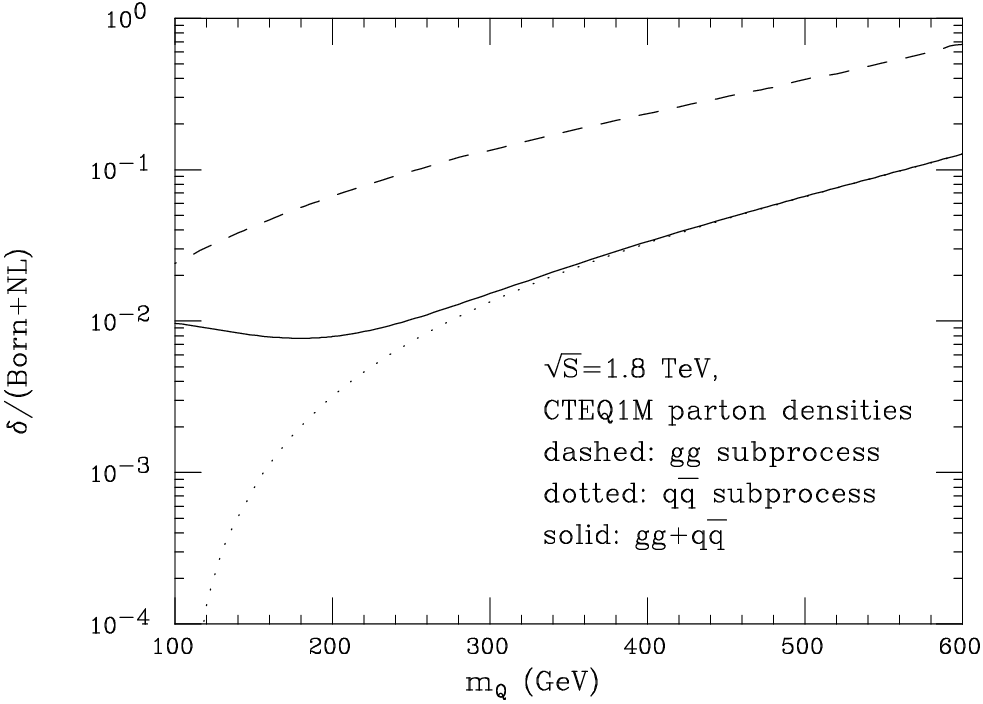,width=0.7\textwidth,clip=}}
\ccaption{}{ \label{frtev}
Contribution of gluon resummation at order $\as^4$ and higher, relative to the
NLO result, for the individual subprocesses and for the total,
as a function of the quark mass in $p\bar p$ collisions at 1.8 TeV. }
\end{figure}

\begin{figure}
\centerline{\epsfig{figure=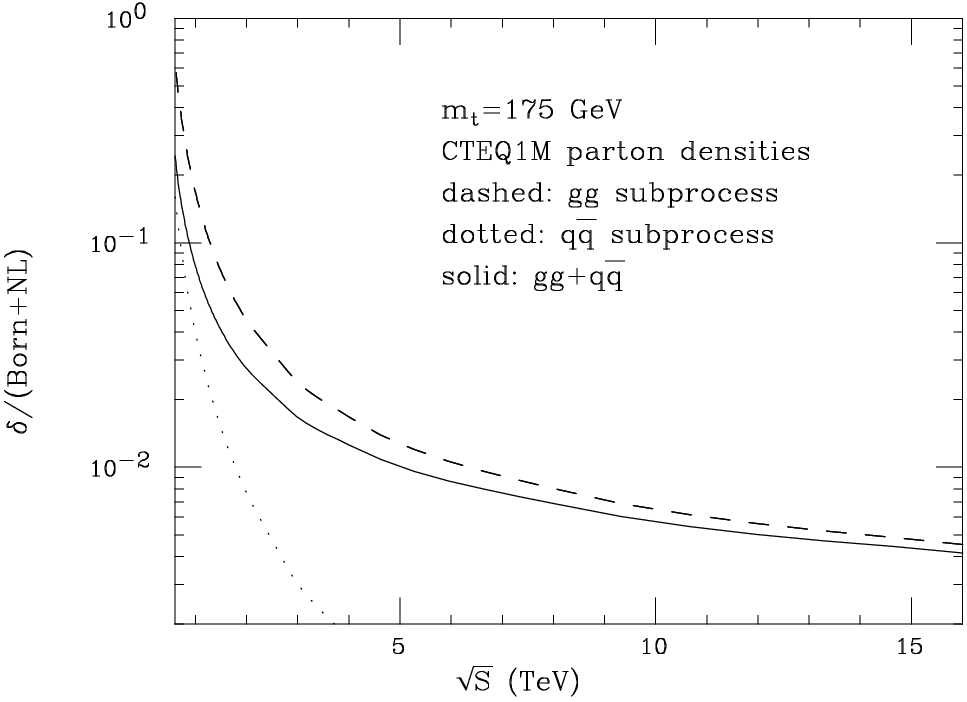,width=0.7\textwidth,clip=}}
\ccaption{}{ \label{frlhc}
Contribution of gluon resummation to the top cross section
at order $\as^4$ and higher, relative to the
NLO result, for the individual channels and for the total,
as a function of the CM energy in $pp$ collisions. }
\end{figure}

As stated earlier, when using the MP approach, one finds
negligible resummation effects in most experimental configurations
of interest.
In figs.~\ref{bot-kfac}, \ref{frtev} and \ref{frlhc},
we plot the quantities
\beq \label{deltadef} \frac{\delta_{\rm
    gg}}{\sigma^{(gg)}_{\rm NLO}}\,,\quad \frac{\delta_{\rm
    q\bar{q}}}{\sigma^{(q\bar{q})}_{\rm NLO}}\,,\quad
\frac{\delta_{\rm gg}+\delta_{\rm q\bar{q}}}{\sigma^{(gg)}_{\rm
    NLO}+\sigma^{(q\bar{q})}_{\rm NLO}}\;.
\eeq
Here $\delta$ is equal
to the MP-resummed hadronic cross section in which the terms of order
$\as^2$ and $\as^3$ have been subtracted, and $\sigma_{\rm NLO}$ is
the full hadronic NLO cross section. Thus, these plots show how important
the impact of resummation is beyond the already computed NLO terms.
The results for $b$ production at the Tevatron can easily be inferred
from fig.~\ref{bot-kfac}, since the $q\bar{q}$ component is negligible
at Tevatron energies.

For top production (figs.~\ref{frtev} and \ref{frlhc}),
we see that in most configurations of practical
interest, the contribution of resummation is very small, being of the order
of 1\% at the Tevatron.
A complete review of top-quark production at the Tevatron,
based upon these findings, has been given in
ref.~\cite{Catani96}. We also observe that, for top production at the LHC,
soft-gluon resummation effects are equally negligible.
Of course, in this last case,
there are other corrections, not included here, that may need to be
considered. Typically, since the values of $x$
involved are small in this configuration, one may have to worry
about the resummation of small-$x$ logarithmic effects \cite{Ellis90,Catani90,Catani91,Collins91}.

We see from the figures that in most experimental configurations of
interest these effects are fully negligible. One notable
exception is $b$ production at HERA-B, at $\sqrt{S}=39.2$~GeV, where
we find a 12\% increase in the cross section.
%
%
%
This correction is however well below the uncertainty due to higher-order
radiative effects. For example, from the NLO calculation
with the MRSA$^\prime$ \cite{Martin94} parton
densities and $m_b=4.75\,$GeV, we get
$\sigma_{b\bar{b}}=10.5{ +8.2  \atop -4.7}\,$nb,
a range obtained by varying the renormalization
and factorization scales from $m_b/2$ to $2m_b$. Thus
the upper bound is 80\% higher than the central value, to be compared with
a 10\% increase from the
resummation effects. This result is much less dramatic than
the results of ref.~\cite{Kidonakis95}, obtained using the
resummation prescription outlined in \cite{Laenen92}.
\section{Heavy-flavour production in $e^+e^-$ collisions}
\sethead{Heavy-flavour production in $e^+e^-$ collisions}
\subsection{Preliminaries}
Heavy-flavour physics is a considerable part of the
physics programme in $e^+e^-$ collisions, both at LEP and the SLC.
An entire chapter of this book is devoted to this topic \cite{Kuhn97}.
We shall limit ourselves here to
some QCD aspects of the production dynamics. In particular we will discuss
the most recent studies of the $b$ fragmentation function,
the quark--antiquark correlations, and the gluon-splitting
production mechanism.

We will not discuss specific topics related to the LEP2 programme.
We would like, however, to point out that LEP2 offers the possibility
of studying heavy-flavour production in photon--photon collisions.
Relevant NLO computations have been performed
for both the $\gamma\gamma$ \cite{Drees93,Kraemer96} and the $e\gamma$ cases
\cite{Laenen96}. Some phenomenological studies have already begun
\cite{Aurenche96}.
A calculation of heavy-flavour production at high transverse momentum
has also been presented \cite{Cacciari96a}.
\subsection{Fragmentation function}
\begin{figure}[htb]
\centerline{\epsfig{figure=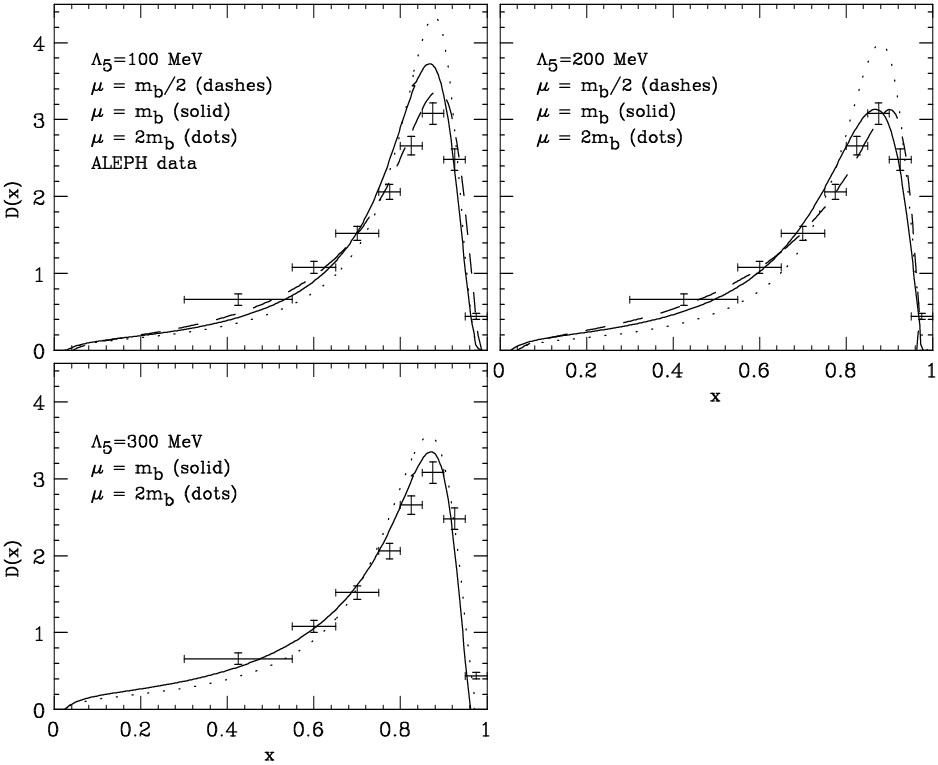,width=0.9\textwidth,clip=}}
\ccaption{}{ \label{fig:bfrlep}
Theoretical prediction for the $b$ fragmentation functions at LEP
together with ALEPH data. Only the statistical error is shown.}
\end{figure}

\begin{figure}[htb]
\centerline{\epsfig{figure=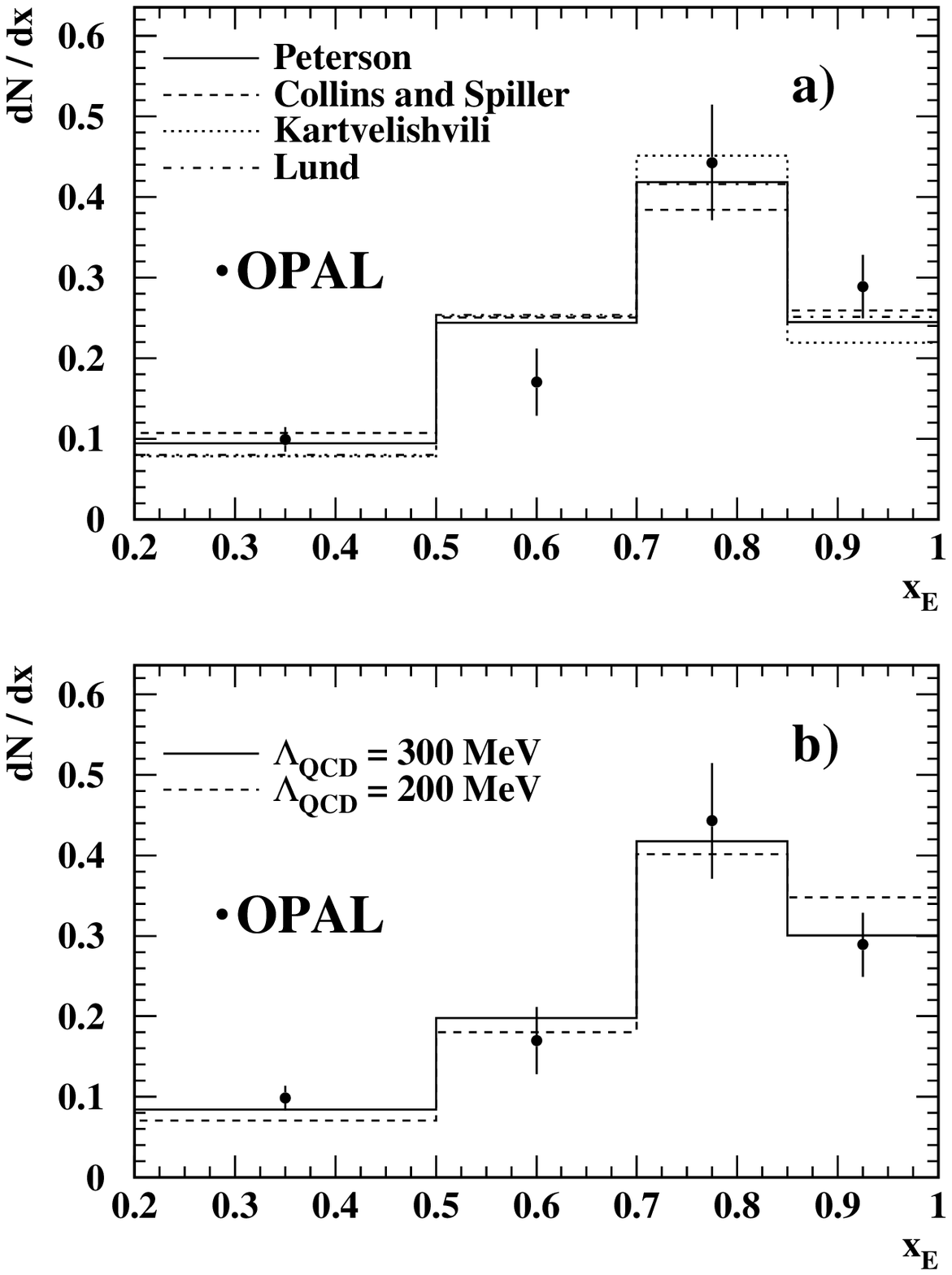,width=0.6\textwidth,clip=}}
\ccaption{}{ \label{fig:Opalbfrag}
OPAL results for the $b$ fragmentation function at LEP.}
\end{figure}

Theoretical predictions for fragmentation functions rely essentially
upon two ingredients: perturbation theory, and a model for
non-perturbative fragmentation effects.
A comprehensive discussion of these topics is given in ref.~\cite{Nason92}.
Leading-order formulae have been available for a long time
\cite{Azimov82,Azimov84}, while the computation of the NLO
perturbative corrections is given in refs.~\cite{Mele90,Mele91}.
Theoretical studies of the effects of non-perturbative physics are
given in refs.~\cite{Colangelo92,Randall95,Nason96a}.
Here we would like to call
attention to the study of ref. \cite{Colangelo92},
where a definite prediction
was made for the $b$-fragmentation function at LEP for various values
of $\LambdaQCD$. In fig.~\ref{fig:bfrlep} we report
the predictions of ref.~\cite{Colangelo92} together with the data
of ref.~\cite{Buskulic95}. Observe that the same figures were
reported in ref.~\cite{Nason92}, together with some unpublished L3
data, and at that time the agreement with theory seemed quite poor.
Interestingly enough, the position of the peak is well reproduced by the QCD
calculation. In general, a value of $\Lambda^{(5)}_{\sss QCD}$
between $200$ and $300\;$MeV gives an adequate fit of the data.
In ref.~\cite{Alexander96} a more detailed fit was performed, using
the calculation of ref.~\cite{Colangelo92}. The results are shown
in fig.~\ref{fig:Opalbfrag}.
There we can see that commonly used parametrizations fail to reproduce
the slope of the fragmentation function, while the QCD result
is quite consistent with the data.
It also turns out that for the larger values of $\Lambda$ the non-perturbative
part of the initial condition is strongly peaked towards large values of
$x$. This means that, if we can assume
that $\Lambda^{(5)}_{\sss QCD}\ge 0.25\;$GeV, the fragmentation function
is largely perturbative, that is to say, is mostly made up of QCD evolution
and soft-gluon emission effects \cite{Mele91}.
The results of LEP measurements of the fragmentation functions
are important also in view of possible applications to collider physics.
The problems that are encountered in large transverse momentum
heavy-flavour production at hadron colliders may be partially due
to a lack of proper treatment of the fragmentation effects.
The study of the fragmentation function has concentrated
in the past on the determination of its second moment (i.e.
the average momentum of the $B$ meson), as measured in $e^+e^-$
collisions.
On the other hand, as pointed out in subsection
\ref{Inclusive-Bottom-Production},
$b$ production at hadron colliders is also sensitive
to higher moments of the fragmentation function.
Since the new data from LEP experiments are now also sensitive
to higher moments, it would
be interesting to perform a new computation of the heavy-flavour
production cross section at large transverse
momentum, using fragmentation functions that fit the LEP data adequately.
\subsection{Heavy-quark production via gluon splitting}
The importance of the measurement of $\Gamma_b$ (the width of the $Z^0$ into
$b\bar{b}$) has prompted several experimental and theoretical
studies of $b$-production characteristics
(see, for instance, \cite{Kuhn97}).
One important issue is to understand how often $b\bar{b}$ or
$c\bar{c}$ pairs are produced indirectly, via a gluon splitting mechanism.
Several theoretical studies are available on this topic
\cite{Nason92,Mueller86,Mangano92a,Seymour95}.
Experimental studies on charm production via gluon splitting
have been presented in refs. \cite{Akers95,Akers95a,Hansper96},
and a first measurement of $g\to b\bar{b}$ has been given in
\cite{Branchini96}.
The reported values are given in table~\ref{tab:gqq}.
\begin{table}
\begin{center}
\begin{tabular}{|c|c|c|} \hline
& $\bar{n}_{g\to c\bar{c}}$ (\%) &   $\bar{n}_{g\to b\bar{b}}$  (\%)\\ \hline
 \cite{Akers95a} &  $  2.27 \pm 0.28 \pm 0.41 $ & \\ \hline
\cite{Hansper96} &  $ 2.65 \pm 0.74 \pm  0.51 $ & \\ \hline
\cite{Branchini96} & &  $ 0.22 \pm 0.10 \pm 0.08 $ \\ \hline
\cite{Seymour95} & & \\
$\lambdamsb=150\,$MeV & \addsp{$ 1.35{+0.48 \atop -0.30} $} & $ 0.20 \pm 0.02$
\\
$\lambdamsb=300\,$MeV & \addsp{$ 1.85{+0.69 \atop -0.44} $} & $ 0.26 \pm 0.03$
\\ \hline
\end{tabular}
\ccaption{}{\label{tab:gqq}
Fraction of events containing $g \to c\bar{c}$ and $g \to b\bar{b}$
subprocesses in $Z$ decays, as measured by the various collaborations,
compared with theoretical predictions.
The central values for the theoretical predictions are obtained
with $m_c=1.5$ and $m_b=4.75\,$GeV, the upper limits
with  $m_c=1.2$ and $m_b=4.5\,$GeV, and the lower limits
with  $m_c=1.8$ and $m_b=5\,$GeV.
}
\end{center}
\end{table}
In ref.~\cite{Seymour95} an explicit calculation
of these quantities has been performed.
Using these results\footnote{We thank M. Seymour for providing us with
the relative FORTRAN code.} we computed the charm and bottom
multiplicities for different values of the masses and of $\lambdamsb$.
We report them in table~\ref{tab:gqq}. As can be seen,
the averaged experimental result of $2.38\pm 0.48$\% \cite{Przysiezniak96}
is consistent with the upper range of the theoretical prediction,
preferring lower values of the quark mass, and larger values of $\lambdamsb$.
A similar problem, although much more severe, was mentioned in
subsection \ref{sub:hqjet}, in relation to charm jets in hadronic
collisions.

As reported in ref.~\cite{Seymour95},
Monte Carlo models are in qualitative agreement with these results,
although the spread of the values they obtain is somewhat larger
than the theoretical error estimated by the direct calculation.
In particular, one finds that while HERWIG \cite{Marchesini92}
and JETSET \cite{Sjostrand87} agree quite
well with the theoretical calculation, ARIADNE \cite{Lonnblad92} is higher by
roughly a factor of 2, and thus is in better agreement with data.
\subsection{Correlations}
\begin{figure}
\centerline{\epsfig{figure=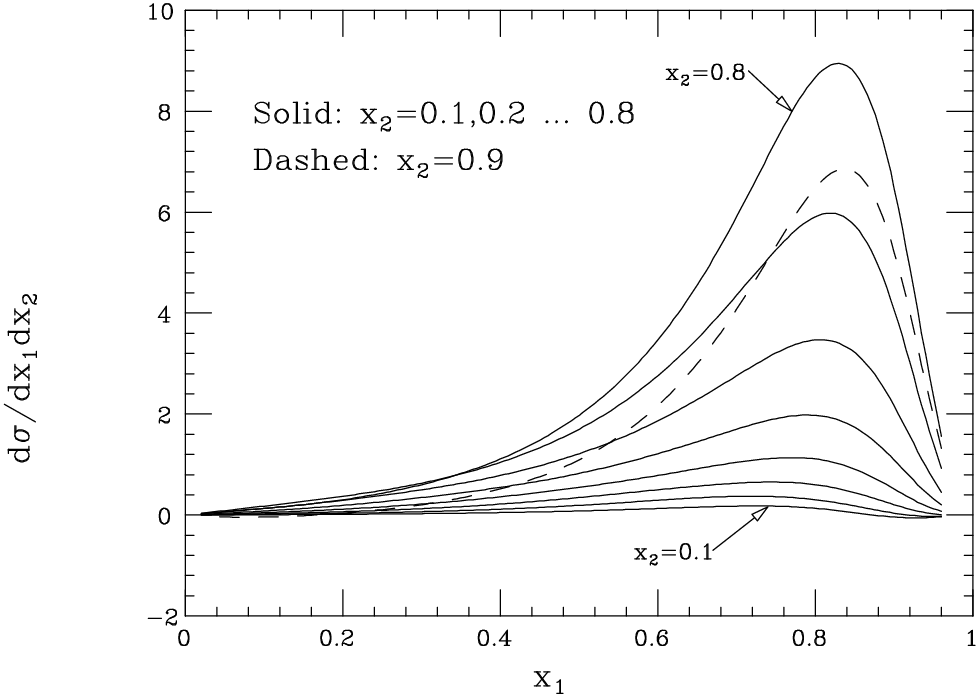,width=0.7\textwidth,clip=}}
\ccaption{}{ \label{sigdx1x2}
Double inclusive cross section $d \sigma/dx_1\,dx_2$, plotted as
a function of $x_1$ for several values of $x_2$. }
\end{figure}

Correlations in the kinematical properties of the $b$ and $\bar{b}$
also affect the systematics of the measurement of $\Gamma_b$.
The most important effect is the correlation in the momentum
of the $b$ and $\bar{b}$ quark, since detection efficiency
is often strongly momentum-dependent. In ref.~\cite{Nason96}
an explicit leading-logarithmic calculation of the momentum correlation
is given. The main result of ref.~\cite{Nason96} is the ${\cal O}(\as)$
formula for the double-inclusive distribution
\beqn &&
\frac{d\sigma}{dx_1 d x_2}\;=\;
D(x_1)D(x_2)+\frac{2\as}{3\pi}
\int_0^1 dy_1 dy_2
\;\theta(y_1+y_2-1)\;\frac{y_1^2+y_2^2}{(1-y_1)(1-y_2)}\;\times
\nonumber \\ &&
\Bigg[ D\left(\frac{x_1}{y_1}\right)D\left(\frac{x_2}{y_2}\right)
\frac{1}{y_1 y_2}
\,\theta(y_1-x_1)\,\theta(y_2-x_2)
-D(x_1)D\left(\frac{x_2}{y_2}\right)\frac{1}{y_2}\,\theta(y_2-x_2)
\nonumber \\ &&
-D\left(\frac{x_1}{y_1}\right)D(x_2)\frac{1}{y_1}\,\theta(y_1-x_1)
+D(x_1)D(x_2)\Bigg]\,,
\eeqn
where $D(x)$ is the total heavy-flavour fragmentation function,
and $x_1$ and $x_2$ are the energy fractions of the two heavy flavoured
hadrons ($x=2E/\sqrt{S}$).
As an illustration, we plot in fig.~\ref{sigdx1x2}
the double-inclusive cross section
$d \sigma/dx_1\,dx_2$ as
a function of $x_1$ for several values of $x_2$.
We use the Peterson parametrization given in eq.~(\ref{Peterson-form})
with $\ep=0.04$, which gives $\VEV{x}=0.70$
and $\as=0.12$.
The positive momentum correlation is quite visible in fig.~\ref{sigdx1x2}.
As $x_2$ increases, the peak of the distribution in $x_1$ also
moves towards larger values.
A particularly interesting quantity is the average momentum
correlation, defined as
\begin{equation}
r=\frac{\langle x_1 x_2\rangle - \langle x\rangle^2}{\langle x\rangle^2}\; .
\end{equation}
This quantity is independent of the fragmentation function, and has therefore
an expansion in the strong coupling with finite coefficients,
starting at order $\as$. In fact, one gets (in the limit $m^2/S\to 0$)
\begin{equation}
r\;=\;0.1\as + {\cal O}(\as^2)\; .
\end{equation}
Since the correlation is so small, non-perturbative
effects may be competitive with the perturbative ones.
For example, if corrections
of the order of $\Lambda_{\rm \small QCD}/\sqrt{S}$ were present,
this would be the case at LEP energies.
In ref.~\cite{Nason96a} it was shown that in the renormalon
approach power corrections to correlations are suppressed
at the level of $\Lambda^2_{\rm \small QCD}/S$ at least.
This result, being based upon the renormalon approach,
is not fully conclusive. It is however encouraging, since it
suggests that the perturbative calculation should be very reliable
at LEP energies.

\section{Conclusions and outlook}
We collect here the main conclusions
of the various sections, and present our outlook on future progress in
the field.

The experimental results on total cross sections for charm production
at fixed target are in good agreement with NLO QCD.  In the case of
hadroproduction the experimental accuracy is far superior to the
theoretical accuracy, and might ultimately help in better pinning down
the values of the theoretical parameters. For example, current data in
pion- and proton-induced reactions strongly disfavour a
charm mass value as large as 1.8~GeV.  Similar conclusions could not
be drawn from a study of the charm photoproduction data. This is
because, although the general precision of the data is better than the
overall theoretical uncertainty, data from different experiments have
a large spread, not obviously compatible with the statistical and
systematic uncertainties quoted by the single experiments. Within the
large theoretical and experimental uncertainties, the agreement with
theoretical expectations is nevertheless good, and consistent with a
charm mass value of 1.5~GeV, and with parameters such as $\lambdamsb$
and PDF sets favoured by the hadroproduction studies.

Bottom production in fixed-target experiments is affected by the low
statistics available. Even within the statistical uncertainties, there
are however conflicting results in the present data. The agreement
with theory, afflicted by equally large uncertainties, is therefore at
this time not particularly enlightening. These conflicts should be
resolved by the current and next generation of high-sensitivity
experiments (at FNAL and HERA-B).

While the theoretical description of the total production rates is
expected to be insensitive to possible non-perturbative aspects of the
production mechanisms, differential distributions for charmed hadrons
produced in fixed-target experiments provide an important probe of
these phenomena. The overall conclusions of the studies we presented
can be summarized as follows.  The inclusive $\pt$ spectra predicted
by NLO QCD for charm photoproduction are too hard to agree with the
data; they show clear evidence of a non-perturbative fragmentation
mechanism that slows down the charmed hadrons. If this fragmentation
is then applied to charm hadroproduction, the spectra turn out to be
far too soft. The apparent inconsistency can be solved by invoking the
effect of an intrinsic non-perturbative $k_{\sss T}$ kick of the
partons inside the hadron. This $k_{\sss T}$ kick makes the $\pt$
spectrum harder, an effect that is stronger in hadroproduction than
in photoproduction because of the presence of two hadrons instead of
one in the initial state. The need for a substantial $k_{\sss
  T}$ kick, of the order of 0.7 to 1~GeV, is corroborated by the study
of azimuthal correlations of charm pairs, both in photo- and in
hadroproduction. There remain indications that a non-perturbative
fragmentation slightly softer than the standard Peterson
parametrization would provide a better agreement with data.
Nevertheless the overall picture is that once these two main effects
are accounted for, the qualitative and quantitative features of most
of the available distributions are properly described by the theory,
independently of the nature of the beam.

Additional non-perturbative effects, such as colour-drag from the
beam, should be invoked to properly describe the $x_{\sss F}$
distribution at large $x_{\sss F}$ and $D/\overline{D}$ asymmetries.
The modelling of colour-drag effects contained in the available Monte
Carlo programs can be properly tuned to provide an acceptable
description of the data.

Heavy-quark photo- and electroproduction studies at HERA are still in
an early stage, with much more statistics waiting to become available
in the near future. The current data on charm total cross sections
agree with the NLO QCD calculations for choices of parameters
consistent with low-energy fixed-target results. However, the current
HERA experimental accuracy and the spread of the low-energy
photoproduction data are not sufficient as yet to improve our
knowledge of the gluon densities of either the proton or the photon.
Differential distributions in $\pt$ and in rapidity have also been
presented, and a general agreement with the theory was found. Some
discrepancies, as found for example in the rapidity distributions, are
not statistically compelling as yet.

Progress in the study of the gluon density of the proton will come
when higher statistics, and in particular large samples of charm pairs,
will be available.  Improved efficiencies at large rapidity and higher
statistics will likewise help the understanding of the gluon structure
of the photon.

The study of bottom production at the Tevatron shows that there is
good agreement between the shape of the $b$-quark \pt\ distribution
predicted by NLO QCD and that observed in the data for central
rapidities. A similar conclusion can be reached in the case of the
shape of the azimuthal correlations between $b$ quarks.  Although the
data rate is higher by a factor of approximately 2 relative to the
default choice of theoretical parameters, acceptable choices of
$\lambdamsb$ and of renormalization and factorization scales bring the
theory in perfect agreement with the data of UA1 and D0, and within
30\% of the CDF measurements.  It is encouraging that studies of
higher-order logarithmic corrections favour the choice of low values
for $\muR$ and $\muF$.  The CDF measurements at 630 and 1800~GeV
indicate that theory correctly predicts the scaling of the
differential $b$-quark \pt\ distribution between 630 and 1800~GeV, a fact
that had often been questioned in the past and now finds strong support.

More theoretical studies should be devoted to the understanding of the
non-perturbative fragmentation function for heavy quarks. We showed
that the interpretation of the experimental data on the $\pt$ spectrum
depends very strongly on the assumed shape of the fragmentation
function.  Measurements of the $b$-quark content of high-\et\ jets, a
quantity that is rather independent of the fragmentation properties
of the $b$ quark, indicate a good agreement between data and theory.
This suggests that residual discrepancies in the comparison of theory
and data for the $\pt$ spectra, and possibly for the azimuthal
correlations, could be resolved with a better understanding of the
fragmentation phenomena.

No surprise has so far come up in the comparison of the data
on top production with the predictions of NLO perturbative QCD.
Given the claimed accuracy of the theoretical predictions, which in the case
of the total cross section is of the order of 10\%, only additional
statistics will make these comparisons more compelling.
We remark here that, on the theoretical side,
soft-gluon resummation effects have been found to be much less
important than previously thought. Furthermore, the $\as$ dependence
of the top cross section is quite small, because of a compensation
mechanism due to structure-function evolution effects.
Thus, the theoretical error on the cross section is quite credible
in this case.
Detailed features of the structure of the final state in top
production, such as studies of the jet activity, have so far been
probed only indirectly in the context of the mass measurements. It is
likely that a lot will be learned in the future from these more
detailed analyses, and that significant progress will be made in their
MC modelling.

Considerable progress has been made in the study of heavy-flavour
production in $e^+e^-$ annihilation. The measurement of the $B$
fragmentation function at LEP has reached a remarkable level of accuracy.
Some details of the production mechanism, particularly important
for the determination of $\Gamma_b$ have received
special attention. Thus, the heavy-flavour production by
the gluon-splitting mechanism has been studied both experimentally
and theoretically, and analytical calculations of the correlations
of the heavy-flavour pair have been performed. Besides their intrinsic
value, $e^+e^-$ studies could clarify several aspects of the production
mechanism, which are important also at hadron colliders.

\par\noindent {\bf Acknowledgements}
\\
We wish to thank J. Appel, D. Barberis, T. Carter, R. Eichler, C. Grab,
C. Harris, D. Harris, K. Harrison, P. Koehn, E. Laenen, B. Osculati,
H. Przysiezniak, L. Rossi and M. Seymour for useful conversations,
for information relative to experimental results, for suggestions,
and for providing us with useful computer programs.
Part of the material presented in this review is taken from work
done by some of us in collaboration with S. Catani, G. Colangelo,
C. Oleari and L. Trentadue.
\eject

\end{document}